\documentclass[11pt]{article}
\usepackage[bottom]{footmisc}
\usepackage[T1]{fontenc}
\usepackage{microtype}
\usepackage{amsmath,amsthm}
\usepackage{aliascnt}
\allowdisplaybreaks[1]
\usepackage{newpxtext,newpxmath}
\usepackage[margin=15pt,font=small,labelfont={bf}]{caption}
\usepackage[margin=1in]{geometry}
\usepackage[utf8]{inputenc}
\usepackage{array}

\usepackage{csquotes}
\usepackage[english]{babel}
\usepackage[backend=biber,style=alphabetic,maxalphanames=3,minalphanames=3,maxbibnames=99,minbibnames=99]{biblatex}
\usepackage{graphicx}
\usepackage{tikz}
\usepackage{algorithm}
\usepackage[noend]{algpseudocode}

\AtBeginDocument{
  }
\usepackage{xcolor}
\usepackage[colorlinks=true,
    citecolor=darkgreen,
    linkcolor=darkred
]{hyperref}

\definecolor{darkgreen}{rgb}{0,0.5,0.5}
\definecolor{darkred}{rgb}{0.5,0,0}
\definecolor{sdpgreen}{rgb}{0,0.5,0}

\newcommand\drop[1]{}
\newcommand*{\newaliastheorem}[3]{%
  \newaliascnt{#1}{#2}%
  \newtheorem{#1}[#1]{#3}%
  \aliascntresetthe{#1}%
  \expandafter\providecommand\csname #1autorefname\endcsname{#3}%
}
\newtheorem{theorem}{Theorem}[section]
\newaliastheorem{proposition}{theorem}{Proposition}
\newaliastheorem{lemma}{theorem}{Lemma}
\newaliastheorem{corollary}{theorem}{Corollary}
\newaliastheorem{definition}{theorem}{Definition}
\newaliastheorem{conjecture}{theorem}{Conjecture}
\newaliastheorem{example}{theorem}{Example}
\newaliastheorem{remark}{theorem}{Remark}
\newaliastheorem{fact}{theorem}{Fact}
\newaliastheorem{claim}{theorem}{Claim}

\usepackage{thm-restate}

\newcommand{\EE}{\mathbb E}
\newcommand{\PP}{\mathbb P}

\title{A Sub-$3^n$ Upper Bound on the Number of Stable Matchings}
\author{Ken-ichi Kawarabayashi\thanks{National Institute of Informatics \& The University of Tokyo, Tokyo, Japan, \texttt{k\_keniti@nii.ac.jp}.\\ 
    Supported by JSPS Kakenhi 22H05001, JP25K24465, and by JST ASPIRE JPMJAP2302.}
    \and 
    Hikaru Wada\thanks{The University of Tokyo, Tokyo, Japan, \texttt{wada-hikaru@is.s.u-tokyo.ac.jp}.\\ 
    Supported by JSPS JP25K24465 and JST ASPIRE JPMJAP2302, and by Hirose Foundation Scholarship.}
    \and 
    Masataka Yoneda\thanks{The University of Tokyo, Tokyo, Japan, \texttt{e869120@gmail.com}.\\ 
    Supported by JSPS JP25K24465 and JST ASPIRE JPMJAP2302.}}
\date{}

\begin{document}
\maketitle

\pagenumbering{gobble}

\begin{abstract}
    We study the problem of determining the maximum number of stable matchings. This classical counting problem was posed by Knuth~\cite{knuth1976mariages} and has been studied for five decades.

    We obtain a new $2.999^n$ upper bound on the number of stable matchings with $n$ men and $n$ women. This improves the best-known previous result by Palmer and P{\'a}lv{\"o}lgyi~\cite{palmer2022most}. This result gets much closer to the well-known lower bound $\frac{2.28^n}{\text{poly}(n)}$.

    We refine their entropy-based framework by capturing broader structural relationships among stable matchings. To capture these relationships, we introduce new frameworks, including cones and supports. This establishes the $3.332^n$ bound. We also exploit relationships among local structures and use a charging argument to improve the bound to $2.999^n$.
\end{abstract}

\newpage
{
    \setcounter{tocdepth}{2}
    \addtocontents{toc}{\protect\linespread{1.08}\protect\selectfont}
    \tableofcontents
    \clearpage
}
\pagenumbering{arabic}
\setcounter{page}{1}

\section{Introduction}\label{sec:introduction}

\begin{figure}[!b]
  \centering
  \includegraphics[width=\linewidth]{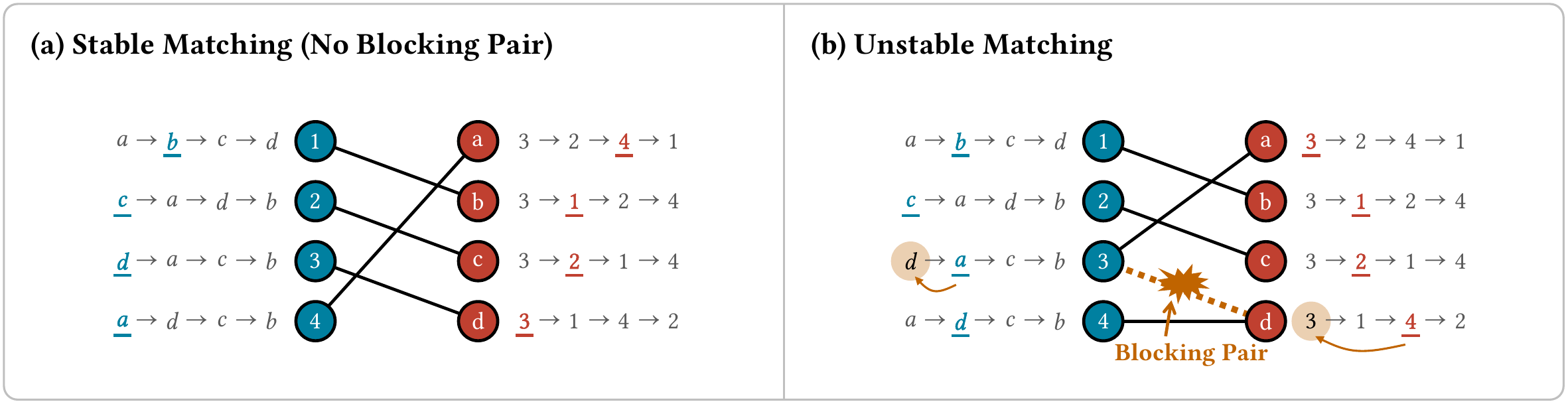}
  \caption{Examples of a stable matching and an unstable matching containing a blocking pair.}
  \label{fig:001}
\end{figure}

An instance of the \emph{stable matching problem} consists of two equally sized sets of individuals, conventionally called men and women, together with each individual's preference order over the members of the opposite set. A \emph{perfect matching} pairs men with women so that every man and every woman belongs to exactly one pair. A \emph{blocking pair} for a perfect matching is a man and a woman who each prefer the other to their current partner. A \emph{stable matching} is a perfect matching with no blocking pair. \autoref{fig:001} shows examples of stable and unstable matchings.

In 1962, Gale and Shapley proved that a stable matching always exists and proposed an algorithm to find one~\cite{gale1962college}. With $n$ men and $n$ women, the Gale--Shapley algorithm runs in $O(n^2)$ time.

\subsection{Research Background}

The stable matching problem models finding a good matching between two sets according to their members' preferences. It has many real-world applications, including matching medical residents to hospitals~\cite{roth1984evolution}. Shapley and Roth jointly received the 2012 Nobel Prize in Economic Sciences for their contributions to the theory of stable allocations and the practice of market design~\cite{nobel2012press}.

However, an instance may admit multiple stable matchings. This leads to the natural combinatorial problem of determining the maximum number of stable matchings in an instance with $n$ men and $n$ women. Knuth~\cite{knuth1976mariages} posed this classical problem, and it is one of the twelve open problems listed in Gusfield and Irving's book~\cite {gusfield1989stable}.
\subsection{Maximum Number of Stable Matchings}\label{subsec:background-related-work}

Let $f(n)$ denote the maximum number of stable matchings in an instance with $n$ men and $n$ women. In principle, $f(n)$ can be determined by enumerating all preference profiles for $n$ men and $n$ women and counting the stable matchings in each instance. However, there are $(n!)^{2n}$ preference profiles to examine, making this approach impractical as $n$ grows.

\paragraph{Maximum values for small \texorpdfstring{$n$}{n}.} For $n=4$, Knuth (1976) gave an instance with ten stable matchings \cite{knuth1976mariages}, and Eilers (1999) established $f(4)=10$ through an exhaustive computer search \cite{eilers1999irvine}. For $n=5$, the maximum is 16 \cite{eilers2022a357269}. In very recent preprints, Jo claims to have established that the maximum is 48 and 85 for $n = 6, 7$, respectively \cite{jo2026ordersix, jo2026orderseven}, and Xiu reports an alternative computational proof for $n=6$, together with a formalization of the entire proof in Lean \cite{xiu2026certificate}. However, for $n\geq 8$, the exact value of $f(n)$ remains unknown.

\paragraph{Asymptotic bounds.} In addition to determining the maximum number of stable matchings for individual values of $n$, researchers have studied the asymptotic behavior of $f(n)$.

\begin{figure}[!b]
  \centering
  \includegraphics[width=\linewidth]{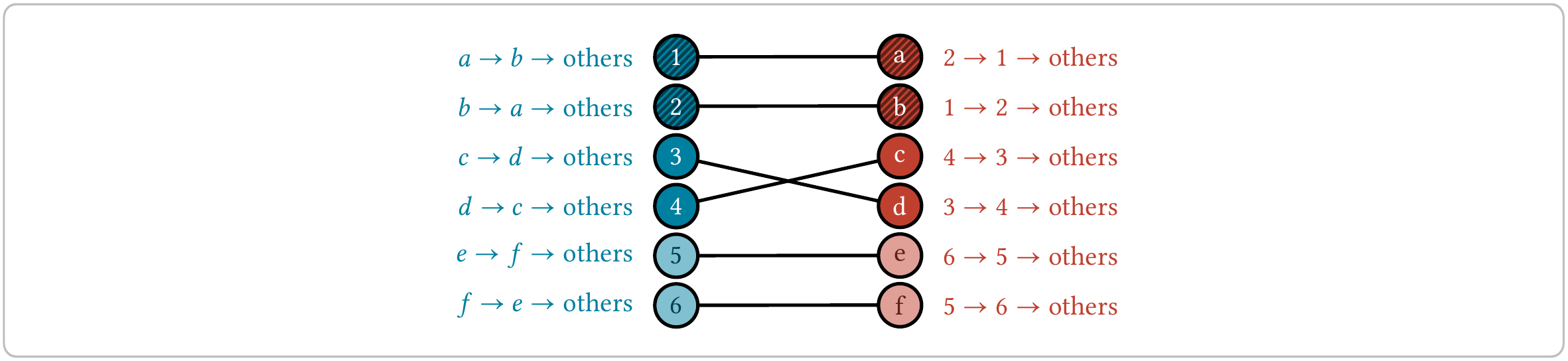}
  \caption{An instance with at least $2^{\lfloor n/2 \rfloor}$ stable matchings.}
  \label{fig:002}
\end{figure}

Several lower bounds on $f(n)$ have long been known. A simple construction (see \autoref{fig:002}) gives $f(n)\ge 2^{\lfloor n/2\rfloor}$. Irving and Leather (1986) constructed instances with many stable matchings when $n$ is a power of two \cite{irving1986complexity}. An analysis by Knuth, presented in the book by Gusfield and Irving, shows that these instances have at least $2.28^n/(1+\sqrt{3})$ stable matchings \cite{gusfield1989stable}. Moreover, Thurber (2002) extends the construction to general $n$, obtaining the lower bound $\frac{2.28^n}{\operatorname{poly}(n)}$ \cite{thurber2002concerning}. All of these lower bounds follow from explicit constructions.

Progress on upper bounds has accelerated more recently. In 2011, Stathopoulos proves that $f(n)=O(n!/c^n)$~\cite{stathopoulos2011variants}, where $c>1$ is a constant. This improves upon the trivial upper bound of $n!$ by an exponential factor. In 2018, Karlin, Oveis Gharan, and Weber prove that, for all sufficiently large $n$, $f(n)$ is at most $2^{17n}$~\cite{karlin2018simply}, the first exponential upper bound. Furthermore, in 2021, Palmer and P{\'a}lv{\"o}lgyi bound the number of stable matchings using the entropy of the uniform distribution on the set of stable matchings, claiming an upper bound of $3.55^n + O(1)$ on $f(n)$~\cite{palmer2022most}. However, at the end of their proof, we identify an error in their numerical evaluation~\cite[Claim~28]{palmer2022most}. Our recalculation yields an exponential base of slightly less than $4.01$, rather than the reported $3.55$; see \autoref{app:previous-numerical-evaluation} for details\footnote{This error is just from the last numerical evaluation, and other arguments seem correct. So we believe the significance of the paper still stands. Indeed, our proof is building on their proof.}. In summary, the current known bounds for $f(n)$ are
\begin{equation*}
    \frac{2.28^n}{\operatorname{poly}(n)} \leq f(n) \leq 4.01^n + O(1),
\end{equation*}
\noindent
which still leaves a multiplicative gap (ratio) between the exponential bases of approximately $1.76$.

\paragraph{Applications.} Studying the maximum number of stable matchings also has algorithmic applications. First, it can be applied to the problem of enumerating all stable matchings. Although counting stable matchings is \#P-complete \cite{irving1986complexity}, Gusfield gave an $O(n^2+n|\Omega|)$-time algorithm, where $\Omega$ denotes the set of stable matchings. Importantly, the running time depends on the number of stable matchings. This means that improving the upper bound on $f(n)$ of stable matchings also improves the time complexity bound for Gusfield's enumeration algorithm; e.g., $f(n) < 3^n$ gives a worst-case running-time bound of $O(n 3^n)$ for this enumeration algorithm.

Second, this enumeration algorithm can also be used to solve optimization problems over stable matchings, such as Sex-Equal Stable Matching. This problem asks for a stable matching that minimizes the absolute difference between the sum of the ranks of the men's partners in their preference lists and the corresponding sum for the women. Whether this problem admits a polynomial-time algorithm was also one of the twelve open problems listed in the book by Gusfield and Irving \cite{gusfield1989stable}, and Kato \cite{kato1993complexity} subsequently proved the NP-hardness. An optimal solution can be found by enumerating all stable matchings and evaluating the objective function for each. Thus, improving the upper bound on the maximum number of stable matchings also improves the worst-case running-time bounds for exact algorithms based on exhaustive enumeration.
\subsection{Our Contributions}\label{subsec:our-contributions}

In this paper, we further improve the upper bound to $2.999^n$. Formally, our main result is the following. 

\begin{theorem}
    \label{thm:intro-main}
    Every stable matching instance with $n$ men and $n$ women admits fewer than $2.999^n$ stable matchings.
\end{theorem}

This improves the corrected $4.01^n + O(1)$ upper bound obtained from \cite{palmer2022most}. The theorem holds for every $n\geq 1$, rather than only for sufficiently large instances, and requires no additional correction term. Combined with Gusfield's enumeration algorithm \cite{gusfield1987three}, the following corollary immediately holds.

\begin{corollary}
    \label{thm:intro-main-2}
    All stable matchings of an instance with $n$ men and $n$ women can be enumerated in $O(n \cdot 2.999^n)$ time in the worst case.
\end{corollary}

Our result also applies to the Stable Roommates problem, which is a non-bipartite variant of the stable matching problem (i.e., each of $n$ participants has a preference ranking of $n-1$ other people). The construction of Dean and Munshi \cite{dean2010faster,karlin2018simply} gives an injection from the stable matchings of a Stable Roommates instance with $n$ participants to those of a Stable Marriage instance with $n$ men and $n$ women. Thus, our result implies the following corollary.

\begin{corollary}
    \label{thm:intro-main-3}
    Every instance of the Stable Roommates problem with $n$ participants admits fewer than $2.999^n$ stable matchings.
\end{corollary}

Furthermore, our result also applies to the Popular Roommate problem, which asks for a matching that does not lose a majority vote against any other matching. This problem is known to be NP-hard even with complete preference lists \cite{cseh2021popular}. Kavitha gave an algorithm that decides whether a Popular Roommates instance with $n$ vertices admits a solution in $O^*((c+2\sqrt{c})^n)$ time, provided that every Stable Marriage instance with $k$ men and $k$ women has at most $c^k$ stable matchings for every $k \geq 1$~\cite{kavitha2025popular}. Therefore, since $c+2\sqrt{c} < 6.463$ at $c = 2.999$, our results imply the following corollary.

\begin{corollary}
    \label{thm:intro-main-4}
    There exists an $O^*(6.463^n)$-time algorithm to decide whether a Popular Roommates instance with $n$ vertices admits a popular matching.
\end{corollary}

Our proof for \autoref{thm:intro-main} builds on the framework of Palmer and P{\'a}lv{\"o}lgyi (2021) for bounding the number of stable matchings via entropy, incorporating additional combinatorial constraints arising from the rotation structure of stable matchings. In addition to bounding the number of candidates for each man separately, we obtain a stronger entropy bound by exploiting the frequencies of local structures that yield small entropy savings, as well as structural relationships among men and across different stable matchings.
\subsection{Technical Overview}

In this subsection, we present a technical overview of the proof of the upper bound $2.999^n$. The proof proceeds through three successive upper bounds: $3.332^n$, $3.060^n$, and $2.999^n$.

\subsubsection{The Random-Reveal Model by PP22}

We begin by introducing the \emph{random reveal model}, which is a technique used in \cite{palmer2022most}. We let $\{m_1, \dots, m_n\}$ be a set of men and $\{w_1, \dots, w_n\}$ be a set of women.

\begin{figure}[!b]
  \centering
  \includegraphics[width=\linewidth]{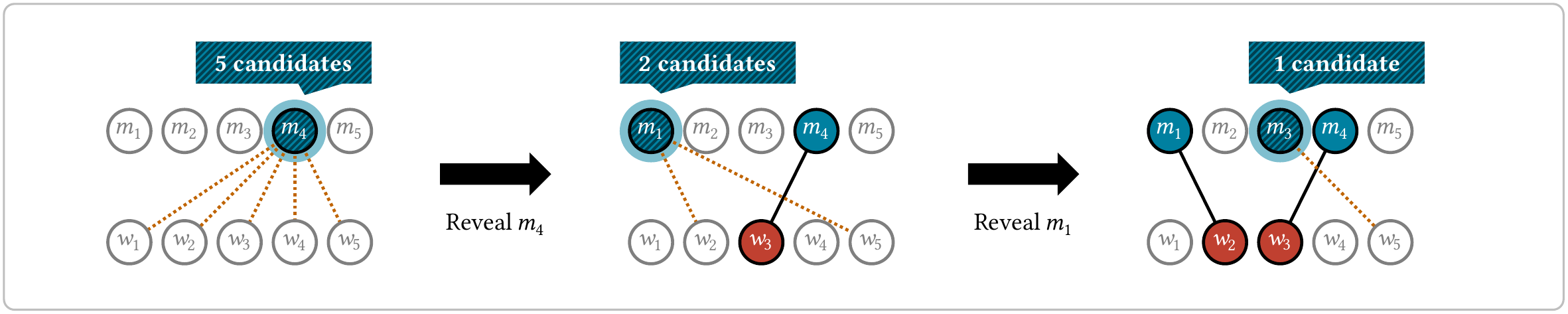}
  \caption{The sketch of the random reveal model. With each partner reveal, the number of remaining candidates decreases dramatically.}
  \label{fig:003}
\end{figure}

\paragraph{Basic idea.} Consider revealing the partner of each man $m$, say $\mu_m$, in random order (see \autoref{fig:003}). When revealing the first man's partner, $\mu_{m_{i_1}}$, at most $n$ women are candidates. However, the number of candidates decreases significantly as the reveal progresses. For example, for the second man $m_{i_2}$, there are at most $n-1$ remaining candidates for the partner $\mu_{m_{i_2}}$, but many of them can be inconsistent with the matching already revealed, because a stable matching should not have any blocking pair. Therefore, for a fixed man $m$, there are fewer candidates in expectation. If the expected logarithm of the number of candidates is less than $\delta$, an entropy argument yields an upper bound of $(e^\delta)^n$ on the number of stable matchings, e.g., $7.39^n$ when $\delta = 2$. However, it is difficult to derive a useful upper bound on $\delta$ directly, so they used the rotation of stable matching, which is well known and already introduced in \cite{gusfield1989stable}.

\begin{figure}[t]
  \centering
  \includegraphics[width=\linewidth]{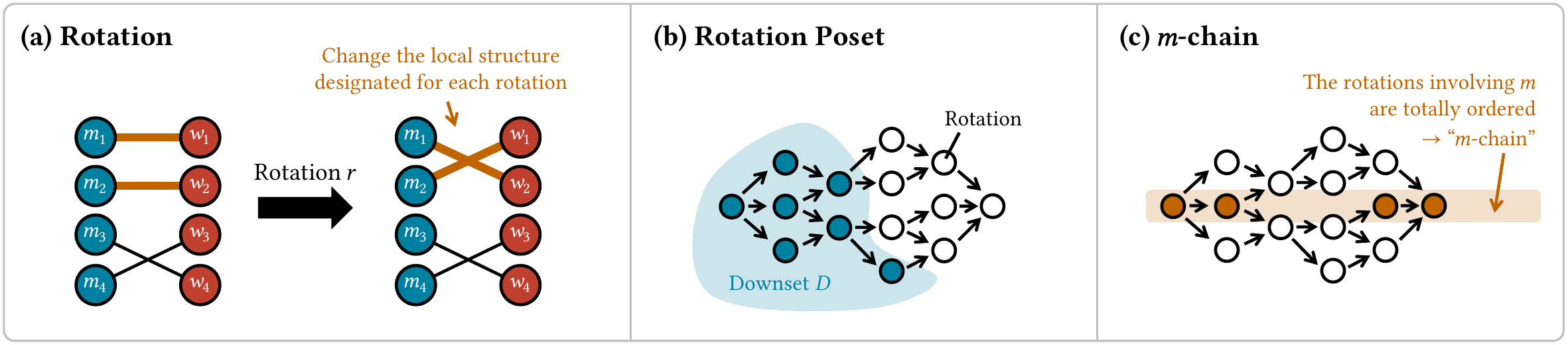}
  \caption{(a) The sketch of a rotation, where $r$ is the rotation operation that changes $(m_1, w_1), (m_2, w_2)$ to $(m_1, w_2), (m_2, w_1)$. (b) The sketch of the rotation poset. (c) The sketch of the $m$-chain.}
  \label{fig:004}
\end{figure}

\paragraph{Rotations in stable matchings.} A rotation is a local operation that changes the local structure of stable matchings; see \autoref{fig:004} (a) for the sketch, and \autoref{def:man-optimal}--\autoref{def:rotation} for a formal definition. Performing a rotation strictly worsens the partners of the men involved in the rotation, while leaving the other men's partners unchanged. The set of rotations forms a partially ordered set (poset) where each relation implies ``a rotation $r_2$ can be performed only after a rotation $r_1$'' (\autoref{fig:004} (b)). In addition, it is known that for each downset\footnote{A subset $(D \subseteq R)$ is a downset of a poset $(R, \preceq)$ if $(u \in D)$ and $(u' \preceq u)$ imply $(u'\in D)$.} of rotation $D$, performing its rotations from the man-optimal stable matching $\mu_{\text{m-opt}}$ in any legal order yields the same stable matching. Furthermore, the downsets and stable matchings form a bijection.

A more useful property holds in rotations. For each fixed man $m$, it is known that rotations involving $m$ form a chain, that is, for any two distinct rotations involving $m$, one must be performed before the other  (\autoref{fig:004} (c)). We call this $m$-chain, and each chain contains less than $n$ rotations because each rotation strictly worsens the partner of $m$.

\paragraph{Using \texorpdfstring{$m$}{m}-chain to derive upper bounds.} We show how to use the $m$-chain for the upper bound. Choose a stable matching $\mu$ uniformly at random, independently of the reveal order, and let $D$ be the corresponding downset (as written above, downsets and stable matchings form a bijection). Let $X_m$ be the number of rotations involving $m$ that are included in $D$. Then, the stable matching $\mu$ can be uniquely determined by $X_{m_1}, \dots, X_{m_n}$. This implies that the number of stable matchings is bounded by the maximum number of possible sequences $(X_{m_1}, \dots, X_{m_n})$.

Now, let us use the random reveal property. Consider revealing $X_{m_1}, \dots, X_{m_n}$ in a random order. For each man $m$, if the expected logarithm of the number of candidates for $X_m$ can be bounded by $\delta$, we obtain an $(e^\delta)^n$ upper bound for the number of stable matchings.

\paragraph{Limitations.} However, bounding the number of candidates for $X_m$ is difficult. Therefore, \cite{palmer2022most} uses both the men's $m$-chains and the women's $w$-chains, where the $w$-chain consists of the rotations involving woman $w$. Our recalculation of the estimate in \cite{palmer2022most} bounds the expected logarithm by $0.694$ for sufficiently large $n$, yielding the bound $(e^{0.694})^{2n} \approx 4.01^n$ because they used $2n$ chains. Palmer and P{\'a}lv{\"o}lgyi~\cite[Section~V]{palmer2022most} noted that reducing the number of components from $2n$ to $n$ is one possible route to improving the bound, although they considered a straightforward reduction unlikely.


\subsubsection{New Techniques for the \texorpdfstring{$3.332^n$}{3.332\string^n} Bound (\autoref{sec:cones-entropy})}

Next, we introduce our idea to reduce the number of chains to $n$ (i.e., using only the $m$-chains) and achieve the $3.332^n$ bound. The core is to introduce the \emph{cone} and the \emph{support} as follows.

\begin{figure}[!b]
  \centering
  \includegraphics[width=\linewidth]{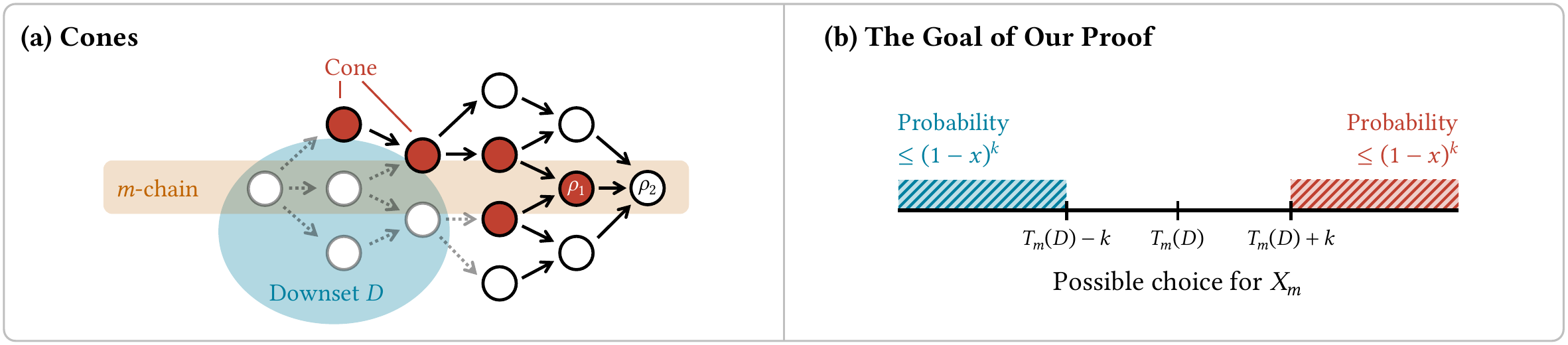}
  \caption{(a) An example of the first principal cone $K_1$, along with $\rho_1$ and $\rho_2$. (b) The sketch of our proof strategy.}
  \label{fig:005}
\end{figure}

\paragraph{Cones and supports.} Fix a downset $D$ and a man $m$. For each $k \geq 1$, let $\rho_k$ be the $k$-th next rotation involving $m$, not included in $D$. Then, for each $k \geq 1$, we define the $k$-th principal \emph{cone} $K_k$ as ``the set of rotations outside $D$ that must be performed to perform $\rho_k$'' (\autoref{fig:005} (a)). Note that $K_k$ includes $\rho_k$. In addition, define the \emph{support} of a cone as the set of men involved in at least one rotation in a cone, and let $s(k)$ be the size of the support of the principal cone $K_k$. Very importantly, $s(k) \geq k + 1$ always holds; see \autoref{lem:support-size} for the proof.

\paragraph{How cones help.} Next, we explain how to bound the number of candidates for $X_m$ using cones. Let $T_m(D)$ denote the actual value of $X_m$ for a downset $D$. Suppose that at least one man other than $m$ in the support of $K_k$ has already been revealed. Then, just before revealing $X_m$, we can rule out any candidate values at least $T_m(D) + k$, only by the previously revealed information. Assign each man $v$ an independent priority $U_v\sim\mathrm{Unif}[0,1]$, independently of $D$, and reveal the values in increasing priority order. Conditional on $D$ and $U_m=x$, each other man has independently been revealed with probability $x$, so
\begin{equation*}
  \mathbb{P}[T_m(D) + k \text{ remains a candidate for } X_m \mid D,U_m=x] \leq (1-x)^{s(k)-1} \leq (1-x)^k
\end{equation*}
\noindent
where the last inequality is due to $s(k) \geq k + 1$ that was explained above. The same also holds for the other side $T_m(D) - k$. Therefore, with some analysis in \autoref{subsec:geometric-benchmark}, we obtain a bound of $1.203572$ on the expected logarithm of the number of candidates for $X_m$, yielding the desired $(e^{1.203572})^n \approx 3.332^n$ bound.

\subsubsection{New Techniques for the \texorpdfstring{$2.999^n$}{2.999\string^n} Bound (\autoref{sec:histories-balance} and \ref{sec:charging})}

The crux of further improving the bound is to show the better support bound than $s(k) \geq k + 1$. For example, if $s(k) \geq k + 2$ holds for all $k \geq 1$, we would theoretically obtain a bound of $2.523^n$. However, $s(k) \geq k + 2$ is not always true. Therefore, we instead use the following strategies utilizing some local structure of cones and supports.

\paragraph {Three-person cones.} To improve the bound to $3.060^n$, we use the fact that if $s(1) = 2$ and $s(2) = 3$ for some men (which yields no savings at these two depths), then there are at least as many men satisfying $s(1) = 3$ and $s(2) \geq 4$. We prove this fact as follows. Let $m$ be a man that satisfies $s(1) = 2$ and $s(2) = 3$. Then, the only man $m'$ in the support of $K_2$ but not in the support of $K_1$ (with respect to $m$) satisfies $s(1) = 3$ and $s(2) \geq 4$. To complete the proof, we establish the injective correspondence between $m$ and $m'$ (\autoref{lem:image-properties}) and combine it with a bijection relating these structures across stable matchings.

\paragraph{Four-person cones.} To further improve the bound to $2.999^n$, we extend to the cones of support size 4. We consider various patterns such as $s(1) = 2$, $s(2) = 3$, and $s(3) = 4$. Then, we analyze the relationships among each case, as we did in the three-people case above. The key idea of the proof is a \emph{charging argument}. We assign each man a \emph{saving credit} that quantifies his contribution to the entropy savings (\autoref{lem:additional-savings}). Then, we charge the credit requirements of men with insufficient credits to men with sufficient available credits (\autoref{lem:charging-simple}--\autoref{lem:charging-image}), and show that all the required credits can be covered without exceeding the available credits (\autoref{lem:load-bound}--\autoref{lem:additional-credits-sum}).

\section{Preliminaries}\label{sec:preliminaries}

This section presents the basic properties of stable matching. For a comprehensive treatment of standard theory, see \cite{gusfield1989stable}. Note that when we use different definitions compared to \cite{gusfield1989stable}, we also cite the original papers.
\subsection{The Stable Matching}

First, we introduce the definition of stable matching.

\begin{definition}[Matching]
    Let $M = \{m_1, \dots, m_n\}$ be a set of men, and $W = \{w_1, \dots, w_n\}$ be a set of women. A perfect matching is a subset $\mu \subseteq M \times W$ in which each man and each woman belong to exactly one pair. We identify a perfect matching $\mu$ with a bijection from $M$ to $W$. We denote the partners of a man $m \in M$ and a woman $w \in W$ by $\mu(m)$ and $\mu^{-1}(w)$, respectively.
\end{definition}

\begin{definition}[Stable matching]
    To define stability, we first specify each participant's preference ranking. Each man has a strict total order $\prec_m$ on $W$, where $w \prec_m w'$ means $m$ prefers $w$ to $w'$. Each woman also has a strict total order $\prec_w$ over $M$, where $m \prec_w m'$ means $w$ prefers $m$ to $m'$. The corresponding weak relations are denoted by $\preceq_m$ and $\preceq_w$, where $w \preceq_m w' \ \Leftrightarrow (w \prec_m w' \text{ or } w = w')$ and $m \preceq_w m' \ \Leftrightarrow (m \prec_w m' \text{ or } m = m')$.
    
    Given these preference rankings, a blocking pair for a perfect matching $\mu$ is a pair $(m, w) \in M \times W$ that satisfies both $w \prec_m \mu(m)$ and $m \prec_w \mu^{-1}(w)$, which means that both $m$ and $w$ prefer each other to their current partners. A perfect matching $\mu$ is stable if it has no blocking pair.
\end{definition}

\begin{definition}\label{def:man-optimal}
    For stable matchings $\mu$ and $\mu'$, we write $\mu \preceq \mu'$ if $\mu(m) \preceq_m \mu'(m)$ for every man $m \in M$. A man-optimal stable matching $\mu_{\text{m-opt}}$ is a matching such that $\mu_{\text{m-opt}} \preceq \mu$ for every stable matching $\mu$.
\end{definition}

\begin{theorem}[{\cite[Theorems~1--2]{gale1962college}}]
    \label{thm:GS}
    Every stable matching instance admits a unique man-optimal stable matching. It can be constructed by the men-proposing version of the Gale--Shapley algorithm \cite{gale1962college}.
\end{theorem}

\subsection{Rotations in Stable Matching}\label{sec:2.2}

Next, we consider an important operation on stable matching, called \emph{rotation elimination}. In the following sections, we denote $\Omega$ as the set of stable matchings, where $(\Omega, \preceq)$ is a poset and $\preceq$ is the order between two matchings defined in \autoref{def:man-optimal}.

\begin{definition}[Cover relation, {\cite{aprile2018twolevel}}]
    \label{def:cover-relation}
    In the poset $(\Omega, \preceq)$, we say that $\mu'$ covers $\mu$ if $\mu \prec \mu'$ and there exists no $\mu'' \in \Omega$ such that $\mu \prec \mu'' \prec \mu'$. We write $\mu \lessdot \mu'$ in this case.
\end{definition}

\begin{figure}[!b]
  \centering
  \includegraphics[width=\linewidth]{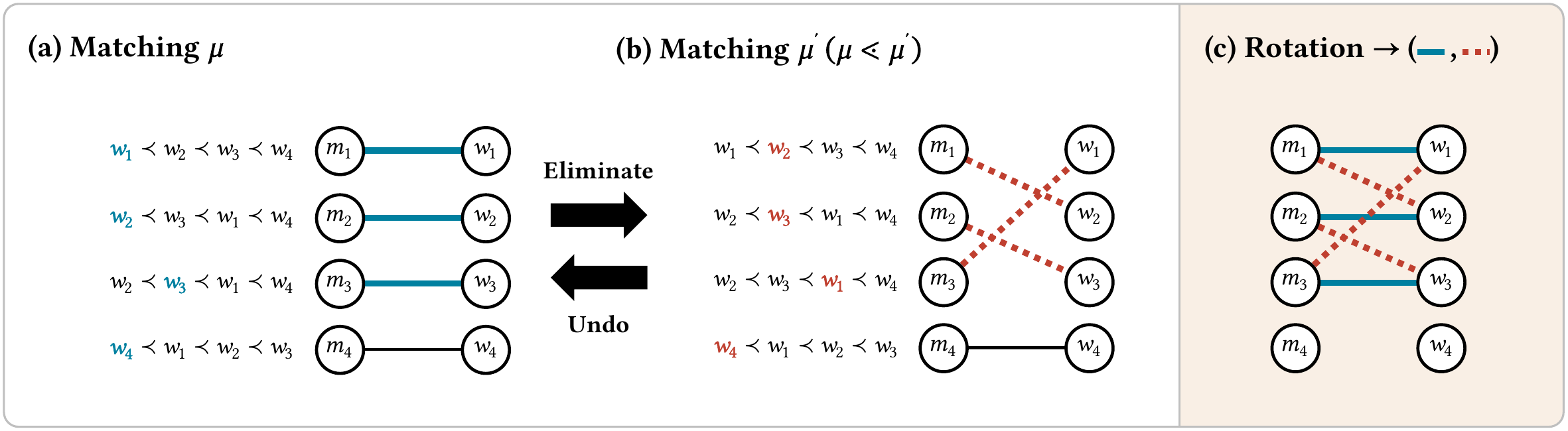}
  \caption{An example of rotation and the elimination operation. The rotation is represented by a pair $(\mu \setminus \mu', \mu' \setminus \mu)$ for stable matchings $\mu, \mu'$ $(\mu \lessdot \mu')$.}
  \label{fig:101}
\end{figure}

\begin{definition}[Rotation, {\cite{aprile2018twolevel}}]
    \label{def:rotation}
    For stable matchings $\mu$ and $\mu'$ with $\mu \lessdot \mu'$, the pair $\rho = (\mu \setminus \mu', \mu' \setminus \mu)$ is called a rotation; see \autoref{fig:101}. The length of a rotation is defined as $|\rho| = |\mu \setminus \mu'| = |\mu' \setminus \mu|$.
\end{definition}
Every rotation has length at least two, since two distinct perfect matchings cannot differ in just one matched pair.

\begin{definition}[Elimination]
    Let $\mu \lessdot \mu'$ and let $\rho = (\mu \setminus \mu', \mu' \setminus \mu)$ be the rotation. We say $\rho$ is exposed in $\mu$. The operation that transforms $\mu$ into $\mu'$ is called eliminating rotation $\rho$. The reverse operation is called undoing the elimination of $\rho$. See the arrows in \autoref{fig:101}.
\end{definition}

Since $\mu \lessdot \mu'$, eliminating the rotation cyclically permutes the current partners of the men involved. Formally, it transforms some matched pairs as
\begin{equation*}
    (m_{i_1}, w_{j_1}), \dots, (m_{i_{|\rho|}}, w_{j_{|\rho|}}) \to (m_{i_1}, w_{j_2}), \dots, (m_{i_{|\rho|}}, w_{j_{1}})
\end{equation*}
\noindent
for some $i_1, j_1, \dots, i_{|\rho|}, j_{|\rho|}$. In addition, different pairs of stable matchings $(\mu_1, \mu_1')$ and $(\mu_2, \mu_2')$ can have the same rotation. Next, we define the rotation sequence.

\begin{definition}[Rotation sequence]
    Suppose that we transform a stable matching from $\mu$ to $\mu'$ $(\mu \preceq \mu')$ by eliminating rotations $\rho_1, \rho_2, \dots, \rho_k$ in order. We call $(\rho_1, \dots, \rho_k)$ a rotation sequence from $\mu$ to $\mu'$.
\end{definition}

\begin{definition}[Rotation poset and downset]
    Let $\mathcal{R}$ be the set of rotations. For $\rho, \rho' \in \mathcal{R}$, we write $\rho \prec \rho'$ when $\rho$ precedes $\rho'$ in every rotation sequence in which both $\rho$ and $\rho'$ are eliminated. Define $\rho \preceq \rho'$ as $\rho = \rho'$ or $\rho \prec \rho'$. Then, the poset $(\mathcal{R}, \preceq)$ is called a rotation poset. In addition, a subset $D$ of the poset $(\mathcal{R}, \preceq)$ is a downset if, for every $\rho \in D$ and $\rho' \in \mathcal{R}$ ($\rho' \preceq \rho$), $\rho' \in D$ holds.
\end{definition}
\subsection{Structures in Rotation Poset}\label{subsec:stable-matchings-rotations}

Next, we state key structures of the rotation poset. In this section, we let $D(\mu)$ be the set of eliminated rotations from $\mu_{\text{m-opt}}$ to $\mu$; it is unique due to \autoref{thm:rotation-representation} (i).

\begin{theorem}[Rotation representation]
    \label{thm:rotation-representation} The following four statements hold \cite{gusfield1989stable}.
    \begin{enumerate}
        \item[(i)] For every $\mu \in \Omega$, the set $D(\mu)$ is uniquely determined and $D(\mu)$ is a downset of the poset $(\mathcal R,\preceq)$.
        \item[(ii)] For any downset $D$, eliminating its rotations from $\mu_{\text{m-opt}}$ along any linear extension $(\rho_1, \rho_2, \ldots, \rho_k)$ yields a unique stable matching $\mu_D$.\footnote{A linear extension of a poset $(\mathcal{R}, \preceq)$ is a total order $\leq$ on $\mathcal{R}$ such that $x \preceq y$ implies $x \leq y$ for all $x, y \in \mathcal{R}$.}
        \item[(iii)] The maps $\mu\mapsto D(\mu)$ and $D\mapsto\mu_D$ are mutually inverse bijections between $\Omega$ and the downsets of $(\mathcal R,\preceq)$.
        \item[(iv)] Eliminating a rotation $\rho$ from $\mu$ corresponds to adding $\rho$ to $D(\mu)$.
    \end{enumerate}
\end{theorem}

\begin{figure}[t]
  \centering
  \includegraphics[width=\linewidth]{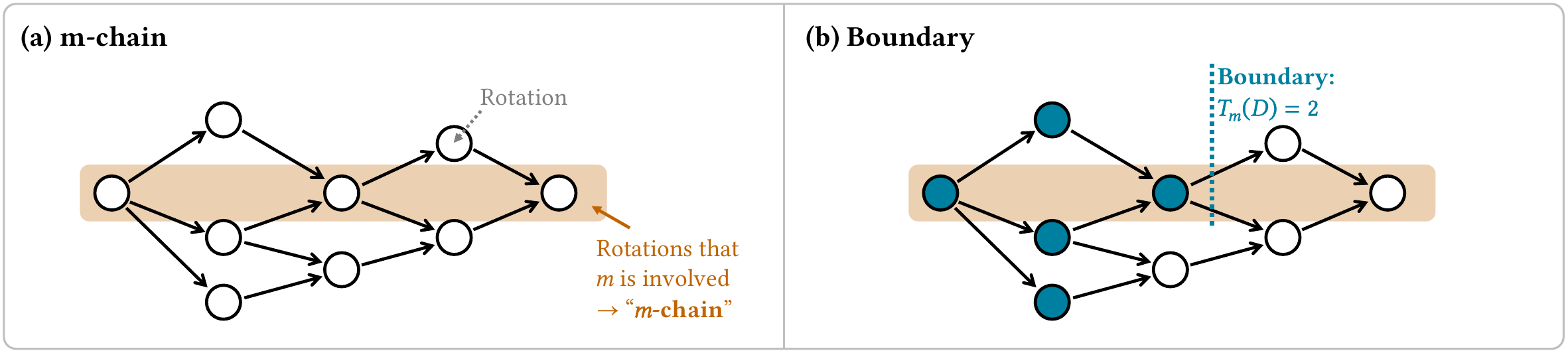}
  \caption{(a) The sketch of $m$-chain $C_m$, which is a total order. (b) An example of boundary $T_m(D)$, where the set of blue vertices corresponds to $D$.}
  \label{fig:102}
\end{figure}

More useful properties hold when we fix a man $m$. For a rotation $\rho=(\mu\setminus\mu',\mu'\setminus\mu)$, define the set of participating men by $M(\rho) = \{ m \in M \mid (m, w) \in \mu\setminus\mu' \text{ for some } w \in W \}$. We also write $C_m = \{\rho \in \mathcal{R} : m \in M(\rho)\}$ for the set of rotations involving a man $m$. Then, surprisingly, a poset $(C_m, \preceq)$ forms a total order. Formally, the following holds:

\begin{theorem}[Uniqueness of rotations and $m$-chains]
    \label{thm:rotation-uniqueness} The following three statements hold.
    \begin{enumerate}
        \item[(i)] For every pair $(m, w)$, there is at most one rotation $\rho=(\mu\setminus\mu', \mu'\setminus\mu)$ such that $(m, w) \in \mu\setminus\mu'$~\cite[Lemma~4.7]{irving1986complexity}.
        \item[(ii)] For every man $m$, $C_m$ forms a chain: for any $\rho, \rho' \in C_m$, either $\rho \preceq \rho'$ or $\rho' \preceq \rho$~\cite[Claim~4.6, p.~924]{karlin2018simply} (\autoref{fig:102} (a)).
        \item[(iii)] Eliminating a rotation changes the partner of every man in $M(\rho)$ to a strictly less preferred woman and leaves the partners of men outside $M(\rho)$ unchanged~\cite[Sections~3--4]{gusfield1987three}.
    \end{enumerate}
\end{theorem}

We call the chain $C_m$ the $m$-chain. Motivated by the representation of downsets through cuts in chains in Palmer--P{\'a}lv{\"o}lgyi~\cite[Section~III-B]{palmer2022most}, we define boundaries as follows.

\begin{definition}[Boundary of a downset]
    \label{thm:downset-boundary}
    For a downset $D$ of the rotation poset $(\mathcal R, \preceq)$, the boundary of $D$ at a man $m$ is $T_m(D)=|D\cap C_m|$.
\end{definition}

Writing $C_m$ as $\{\rho_{m,1} \prec \rho_{m,2} \prec \cdots \prec \rho_{m,k_m}\}$, we have
\begin{equation*}
    D\cap C_m = \{\rho_{m,1}, \ldots, \rho_{m,T_m(D)}\}.
\end{equation*}
\noindent
See \autoref{fig:102} (b) for an example. When $T_m(D)=0$, the set on the right-hand side is understood to be empty.

\begin{lemma}[Boundary--partner correspondence]
    \label{lem:boundary-partner-correspondence}
    For any downsets $D, D'$ and man $m$, $T_m(D) = T_m(D')$ if and only if $\mu_D(m) = \mu_{D'}(m)$. Likewise, $T_m(D) < T_m(D')$ if and only if $\mu_D(m) \prec_m \mu_{D'}(m)$.
\end{lemma}

\begin{proof}
    By \autoref{thm:rotation-uniqueness} (iii), the partner of $m$ strictly worsens for each elimination of rotation involving $m$. In addition, let $C_m = \{\rho_1, \rho_2, \dots\}$ be the $m$-chain where $\rho_1 \prec \rho_2 \prec \cdots$. Then, by \autoref{thm:downset-boundary}, the set of rotations involving $m$ eliminated to obtain $\mu_D$ is exactly $\{\rho_1, \dots, \rho_{T_m(D)}\}$. Therefore, as $T_m(D)$ increases, the partner of $m$ strictly worsens, leading to \autoref{lem:boundary-partner-correspondence}.
\end{proof}
\subsection{Outward Orders}\label{subsec:outward-orders}

Finally, we define the outward orders. In this section, we consider both \emph{forward order} that eliminates the rotations following the order of a poset, and \emph{backward order} that undoes the elimination of rotations following the reverse order of a poset.

\begin{figure}[t]
  \centering
  \includegraphics[width=\linewidth]{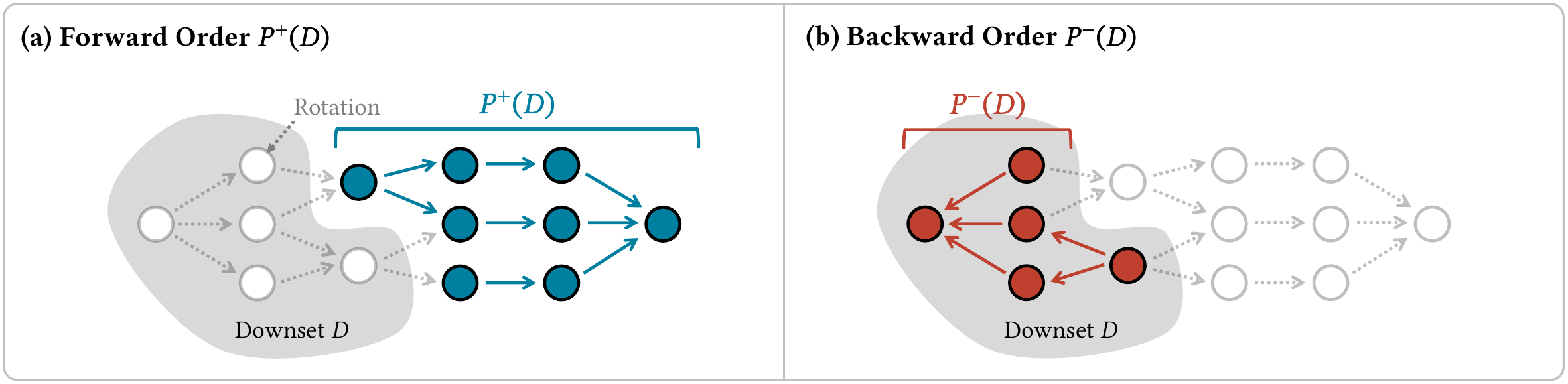}
  \caption{The sketch of (a) forward order and (b) backward order.}
  \label{fig:103}
\end{figure}

\begin{definition}[Outward orders]
    \label{def:outward-order}
    For a downset $D$ of the rotation poset $(\mathcal{R}, \preceq)$, we define the forward order $P^+(D)$ as the order induced by $\preceq$ on  $\mathcal{R} \setminus D$ (see \autoref{fig:103} (a)). In addition, we define a backward order $P^-(D)$ as the reverse of the order induced by $\preceq$ on $D$ (see \autoref{fig:103} (b)). Both are posets, and we call them collectively as ``outward orders from $D$''.
\end{definition}
\begin{lemma}[Executing downsets in outward orders]
    \label{lem:outward-order-execution}
    Let $D$ be a downset of the rotation poset $(\mathcal R, \preceq)$. 
    \begin{itemize}
        \item[(i)] For any downset $A$ of $P^+(D)$, eliminating the elements of $A$ from $\mu_D$ along any linear extension of $A$ yields $\mu_{D\cup A}$.
        \item[(ii)] For any downset $B$ of $P^-(D)$, undoing the eliminations of the elements of $B$ from $\mu_D$ along any linear extension of $B$ yields $\mu_{D\setminus B}$.
        \item[(iii)] In both cases, every intermediate state is a stable matching.
    \end{itemize}
\end{lemma}

\begin{proof}
    Since $A$ is a downset of $P^+(D)$, $D \cup A$ is also a downset of $\mathcal{R}$. Since $B$ is a downset of the reversed order $P^-(D)$, it is upward closed within $D$ in the original order, so $D \setminus B$ is also a downset of $\mathcal{R}$. Every prefix $A'$ of a linear extension of $A$ is itself a downset of $P^+(D)$, and likewise each prefix $B'$ is a downset of $P^-(D)$. The same reasoning therefore shows that $D\cup A'$ and $D\setminus B'$ are downsets of $\mathcal{R}$. By \autoref{thm:rotation-representation}, every downset of $\mathcal{R}$, say $D$, represents a stable matching $\mu_D$. Thus, every intermediate state is a stable matching, and the final states are $\mu_{D \cup A}$ and $\mu_{D \setminus B}$, respectively.
\end{proof}

For a poset $P$, let $\min P$ denote the set of its minimal elements.

\begin{lemma}[Non-revisiting property]
    \label{lem:non-retrogression}
    Let $D$ be a downset of the rotation poset $(\mathcal R, \preceq)$ and let $m$ be a man. Let $1_{\rho\in C_m}$ be the indicator equal to one if $\rho\in C_m$ and zero otherwise. Then, the following holds.
    \begin{align*}
        T_m(D\cup\{\rho\})
         & = T_m(D) + 1_{\rho\in C_m}
        \qquad (\rho\in\min P^+(D)), \\
        T_m(D\setminus\{\rho\})
         & = T_m(D) - 1_{\rho\in C_m}
        \qquad (\rho\in\min P^-(D)).
    \end{align*}
    \noindent
    Consequently, during an execution of rotations in a single direction, either forward or backward, no man returns to a woman after leaving her.
\end{lemma}
\begin{proof}
   Both equalities follow directly from the definition $T_m(D) = |D \cap C_m|$. Therefore, the boundary $T_m(D)$ must not decrease in the forward case, and must not increase in the backward case. By \autoref{lem:boundary-partner-correspondence}, no man returns to a woman after leaving her.
\end{proof}

\begin{remark}
    When distinguishing the orders explicitly, we write $\preceq_{\mathcal R}$ for the original order of the (original) rotation poset $\mathcal R$. For example, since the order is reversed in $P^-(D)$, the precedence of $\rho_1$ over $\rho_2$ in this order means $\rho_2 \prec_{\mathcal R} \rho_1$ in the original order.
\end{remark}
\subsection{Basic Facts of Entropy}\label{sec:2.5}

We assume that all logarithms are natural. We use the standard definitions and basic properties of entropy; see, for example, Cover--Thomas~\cite[Chapter~2]{cover2006elements}. First, the entropy of a discrete random variable $Z$ is defined by
\begin{equation*}
    H(Z) = -\sum_{z} \PP(Z=z)\log \PP(Z=z).
\end{equation*}

\noindent
Here, $0\log 0$ is defined to be zero. Second, when we denote the support of $Z$ as $\text{supp} Z=\{z \mid \PP(Z=z) > 0\}$, for another discrete random variable $Y$, the conditional entropy is defined by
\begin{equation*}
    H(Z \mid Y) = \sum_{y\in\text{supp} Y} \PP(Y=y) H(Z \mid Y=y).
\end{equation*}
\noindent
Third, for discrete random variables $Z_1, Z_2, \ldots, Z_n$, the chain rule states that
\begin{equation*}
H(Z_1, Z_2, \ldots, Z_n) = \sum_{i=1}^{n} H(Z_i \mid Z_1, \ldots, Z_{i-1}).
\end{equation*}
\noindent
Fourth, Jensen's inequality bounds entropy in terms of the size of the support:
\begin{equation*}
    H(Z) = \sum_{z\in\text{supp} Z} \PP(Z=z) \log \frac{1}{\PP(Z=z)} \leq \log\sum_{z\in\text{supp} Z} 1 = \log\left|\text{supp} Z\right|.
\end{equation*}

\section{The \texorpdfstring{$3.332^n$}{3.332\string^n} Bound}\label{sec:cones-entropy}

In this section, we show the first improvement from the previous $4.01^n$ bound to the $3.332^n$ bound. We begin by introducing the previous approach \cite{palmer2022most} and its limitations.
\subsection{The PP22 Approach and Its Limitation}\label{sec:3.1}

\paragraph{Overview of Strategy.} First, we describe our entropy-based strategy, which builds on the random-reveal framework used in \cite{palmer2022most}. The key observation is that a stable matching $\mu$ is uniquely determined by boundaries $T_{m_1}(D), \dots, T_{m_n}(D)$. Using this observation, we consider the following \emph{random-reveal strategy}.

Choose a stable matching $\mu$ uniformly at random from $\Omega$, and let $D = D(\mu)$. For each man $m_i$, define a random variable $X_{m_i} = T_{m_i}(D)$, and consider revealing the values of $X_{m_1}, \dots, X_{m_n}$ in a uniformly random order among the $n!$ ways (initially, none of these values are revealed). More precisely, assign each man $m$ an independent random variable $U_m$, uniformly distributed on $[0, 1]$, and reveal the men's boundaries in increasing order of $U_m$. We work on the probability-one event that all priorities are distinct.

\begin{figure}[!b]
    \centering
    \includegraphics[width=\linewidth]{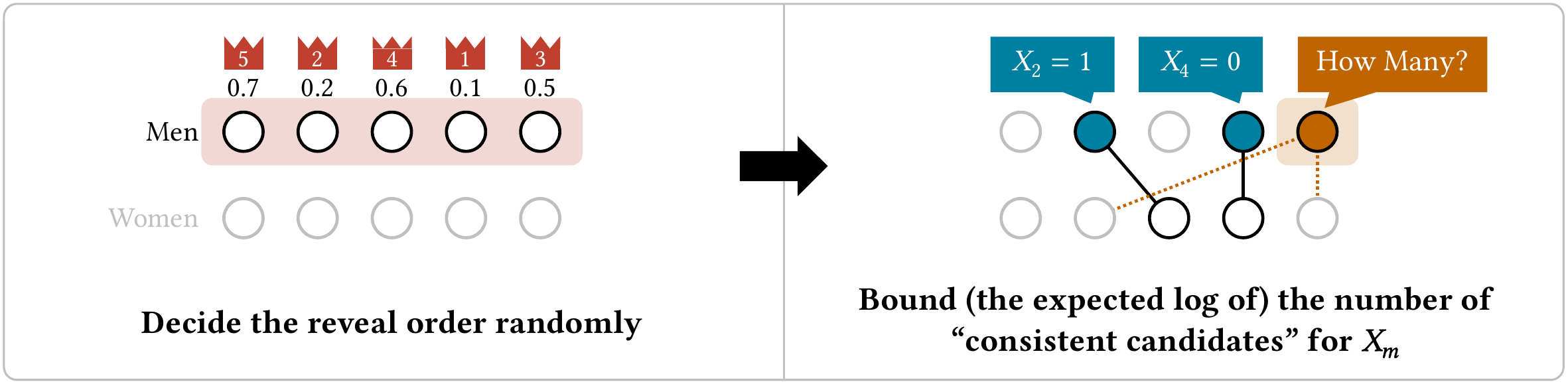}
    \caption{The sketch of our strategy. We bound (the expected logarithm of) the number of possible candidates for $X_m$ when the correct stable matching $\mu$ and the revealed order are chosen randomly.}
    \label{fig:301}
\end{figure}

Let $N_m$ be the number of candidate values for $X_m$ that are consistent with the already revealed information. The core of the strategy is to bound the (expected logarithm of) $N_m$ (see \autoref{fig:301}). If $\mathbb{E}[\log N_m] \leq \delta$ for every man $m$, we also obtain a bound of $|\Omega| \leq (e^\delta)^n$ for the number of stable matchings. For example, if $\delta \approx 1.203572$, we obtain $|\Omega| \leq 3.332^n$.

\paragraph{Formalization.} To write the strategy as a formal lemma, we assume that $\mu$ is chosen uniformly at random from $\Omega$, let $D = D(\mu)$, let $U = [U_{m_1}, \dots, U_{m_n}]$, and let
\begin{equation*}
    N_m(D, U) = \left|\left\{T_m(E) : E \text{ is a downset of } \mathcal{R}, T_v(E) = T_v(D) \text{ for all } v\in M \text{ with } U_v < U_m\right\}\right|.
\end{equation*}
\noindent
Then, the following key lemma holds:

\begin{lemma}[Entropy bound]
    \label{lem:entropy-bound-candidates} $\log |\Omega| \leq \sum_{m\in M} \EE_{D,U}[\log N_m(D, U)]$.
\end{lemma}

\begin{proof}
    Since $\mu \in \Omega$ is uniquely determined only by $X_{m_1}, \dots, X_{m_n}$, $H(\mu) = H(X_{m_1}, \ldots, X_{m_n})$ holds. By the chain rule of entropy,
    \begin{equation*}
        H(\mu) = H(X_{m_1}, \ldots, X_{m_n}) = \sum_{i=1}^{n} H(X_{m_i} \mid X_{m_1}, \ldots, X_{m_{i-1}}).
    \end{equation*}
    \noindent
    Since $U$ is independent of $\mu$, fixing $U$ does not change the distribution of $\mu$. We may therefore randomize the order of the men. For any fixed $U$ with distinct components, we have
    \begin{equation*}
        H(\mu) = \sum_{i=1}^n H(X_{m_i} \mid (X_{m_j})_{j:U_{m_j}<U_{m_i}}) \leq \sum_{i=1}^n \EE_D[\log N_{m_i}(D, U)].
    \end{equation*}
    \noindent
    Here we use the fact that conditional on any reveal history of positive probability, the support of $X_{m_i}$ (see \autoref{sec:2.5} for definition) has size $N_{m_i}(D,U)$; the entropy conditional on that history is thus at most $\log N_{m_i}(D,U)$, and averaging over the reveal history gives the inequality above. Finally, averaging over random $U$ gives
    \begin{equation*}
        H(\mu) \leq \sum_{i=1}^n \EE_{D,U}\log N_{m_i}(D, U). \qedhere
    \end{equation*}
\end{proof}

\paragraph{Limitations of PP22 Approach.} Although the men's boundaries alone determine a stable matching, \cite{palmer2022most} used both the men's boundaries $T_m(D)$ and women's boundaries $T_w(D)$ to obtain sufficiently strong estimates of the number of candidates.

Specifically, they considered a different random model that $2n$ values $T_{m_1}(D)$, $T_{w_1}(D)$, $\dots$, $T_{m_n}(D)$, $T_{w_n}(D)$ are revealed in random order. In \cite[Section~III]{palmer2022most}, they show $\mathbb{E}[\log N_i(D, U)]\leq 1.2038$ for $i \in \{m_1, w_1, \dots \}$ that implies the $(e^{1.2038})^{2n} < 11.11^n$ bound. In addition, our corrected evaluation\footnote{As shown in \autoref{lem:pp-numerical-bounds}, our reevaluation of the numerical expressions used in \cite[Claim 28]{palmer2022most} yields approximately $0.694$, rather than the claimed $0.6331$.} of \cite[Section~IV]{palmer2022most} gives $\mathbb{E}[\log N_i(D, U)]\leq 0.694$ for sufficiently large $n$, implying the $(e^{0.694})^{2n} \approx 4.01^n$ bound. In their concluding remarks, Palmer and P{\'a}lv{\"o}lgyi~\cite[Section~V]{palmer2022most} suggested reducing the number of components from $2n$ to $n$ as a possible route to improvement, while arguing that a straightforward reduction is unlikely to work.

\paragraph{Our Approach.} In our approach, we resolve the limitation of previous research. We reduce the number of chains to $n$ (only men's side) and prove $\mathbb{E}[\log N_m(D, U)]\leq 1.203572$, yielding the $(e^{1.203572})^n \approx 3.332^n$ bound. To reduce the number of chains from $2n$ to $n$, we introduce a new notion of ``cone'' and combine it with the analysis techniques used in \cite{palmer2022most}. In \autoref{subsec:cones-definition} and \autoref{subsec:cones-witnesses}, we introduce the notion of a cone. In \autoref{subsec:geometric-benchmark}, we bound the expected logarithm of $N_m(D, U)$.

\subsection{Cones and Supports}\label{subsec:cones-definition}

As discussed in the previous section, a key step is to bound $N_m(D, U)$, the number of candidate boundary positions consistent with the revealed boundaries. To this end, we analyze the probability that eliminating the next $k$ rotations involving $m$ (from $\mu$), or undoing the last $k$, remains consistent. Precedence constraints may require us to execute additional rotations not involving $m$. To describe these dependencies, we introduce \emph{cones} and their \emph{supports}. The $k$th principal cone in the forward direction contains all rotations that must be eliminated to eliminate the next $k$ rotations involving $m$; in the backward direction, it contains all rotations whose eliminations must be undone to undo the last $k$. In either direction, the principal cone includes those $k$ rotations themselves. In the following sections, we fix a man $m$ and a downset $D$.

\begin{figure}[!b]
  \centering
  \includegraphics[width=\linewidth]{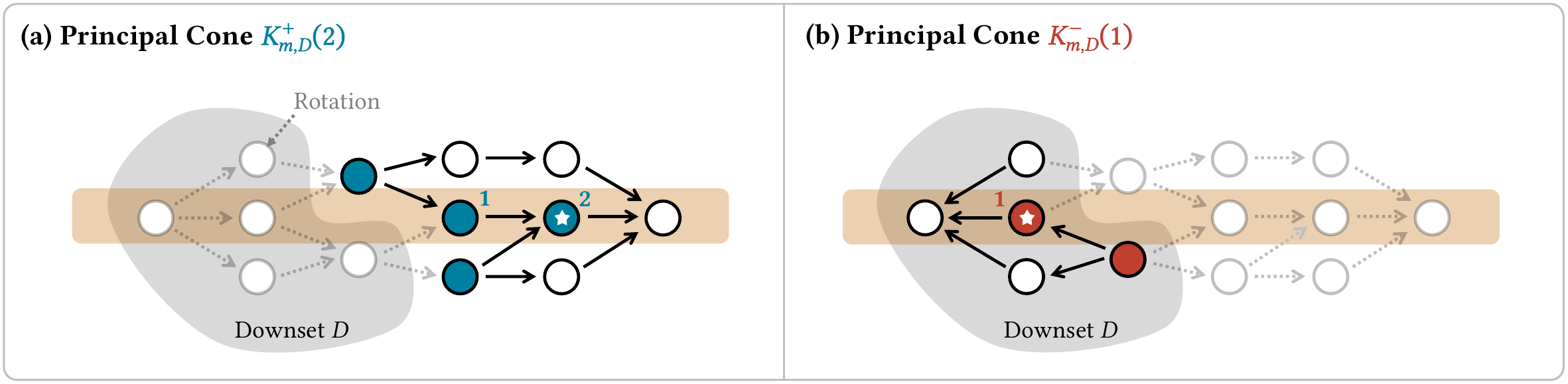}
  \caption{Examples of principal cones $K^+_{m, D}(2)$ and $K^-_{m, D}(1)$. Red and blue rotations are included in each principal cone. Rotations with dark outlines are included in outward order $P^+(D)$ or $P^-(D)$. The starred rotations correspond to $\rho^+_2$ and $\rho^-_1$, respectively.}
  \label{fig:302}
\end{figure}

\begin{definition}[Cones]
    \label{def:cone}
    For a downset $D$ of the rotation poset $(\mathcal R, \preceq)$, a cone is a downset in either outward order (forward or backward; see \autoref{def:outward-order}), $P^+(D)$ or $P^-(D)$. Let $\rho^+_k$ and $\rho^-_k$ be the $k$-th rotation involving $m$ in the forward and backward order, respectively. Then, we define the $k$-th principal cone by
    \begin{align*}
        K^+_{m,D}(k) & = \{\rho \in \mathcal R\setminus D : \rho \preceq_{\mathcal R} \rho^+_k\} \\
        K^-_{m,D}(k) & = \{\rho \in D : \rho^-_k \preceq_{\mathcal R} \rho\}.
    \end{align*}
    \noindent
    Intuitively, $K^+_{m,D}(k)$ consists of all rotations that must be eliminated to eliminate $\rho^+_k$ (\autoref{fig:302} (a)), and $K^-_{m,D}(k)$ consists of all rotations whose eliminations must be undone to undo $\rho^-_k$ (\autoref{fig:302} (b)). If the $k$-th rotation from $D$ does not exist, we leave $K^+_{m,D}(k)$ or $K^-_{m,D}(k)$ undefined.
\end{definition}

\begin{definition}[Support]
    \label{def:cone-support}
    We define the support of a cone $K$ by
    \begin{equation*}
        S(K) = \bigcup_{\rho\in K} M(\rho)
    \end{equation*}
    \noindent
    which represents the set of men involved in at least one rotation in $K$. We write $S^\pm_{m,D}(k) = S(K^\pm_{m,D}(k))$ for its support, and $s^\pm_{m,D}(k) = |S^\pm_{m,D}(k)|$ for its support size. If the $k$-th rotation from $D$ does not exist, we leave $S^\pm_{m,D}(k)$ undefined and set $s^\pm_{m,D}(k) = \infty$. For convenience, set $S^\pm_{m, D}(0) = \{m\}$.
\end{definition}

\begin{remark}[Nesting of Support]
    \label{remark:nesting}
    The following three hold when both sides are defined.
    \begin{equation*}
        K^\pm_{m,D}(k) \subseteq K^\pm_{m,D}(k+1), \ S^\pm_{m,D}(k) \subseteq S^\pm_{m,D}(k+1), \ s^\pm_{m,D}(k) \leq s^\pm_{m,D}(k+1)
    \end{equation*}
\end{remark}

\begin{lemma}[Partner Changes]
    \label{lem:support-size-2}
    Let $k \geq 1$. For any downset $D_A$ with $T_m(D_A) - T_m(D) \geq k$, every man $v$ in $S^+_{m,D}(k)$ satisfies $\mu_{D_A}(v) \neq\mu_D(v)$. Similarly, for any downset $D_B$ with $T_m(D) - T_m(D_B) \geq k$, every man $v$ in $S^-_{m,D}(k)$ satisfies $\mu_{D_B}(v) \neq\mu_D(v)$.
\end{lemma}

\begin{proof}
    Since $D_A$ and $D_B$ are downsets, the assumptions imply $\rho_k^+ \in D_A$ and $\rho_k^- \notin D_B$. Thus, by the definition of principal cone (\autoref{def:cone}), we have
    \begin{equation*}
        K^+_{m,D}(k) \subseteq D_A \setminus D, \ K^-_{m,D}(k) \subseteq D \setminus D_B.
    \end{equation*}
    \noindent
    Therefore, for any $v \in S^+_{m, D}(k)$, at least one rotation involving $v$ is included in $D_A$ but not in $D$, and thus $T_v(D_A) > T_v(D)$. For any $v \in S^-_{m, D}(k)$, at least one rotation involving $v$ is included in $D$ but not in $D_B$, and thus $T_v(D) > T_v(D_B)$. By \autoref{lem:boundary-partner-correspondence}, $\mu_{D_A}(v) \neq\mu_D(v)$ for every $v\in S^+_{m,D}(k)$, and $\mu_{D_B}(v) \neq\mu_D(v)$ for every $v\in S^-_{m,D}(k)$.
\end{proof}
\subsection{Bounding Number of Candidates Using Cones}\label{subsec:cones-witnesses}

\begin{figure}[!b]
    \centering
    \includegraphics[width=\linewidth]{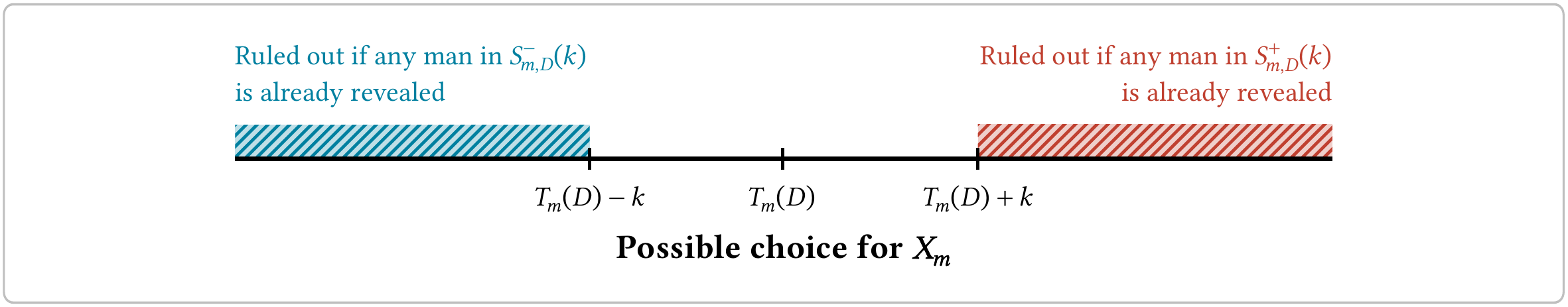}
    \caption{Sketch of our idea for bounding the number of candidates.}
    \label{fig:303}
\end{figure}

Now, consider bounding the number of candidate values for $X_m$. By \autoref{lem:support-size-2}, revealing even one man in $S^+_{m, D}(k) \setminus\{m\}$ rules out all candidate values $\geq T_m(D) + k$, and revealing even one man in $S^-_{m,D}(k) \setminus\{m\}$ rules out all candidate values $\leq T_m(D) - k$ (\autoref{fig:303}). We call each man in $S^\pm_{m,D}(k)\setminus\{m\}$ a witness at depth $k$ in the corresponding direction. To formally write this idea, we introduce random variables based on cones to bound $N_m(D, U)$.

\begin{definition}[Maximum unrevealed depth]
    \label{def:unrevealed-depth}
    For a downset $D$ of the rotation poset $(\mathcal R, \preceq)$ and a man $m$, define the random variables $L_m(D, U)$ and $R_m(D, U)$ by
    \begin{align*}
        L_m(D, U) & = \max\{k\in\mathbb Z_{\geq 0} : U_m < U_v \text{ for all } v\in S^-_{m,D}(k)\setminus \{m\}\}, \\
        R_m(D, U) & = \max\{k\in\mathbb Z_{\geq 0} : U_m < U_v \text{ for all } v\in S^+_{m,D}(k)\setminus \{m\}\}.
    \end{align*}
    Each $\max$ is taken over the nonnegative integers $k$ for which the corresponding support $S^\pm_{m,D}(k)$ is defined.
\end{definition}

\begin{lemma}[Candidate bound via cones]
    \label{lem:candidate-bound-cones}
    For a downset $D$ of the rotation poset $(\mathcal R, \preceq)$ and a man $m$, the number $N_m(D, U)$ is bounded in terms of $L_m(D, U)$ and $R_m(D, U)$ as follows:
    \begin{equation*}
        N_m(D, U) \leq 1 + L_m(D, U) + R_m(D, U).
    \end{equation*}
\end{lemma}
\begin{proof}
    Let $E$ be any downset which is consistent with all revealed boundary values. First, suppose that $r = T_m(E) - T_m(D) > 0$. By \autoref{lem:support-size-2}, every man $v \in S^+_{m,D}(r)$ satisfies $\mu_E(v) \neq \mu_D(v)$. By \autoref{lem:boundary-partner-correspondence}, every man $v \in S^+_{m,D}(r)$ satisfies $T_v(E) \ne T_v(D)$. Since $E$ is consistent with all revealed boundary values, every man in $S^+_{m,D}(r)\setminus\{m\}$ must be unrevealed, implying $r \leq R_m(D,U)$. Similarly, if $r = T_m(D) - T_m(E) > 0$, then $r \leq L_m(D, U)$. Therefore,
    \begin{equation*}
        T_m(D) - L_m(D, U) \leq T_m(E) \leq T_m(D) + R_m(D, U).
    \end{equation*}
    Since $T_m(E)$ is an integer, $N_m(D, U)$ is at most $1 + L_m(D, U) + R_m(D, U)$.
\end{proof}

\paragraph{Bounding the probability.} Finally, we bound the tail probabilities of $L_m(D, U)$ and $R_m(D, U)$ separately. First, we prove an important lemma on the size of the support, as follows.

\begin{lemma}[Support size]
    \label{lem:support-size}
    For any downset $D$ of the rotation poset $(\mathcal R, \preceq)$, man $m$, and positive integer $k$,
    \begin{equation*}
        s^\pm_{m,D}(k) \geq k+1.
    \end{equation*}
\end{lemma}

\begin{proof}
    If the $k$-th rotation does not exist, the claim holds by $s^\pm_{m,D}(k) = \infty \geq k+1$. Otherwise, consider executing the rotations in $K^\pm_{m,D}(k)$ in a suitable order. Since $K^\pm_{m,D}(k)$ contains exactly $k$ rotations involving $m$, during the execution, by \autoref{lem:non-retrogression}, the man $m$ experiences exactly $k+1$ distinct partners in total. On the other hand, each rotation cyclically permutes the current partners of the men it involves. This means that ``the set of women matched to the men in $S^\pm_{m,D}(k)$'' remains unchanged throughout the execution. Therefore, $s^\pm_{m,D}(k) \geq k+1$.
\end{proof}

\begin{lemma}
    \label{lem:geometric-distribution}
    Fix a downset $D$ of the rotation poset $(\mathcal R, \preceq)$ and a man $m$. For every positive integer $k$ and $x\in(0,1)$,
    \begin{align*}
        \PP(L_m(D, U) \geq k \ | \ D, U_m = x) & = (1-x)^{s^-_{m,D}(k)-1} \leq (1-x)^{k}, \\
        \PP(R_m(D, U) \geq k \ | \ D, U_m = x) & = (1-x)^{s^+_{m,D}(k)-1} \leq (1-x)^{k}.
    \end{align*}
    If the $k$-th rotation does not exist and $s^\pm_{m,D}(k) = \infty$, we consider $(1-x)^{s^\pm_{m,D}(k)-1} = 0$.
\end{lemma}
\begin{proof}
    First consider the backward direction. If the $k$-th rotation does not exist, the probability of $L_m(D, U) \geq k$ is zero. Otherwise, by \autoref{def:unrevealed-depth}, $L_m(D, U) \geq k$ holds if and only if $U_v > U_m$ for every man $v \in S^-_{m,D}(k) \setminus \{m\}$. Since $U_{m_1}, \dots, U_{m_n}$ are independently and uniformly chosen from $[0, 1]$,
    \begin{equation*}
        \PP\left(L_m(D, U) \geq k \ | \ D, U_m = x\right) = (1-x)^{s^-_{m,D}(k)-1} \leq (1-x)^k.
    \end{equation*}
    \noindent
    where the last inequality is due to \autoref{lem:support-size}. The bound for $R_m(D, U)$ also holds analogously.
\end{proof}

\subsection{The Geometric Benchmark}\label{subsec:geometric-benchmark}

In this section, we finally show the $3.332^n$ bound on the number of stable matchings, using the results of \autoref{lem:candidate-bound-cones} and \autoref{lem:geometric-distribution}.

The analysis is complicated because the two random variables $L_m(D, U)$ and $R_m(D, U)$ are not independent. Indeed, when two variables are not independent, we cannot directly obtain the desired upper bound on $\mathbb{E}[\log N_m(D, U)]$ from \autoref{lem:candidate-bound-cones} ($N_m(D, U) \leq L_m(D, U) + R_m(D, U) + 1$) and the probability bounds in \autoref{lem:geometric-distribution}. Therefore, we first show the results when two variables are independent and then we remove the dependency.

\paragraph{Independent case.} As in \autoref{sec:3.1}, let $U_m$ be a variable uniformly distributed on $[0, 1]$. Conditional on $x = U_m$, we let $G_1$ and $G_2$ be two independent random variables where $\mathbb{P}(G_1 = k) = \mathbb{P}(G_2 = k) = x (1 - x)^k$ for any non-negative integer $k$. Since $\mathbb{P}(G_1 \geq k) = \mathbb{P}(G_2 \geq k) = (1 - x)^k$ for any non-negative integer $k$, these tail probabilities match the upper bounds for $L_m(D, U)$ and $R_m(D, U)$. Then, the following lemma holds.

\begin{lemma}
    \label{lem:geometric-expectation}
    $\mathbb{E}_{U_m, G_1, G_2}[\log (1 + G_1 + G_2)] \leq 1.203572$.
\end{lemma}

\begin{proof}
    Fix $x$. The variables $G_1$ and $G_2$ are independent and geometric with tails $\PP(G_1 \geq k) = (1-x)^k$ and $\PP(G_2 \geq k) = (1-x)^k$. Therefore,
    \begin{align*}
        \PP(1 + G_1 + G_2 = j)
        & = \sum_{k=0}^{j-1} \PP(G_1 = k)\PP(G_2 = j-1-k) \\
        & = \sum_{k=0}^{j-1} (1-x)^k x (1-x)^{j-1-k} x = j x^2 (1-x)^{j-1}.
    \end{align*}
    \noindent
    Thus, $\EE_{G_1,G_2}[\log(1 + G_1 + G_2) \mid U_m = x] = \sum_{j\geq 1} \PP(1 + G_1 + G_2 = j) \log j = \sum_{j\geq 1} j x^2 (1-x)^{j-1} \log j$. By averaging over $x \in [0, 1]$, the following formula holds.
    \begin{align*}
        \EE_{x,G_1,G_2}[\log(1 + G_1 + G_2)] & = \int_0^1 \sum_{j\geq 1} j x^2 (1-x)^{j-1} \log j \, dx \\
                                               & = \sum_{j\geq 1} j\log j \int_0^1 x^2 (1-x)^{j-1} \, dx  \\
                                               & = \sum_{j\geq 1} j\log j \frac{2}{j(j+1)(j+2)}          = \sum_{j\geq 1} \frac{2\log j}{(j+1)(j+2)}.
    \end{align*}
    \noindent
    By the numerical evaluation in the proof of \autoref{lem:numerical-thresholds}, we obtain $\sum_{j\geq 1} \frac{2\log j}{(j+1)(j+2)} \leq 1.203572$, and therefore \autoref{lem:geometric-expectation} holds. Note that for the remaining sections, we denote $C_0 = \sum_{j\geq 1} \frac{2\log j}{(j+1)(j+2)}$.
\end{proof}


\paragraph{Dependent case.} Next, we reduce the actual dependent case to the independent case to use \autoref{lem:geometric-expectation}. In the following three lemmas, consider fixing the downset $D$, the rotation poset $(\mathcal{R}, \preceq)$, man $m$, and his priority $x = U_m \in (0, 1)$. Write $L = L_m(D, U)$ and $R = R_m(D, U)$. Unless otherwise specified, all probabilities, expectations, and independence statements are conditional on $D$ and $U_m$. In particular, $\EE_U$ averages over the priorities\footnote{We call the random variables $U_{m_1}, \dots, U_{m_n}$ ``priority values''.} other than the fixed $U_m$. Note that we mostly follow the proof techniques described in \cite[Section~IV-B]{palmer2022most}.

\begin{lemma}[Separating a shared witness]
    \label{lem:shared-witness-separation}
    Let $v$ $(\neq m)$ be a witness shared by the forward and backward sides, and let $U'_v$ be uniform on $[0,1]$ and independent of $U_v$ and all other existing random variables. If $R'$ is obtained from $R$ by replacing $U_v$ with $U'_v$ on the $R$ side,
    \begin{equation*}
        \EE_U[\log(1 + L + R)] \leq \EE_{U,U'_v}[\log(1 + L + R')].
    \end{equation*}
    \noindent
    Note that the only difference between $R$ and $R'$ is whether a man $v$ is treated as revealed or not (in $R$ and $R'$). In particular, $R = R'$ holds whenever $U_v, U_v' < U_m$ or $U_m \leq U_v, U_v'$.
\end{lemma}
\begin{proof}
    Since $L$ and $R$ do not depend on $U'_v$, we have $\EE_U[\log(1 + L + R)]=\EE_{U,U'_v}[\log(1 + L + R)]$. Now, we aim to prove $\EE_{U,U'_v}[\log(1 + L + R)] \leq \EE_{U,U'_v}[\log(1 + L + R')]$. Fix all priority values other than $U_v$ and $U'_v$, and define the following values:
    \begin{itemize}
        \item Let $l_0$ and $r_0$ be the values of $L$ and $R$ when $v$ is revealed, and
        \item Let $l_1$ and $r_1$ be the values of $L$ and $R$ when $v$ is unrevealed.
    \end{itemize}
    \noindent
    By definition of $L$ and $R$, we have $l_0 \leq l_1$ and $r_0 \leq r_1$. With all priorities other than $U_v$ and $U'_v$ fixed, the expectation of $\log(1 + L + R)$, over variable $U_v$, is
    \begin{equation*}
        x\log(1 + l_0 + r_0) + (1-x)\log(1 + l_1 + r_1).
    \end{equation*}
    \noindent
    Likewise, the expectation of $\log(1+ L + R')$, over variables $U_v$ and $U_v'$, is
    \begin{align*}
        x^2\log(1 + l_0 + r_0) 
        & + x(1-x)\log(1 + l_0 + r_1) \\
        & + x(1-x)\log(1 + l_1 + r_0) + (1-x)^2\log(1 + l_1 + r_1).
    \end{align*}
    \noindent
    Subtracting the former from the latter gives
    \begin{equation*}
        x(1-x)\log\frac{(1 + l_0 + r_1)(1 + l_1 + r_0)}{(1 + l_0 + r_0)(1 + l_1 + r_1)} = x(1-x)\log\left(1 + \frac{(l_1-l_0)(r_1-r_0)}{(1 + l_0 + r_0)(1 + l_1 + r_1)}\right) \geq 0.
    \end{equation*}
    \noindent
    Thus, with all priorities other than $U_v$ and $U'_v$ fixed, the expectation of $\log(1+ L + R)$ is at most that of $\log(1 + L + R')$. Averaging over the remaining priorities gives \autoref{lem:shared-witness-separation}.
\end{proof}


\begin{lemma}[Independence by separating shared witnesses]
    \label{lem:recursive-shared-witness-separation}
    Let $\widetilde{R}$ have the same distribution as $R$ and be independent of $L$. Then,
    \begin{equation*}
        \EE_U[\log(1 + L + R)] \leq \EE_{U,U'}[\log(1 + L + \widetilde{R})]
    \end{equation*}
    \noindent
    holds, where $U' = \{U_{m_1}', \dots, U_{m_n}'\}$ consists of mutually independent uniform random variables on $[0, 1]$, independent of the original priorities $U$.
\end{lemma}
\begin{proof}
    The core is to apply \autoref{lem:shared-witness-separation} repeatedly. Let $E = \{e_1, \dots, e_k\}$ be the set of witnesses shared by $L$ and $R$. Let $R_0 = R$, and for each $1 \leq i \leq k$, let $R_i$ be the random variable obtained from $R_{i-1}$ by replacing the priority $U_{e_i}$ with $U_{e_i}'$. Then, we apply the previous \autoref{lem:shared-witness-separation} and show
    \begin{equation*}
        \EE_{U, U'}[\log(1 + L + R_{i-1})] \leq \EE_{U, U'}[\log(1 + L + R_i)],
    \end{equation*}
    \noindent
    in the order of $i = 1, 2, \dots, k$. Since $R$ is independent of $U'$, $\EE_{U}[\log(1 + L + R)] = \EE_{U, U'}[\log(1 + L + R)]$ holds. In addition, the distribution of $R_{i-1}$ and that of $R_i$ are the same, and $R_k$ is independent of $L$ since they depend on disjoint families of independent priorities. Thus, setting $\widetilde{R} = R_k$ and chaining the inequalities above proves \autoref{lem:recursive-shared-witness-separation}.
\end{proof}

\begin{lemma}
    \label{lem:geometric-tail-benchmark}
    $\EE_{L,R}[\log(1 + L + R)] \leq \EE_{G_1,G_2}[\log(1 + G_1 + G_2)]$.
\end{lemma}

\begin{proof}
    \autoref{lem:recursive-shared-witness-separation} gives $\EE_{L,R}[\log(1 + L + R)] \leq \EE_{L,\widetilde{R}}[\log(1 + L + \widetilde{R})]$. Here, $\widetilde{R}$ is independent of $L$. By \autoref{lem:geometric-distribution}, the tail probabilities of both $L$ and $R$ are at most $(1-x)^k$. For any nonnegative integer-valued random variable $Z$ and nondecreasing function $f$, we have
    \begin{equation*}
        \EE[f(Z)] = f(0) + \sum_{k\geq 1} \PP(Z \geq k) (f(k) - f(k-1)).
    \end{equation*}
    \noindent
    Set $f(k) = \EE_{\widetilde{R}}[\log(1 + k + \widetilde{R})]$, which is non-decreasing. Since the tail probabilities of $L$ are at most $(1-x)^k$ due to \autoref{lem:geometric-distribution},
    \begin{align*}
        \EE_{L,\widetilde{R}}\left[\log(1 + L + \widetilde{R})\right] & = \EE_L[f(L)]                                             \\
                                                           & = f(0) + \sum_{k\geq 1} \PP(L \geq k) (f(k) - f(k-1))     \\
                                                           & \leq f(0) + \sum_{k\geq 1} (1-x)^k (f(k) - f(k-1))        \\
                                                           & = \EE_{G_1,\widetilde{R}}[\log(1 + G_1 + \widetilde{R})].
    \end{align*}
    Next, set $g(k) = \EE_{G_1}[\log(1 + G_1 + k)]$, which is also nondecreasing. By \autoref{lem:geometric-distribution}, the tail probabilities of $R$ are at most $(1-x)^k$. The same bound holds for $\widetilde{R}$ because it has the same distribution as $R$. Thus,
    \begin{align*}
        \EE_{G_1,\widetilde{R}}\left[\log(1 + G_1 + \widetilde{R})\right] & = \EE_{\widetilde{R}}[g(\widetilde{R})]                           \\
                                                               & = g(0) + \sum_{k\geq 1} \PP(\widetilde{R} \geq k) (g(k) - g(k-1)) \\
                                                               & \leq g(0) + \sum_{k\geq 1} (1-x)^k (g(k) - g(k-1))                \\
                                                               & = \EE_{G_1,G_2}[\log(1 + G_1 + G_2)].
    \end{align*}
    \noindent
    Combining these inequalities gives
    \begin{equation*}
        \EE_{L,R}[\log(1 + L + R)] \leq \EE_{G_1,G_2}[\log(1 + G_1 + G_2)].
    \end{equation*}
    \noindent
    All expectations here are finite: $L,\widetilde{R}$ are bounded, the geometric variables have finite means for fixed $x>0$, and $\log(1+z)\leq z$.
\end{proof}

\begin{lemma}
    \label{thm:geometric-benchmark}
    $|\Omega| < 3.332^n$.
\end{lemma}

\begin{proof}
    By \autoref{lem:geometric-tail-benchmark} and \autoref{lem:geometric-expectation}, averaging over $U_m$, we obtain $\mathbb{E}_{U}[\log(1 + L + R)] \leq 1.203572$ for every fixed downset $D$ and a man $m$. By \autoref{lem:candidate-bound-cones}, we obtain $\mathbb{E}_{D, U}[\log N_m(D, U)] \leq 1.203572$. By \autoref{lem:entropy-bound-candidates}, we obtain $|\Omega| \leq (e^{1.203572})^n < 3.332^n$.
\end{proof}
\section{Frameworks for Further Improvements: Additive Savings}\label{subsec:additive-savings}

In the preceding section, we bounded the entropy using the fact that both $\PP(L\geq k)$ and $\PP(R\geq k)$ are at most $(1-x)^k$, where $x=U_m$. If the support size $s^\pm_{m,D}(k)$ is strictly larger than the lower bound $k+1$, the corresponding tail probability is smaller, further improving the entropy bound. For example, if $s^\pm_{m,D}(k) \geq k+2$ held for all $k \geq 1, m, D$ in both directions, we would obtain a stronger bound of $2.523^n$. Therefore, the basic strategy to obtain a better upper bound is to get a support-size bound better than $s^\pm_{m,D}(k) \geq k+1$.

In this section, we introduce an \emph{additive savings} framework to quantify the entropy savings (and how they contribute to the upper bound of $|\Omega|$) by the larger supports, and show that the entropy savings can be added across different depths and directions.

\paragraph{Tail-probability deficit.} Fix a downset $D$, a man $m$, and his priority $x=U_m\in(0,1)$. Unless otherwise specified, all probabilities, expectations, and independence statements in this section are conditional on these choices. Set $y=1-x$, and write $L=L_m(D,U)$ and $R=R_m(D,U)$. As in the preceding subsection, let $G_1,G_2$ be independent geometric random variables, also independent of the original priorities and all copies used to separate shared witnesses. Their distributions are $\PP(G_1=k)=\PP(G_2=k)=x(1-x)^k$ for $k=0,1,2,\ldots$, and their tail probabilities bound those of $L,R$ from above. We introduce tail-probability deficits to measure how far the tails of $L$ and $R$ fall below $y^k$, together with weights that convert these deficits into entropy savings.

\begin{definition}[Tail-probability deficits and weights]
    \label{def:tail-difference}
    For every positive integer $k$, define
    \begin{align*}
        a_k & = \PP(G_1 \geq k) - \PP(L \geq k) = y^k - \PP(L \geq k), \\
        b_k & = \PP(G_2 \geq k) - \PP(R \geq k) = y^k - \PP(R \geq k), \\
        w_k & = \EE_{G_1}\left[\log\frac{G_1+k+1}{G_1+k}\right].
    \end{align*}
    \ignorespaces
    We call $a_k,b_k$ the ``tail-probability deficits'' and $w_k$ the weight. For fixed $D,m,x$, these are deterministic nonnegative real numbers. For each fixed $k$, the deficits $a_k,b_k$ depend on $D,m,x$, whereas $w_k$ depends only on $x$.
\end{definition}

\begin{lemma}[Entropy bound from tail-probability deficits]
    \label{lem:tail-difference-entropy}
    Under the assumptions stated above, let $a_k,b_k,w_k$ be as in \autoref{def:tail-difference}.
    \begin{equation*}
        \EE_{L,R}[\log(1 + L + R)] \leq \EE_{G_1,G_2}[\log(1 + G_1 + G_2)] - \sum_{k\geq 1} (a_k + b_k) w_k.
    \end{equation*}
\end{lemma}

\begin{proof}
    \autoref{lem:recursive-shared-witness-separation} gives
    \begin{equation*}
        \EE_{L,R}[\log(1 + L + R)] \leq \EE_{L,\widetilde{R}}[\log(1 + L + \widetilde{R})].
    \end{equation*}
    \noindent
    Here, $\widetilde{R}$ is independent of $L$ and has the same distribution as $R$. By the choice of $G_1,G_2$, the variables $L,\widetilde{R},G_1,G_2$ are mutually independent under this conditioning. For fixed $x>0$, the variables $L,\widetilde{R}$ are bounded, and the geometric variables have finite means. Since $\log(1+z)\leq z$, all expectations and tail sums whose differences we take below are finite. For a nonnegative integer-valued random variable $Z$ and a nondecreasing function $f$, we have
    \begin{equation*}
        \EE[f(Z)] = f(0) + \sum_{k\geq 1} \PP(Z \geq k) (f(k) - f(k-1)).
    \end{equation*}
    \noindent
    Set $f(k) = \EE_{\widetilde{R}}[\log(1 + k + \widetilde{R})]$, which is non-decreasing. Then
    \begin{align*}
          & \ \EE_{G_1,\widetilde{R}}[\log(1 + G_1 + \widetilde{R})] - \EE_{L,\widetilde{R}}[\log(1 + L + \widetilde{R})] \\
        = & \ \EE_{G_1}[f(G_1)] - \EE_L[f(L)]                                                                             \\
        = & \ \sum_{k\geq 1} (\PP(G_1 \geq k) - \PP(L \geq k)) (f(k) - f(k-1))                                            \\
        = & \ \sum_{k\geq 1} a_k (\EE_{\widetilde{R}}[\log(1 + k + \widetilde{R}) - \log(1 + k - 1 + \widetilde{R})])     \\
        = & \ \sum_{k\geq 1} a_k \EE_{\widetilde{R}}\left[\log\frac{\widetilde{R}+k+1}{\widetilde{R}+k}\right].                      \\
    \end{align*}
    \noindent
    The function $\log\frac{\widetilde{R}+k+1}{\widetilde{R}+k}$ is decreasing in $\widetilde{R}$. Since $\PP(\widetilde{R} \geq r) \leq \PP(G_1 \geq r)$ for every $r\geq 1$, we have
    \begin{equation*}
        \EE_{\widetilde{R}}\left[\log\frac{\widetilde{R}+k+1}{\widetilde{R}+k}\right] \geq \EE_{G_1}\left[\log\frac{G_1+k+1}{G_1+k}\right] = w_k.
    \end{equation*}
    \noindent
    Thus,
    \begin{align*}
             & \EE_{G_1,\widetilde{R}}[\log(1 + G_1 + \widetilde{R})] - \EE_{L,\widetilde{R}}[\log(1 + L + \widetilde{R})] \\
        \geq & \sum_{k\geq 1} a_k w_k.
    \end{align*}
    \noindent
    Next, set $g(k) = \EE_{G_1}[\log(1 + G_1 + k)]$, which is also nondecreasing. By \autoref{lem:geometric-distribution}, the tail probabilities of $R$ are at most $(1-x)^k$. The same bound holds for $\widetilde{R}$ because it has the same distribution as $R$.
    Hence,
    \begin{align*}
          & \EE_{G_1,G_2}[\log(1 + G_1 + G_2)] - \EE_{G_1,\widetilde{R}}[\log(1 + G_1 + \widetilde{R})] \\
        = & \EE_{G_2}[g(G_2)] - \EE_{\widetilde{R}}[g(\widetilde{R})]                                   \\
        = & \sum_{k\geq 1} (\PP(G_2 \geq k) - \PP(\widetilde{R} \geq k)) (g(k) - g(k-1))                \\
        = & \sum_{k\geq 1} b_k (\EE_{G_1}[\log(1 + G_1 + k) - \log(1 + G_1 + k - 1)])                   \\
        = & \sum_{k\geq 1} b_k \EE_{G_1}\left[\log\frac{G_1+k+1}{G_1+k}\right]                                     \\
        = & \sum_{k\geq 1} b_k w_k.
    \end{align*}
    \noindent
    Combining these bounds gives
    \begin{align*}
        \EE_{L,R}[\log(1 + L + R)] & \leq \EE_{L,\widetilde{R}}[\log(1 + L + \widetilde{R})]                              \\
                                   & \leq \EE_{G_1,\widetilde{R}}[\log(1 + G_1 + \widetilde{R})] - \sum_{k\geq 1} a_k w_k \\
                                   & \leq \EE_{G_1,G_2}[\log(1 + G_1 + G_2)] - \sum_{k\geq 1} (a_k + b_k) w_k.
    \end{align*}
\end{proof}

\paragraph{Introducing additive savings.} The saving relative to the benchmark $\EE_{G_1,G_2}[\log(1 + G_1 + G_2)]$ is bounded below by a nonnegative weighted sum of the tail-probability deficits $a_k,b_k$. Since a single inequality includes the terms for both directions and all depths, the savings from these deficits can be added simultaneously.

\begin{definition}
    \label{def:delta-d}
    For every positive integer $d$, define
    \begin{equation*}
        \delta_d = \sum_{j\geq d} \frac{2\log \frac{j+1}{j}}{(j+1)(j+2)(j+3)}.
    \end{equation*}
\end{definition}

Every term is positive, and $\log\frac{j+1}{j} < \frac{1}{j}$ implies that $\delta_d$ is finite. Write $w_d(x)$ to make the dependence of $w_d$ on $x$ explicit.

\begin{lemma}[Integrated savings]
    \label{lem:integrated-savings}
    Let $d$ be a positive integer.
    \begin{align*}
        \int_0^1 x(1-x)^d w_d(x) \, dx     & = \delta_d,                    \\
        \int_0^1 x(1-x)^{d+1} w_d(x) \, dx & \geq \frac{d+1}{d+4} \delta_d.
    \end{align*}
\end{lemma}

\begin{proof}
    By definition, $w_d(x) = \EE_{G_1}[\log\frac{G_1+d+1}{G_1+d}]$. The nonnegativity of the terms justifies all interchanges of sums and integrals below. Since $G_1$ is geometric,
    \begin{align*}
        w_d(x) & = \sum_{j\geq 0} \PP(G_1=j) \log\frac{j+d+1}{j+d} \\
               & = \sum_{j\geq 0} x(1-x)^j \log\frac{j+d+1}{j+d}.
    \end{align*}
    \noindent
    It follows that
    \begin{align*}
        \int_0^1 x(1-x)^d w_d(x) \, dx & = \int_0^1 x(1-x)^d \sum_{j\geq 0} x(1-x)^j \log\frac{j+d+1}{j+d} \, dx      \\
                                       & = \sum_{j\geq 0} \log\frac{j+d+1}{j+d} \int_0^1 x^2 (1-x)^{d+j} \, dx        \\
                                       & = \sum_{j\geq 0} \log\frac{j+d+1}{j+d} \cdot \frac{2}{(d+j+1)(d+j+2)(d+j+3)} \\
                                       & = \sum_{j\geq d} \log\frac{j+1}{j} \cdot \frac{2}{(j+1)(j+2)(j+3)}           \\
                                       & = \delta_d.
    \end{align*}
    \noindent
    Similarly,
    \begin{align*}
        \int_0^1 x(1-x)^{d+1} w_d(x) \, dx & = \int_0^1 x(1-x)^{d+1} \sum_{j\geq 0} x(1-x)^j \log\frac{j+d+1}{j+d} \, dx  \\
                                           & = \sum_{j\geq 0} \log\frac{j+d+1}{j+d} \int_0^1 x^2 (1-x)^{d+j+1} \, dx      \\
                                           & = \sum_{j\geq 0} \log\frac{j+d+1}{j+d} \cdot \frac{2}{(d+j+2)(d+j+3)(d+j+4)} \\
                                           & = \sum_{j\geq d} \log\frac{j+1}{j} \cdot \frac{2}{(j+2)(j+3)(j+4)}.
    \end{align*}
    \noindent
    For $j\geq d$, we have
    \begin{equation*}
        \frac{2}{(j+2)(j+3)(j+4)} = \frac{2}{(j+1)(j+2)(j+3)} \cdot \frac{j+1}{j+4} \geq \frac{d+1}{d+4} \cdot \frac{2}{(j+1)(j+2)(j+3)}.
    \end{equation*}
    \noindent
    Therefore,
    \begin{equation*}
        \int_0^1 x(1-x)^{d+1} w_d(x) \, dx \geq \frac{d+1}{d+4} \sum_{j\geq d} \log\frac{j+1}{j} \cdot \frac{2}{(j+1)(j+2)(j+3)} = \frac{d+1}{d+4} \delta_d.
    \end{equation*}
\end{proof}

Finally, we use $\delta_d$ to quantify savings from the baseline bound $C_0$.

\begin{lemma}[Savings from support sizes]
    \label{lem:delta-d-savings}
    Fix a downset $D$, a man $m$, and a positive integer $d$. The expectations below are over all priorities, including $U_m$, with $D$ fixed. If $s^\pm_{m,D}(d) \geq d + 2$ in either direction,
    \begin{equation*}
        \EE_U[\log N_m(D,U)\mid D] \leq C_0-\delta_d.
    \end{equation*}
    Moreover, if $s^\pm_{m,D}(d) \geq d + 3$ in either direction,
    \begin{equation*}
        \EE_U[\log N_m(D,U)\mid D] \leq C_0-\left(1+\frac{d+1}{d+4}\right)\delta_d.
    \end{equation*}
    \noindent
    For the same $D,m$, if these conditions hold for distinct direction--depth pairs, the savings from any finite collection of such pairs may be summed, and the total subtracted from $C_0$. If a single direction--depth pair satisfies both conditions, only the stronger saving is used, once.
\end{lemma}

\begin{proof}
    First consider the backward direction. If $s^-_{m,D}(d) \geq d + 2$, then \autoref{lem:geometric-distribution} gives, for every $x\in(0,1)$,
    \begin{equation*}
        a_d=\PP(G_1 \geq d) - \PP(L \geq d) = y^d - \PP(L \geq d) \geq y^d - y^{d+1} = y^d (1-y) = x(1-x)^d.
    \end{equation*}
    \noindent
    Average \autoref{lem:tail-difference-entropy} over $x$. Using the fact that the benchmark averages to $C_0$ and applying \autoref{lem:integrated-savings}, we obtain
    \begin{align*}
        C_0-\EE_U[\log(1+L_m(D,U)+R_m(D,U))\mid D]
         & \geq \int_0^1 \sum_{k\geq 1}(a_k+b_k)w_k(x)\,dx \\
         & \geq \int_0^1 a_d w_d(x)\,dx                    \\
         & \geq \int_0^1 x(1-x)^d w_d(x)\,dx               \\
         & = \delta_d.
    \end{align*}
    Here, $a_k,b_k$ also depend on the integration variable $x$. If, in addition, $s^-_{m,D}(d) \geq d + 3$, then
    \begin{equation*}
        a_d=\PP(G_1 \geq d) - \PP(L \geq d) = y^d - \PP(L \geq d) \geq y^d - y^{d+2} = x(1-x)^d + x(1-x)^{d+1}.
    \end{equation*}
    \noindent
    Similarly, \autoref{lem:tail-difference-entropy} and \autoref{lem:integrated-savings} give
    \begin{align*}
        C_0-\EE_U[\log(1+L_m(D,U)+R_m(D,U))\mid D]
         & \geq \int_0^1 \sum_{k\geq 1}(a_k+b_k)w_k(x)\,dx           \\
         & \geq \int_0^1 a_d w_d(x)\,dx                              \\
         & \geq \int_0^1 \bigl(x(1-x)^d+x(1-x)^{d+1}\bigr)w_d(x)\,dx \\
         & \geq \delta_d+\frac{d+1}{d+4}\delta_d.
    \end{align*}
    \noindent
    In the forward direction, $s^+_{m,D}(d) \geq d + 2$ gives $b_d \geq x(1-x)^d$, and $s^+_{m,D}(d) \geq d + 3$ gives $b_d \geq x(1-x)^d+x(1-x)^{d+1}$. Thus, the same savings apply. In either direction, if $s^\pm_{m,D}(d)=\infty$, the $d$th rotation does not exist, so the probability that the unrevealed depth in that direction is at least $d$ is $0$, and the corresponding deficit $a_d$ or $b_d$ equals $y^d-0=y^d$. For $x\in(0,1)$ and $y=1-x$, the identity $x(1+y)=1-y^2\leq1$ gives both $y^d\geq x(1-x)^d$ and $y^d\geq x(1-x)^d+x(1-x)^{d+1}$. Thus, both savings bounds above also apply when the support size is infinite. The bound $N_m(D,U)\leq1+L_m(D,U)+R_m(D,U)$ from \autoref{lem:candidate-bound-cones} yields the claimed upper bounds on the expected logarithm of the number of candidates.

    In the sum above, depth $d$ in the backward direction corresponds to $a_dw_d(x)$, and depth $d$ in the forward direction corresponds to $b_dw_d(x)$. All terms are nonnegative, so keeping the terms for distinct direction--depth pairs and integrating gives the sum of their respective savings.
\end{proof}

In the following sections, we examine how large $s^\pm_{m,D}(d)$ is at each depth and add the resulting savings.

\section{The \texorpdfstring{$3.060^n$}{3.060\string^n} Bound}\label{sec:histories-balance}
In this section, we study the support sizes $s^\pm_{m,D}(d)$ at depths $d$ equal to one or two.
To this end, we analyze cones involving only a few men.
We use a bijection that changes the reference stable matching to bound how frequently configurations with small local savings can occur. Combining this frequency bound with the preceding entropy estimate gives fewer than $3.060^n$ stable matchings.

\subsection{Non-Revisiting Rotation Sequences on Three People}\label{subsec:three-person-lemma}
Fix a reference downset $D$ and one direction, either backward or forward. We analyze rotation sequences in cones whose supports consist of exactly three men.
In the next subsection, we will use this classification to identify the first two rotations of a man whose first and second principal cones have the minimum support sizes of two and three.
We will use these histories to associate each such man with a distinct other man who yields savings from tail-probability deficits at both depths.
An operation exchanging the partners of two men is called a swap, and an operation cyclically permuting the partners of three men is called a 3-cycle.
We denote a swap between men $a,b$ by $(a,b)$.
We denote a 3-cycle involving men $a,b,c$ by $(a,b,c)$.
The operation $(a,b,c)$ assigns the partners held by $b,c,a$ immediately before the operation to $a,b,c$, respectively.
When a rotation is specified by the context, we use notation such as $\tau=(a,b)$ to describe the permutation of its participants' partners when it is executed in the chosen direction.
Here, execution means eliminating the rotation in the forward direction and undoing its elimination in the backward direction.
The orientation of a 3-cycle describes the operation in this execution direction; eliminating a rotation and undoing its elimination have opposite orientations.
\begin{lemma}
    \label{lem:three-person-cone}
    Let $K$ be a cone with $|S(K)|=3$.
    \ignorespaces
    \begin{enumerate}
        \item The cone $K$ contains at most two rotations.
        \item If $K$ contains two rotations, their execution consists of either two swaps on distinct pairs of men or two 3-cycles with the same cyclic orientation relative to a fixed ordering of the three men.
    \end{enumerate}
\end{lemma}
\begin{proof}
    Let $\mu_D$ be the reference state of the outward order containing $K$, and execute the rotations of $K$ along any linear extension.
    This execution is legal by \autoref{lem:outward-order-execution}.
    Write $S(K)=\{u,v,w\}$, and let $A,B,C$ be their respective partners in the reference state.
    Since every rotation involves only men in $S(K)$, the set of their partners remains $\{A,B,C\}$ throughout the execution.
    By \autoref{lem:non-retrogression}, no man returns to a partner he has left.

    Each rotation acts as either a swap or a 3-cycle.
    The following table gives representative states up to relabeling the men and their initial partners.
    Each tuple lists the partners of $u,v,w$ in that order. The last column gives a state consistent with the non-revisiting property after a second operation, if one exists.
    \begin{center}
        \begin{tabular}{c|ccc}
            First operation & Initial state & After the first & After the second \\
            \hline
            swap            & $(A,B,C)$     & $(B,A,C)$       & $(C,A,B)$        \\
            3-cycle         & $(A,B,C)$     & $(B,C,A)$       & $(C,A,B)$
        \end{tabular}
    \end{center}
    For each row, we determine the possible second operation and show that no third operation is possible.

    If the first operation is a swap, relabel the men so that it is $(u,v)$, giving the state $(B,A,C)$.
    Swapping the same pair again returns $u,v$ to their initial partners.
    A 3-cycle in either orientation also returns $u$ or $v$ to his initial partner.
    Thus, if a second operation exists, it is a swap on one of the other two pairs.
    By symmetry between $u,v$, we may take the second operation to be the swap $(u,w)$, yielding the state $(C,A,B)$ in the table.
    At this point, $u$ has had all of $A,B,C$ as partners, so he cannot participate in any subsequent rotation.
    The only operation not involving $u$ is the swap $(v,w)$, but it would return $v$ to his initial partner $B$ and is therefore impossible.
    Hence no third operation is possible.

    If the first operation is a 3-cycle, relabel the men so that the resulting state is $(B,C,A)$.
    From this state, every swap returns one man to his initial partner, and a 3-cycle in the opposite orientation returns all three men to their initial partners.
    Thus, if a second operation exists, it is a 3-cycle in the same orientation, yielding the state $(C,A,B)$ in the table.
    All three men have now had all of $A,B,C$ as partners, so no subsequent rotation is possible.
    In either case, there are at most two rotations, and if there are two, their operations have one of the stated forms.
\end{proof}

\subsection{Simple Men and Their Images}\label{subsec:simple-pairs-images}
We study the support sizes of the first and second principal cones.
Fix a downset $D$ and a direction. Write $K_m(k), S_m(k), s_m(k)$ for the $k$th principal cone of $m$, its support, and its support size, respectively.
\begin{definition}[Simple, B, C]
    \label{def:simple-b-c}
    A man $m$ is simple if $s_m(1)=2$.
    Define the sets of men $B,C$ by
    \begin{align*}
        C & = \{m : s_m(1) = 2, s_m(2) = 3\},    \\
        B & = \{m : s_m(1) = 2, s_m(2) \geq 4\}.
    \end{align*}
    \ignorespaces
\end{definition}
By \autoref{lem:support-size}, $s_m(2)\geq3$, so the set of simple men is $B \cup C$.
A simple man with no second rotation from the boundary belongs to $B$, since then $s_m(2)=\infty$.
We next consider the first principal cone of a simple man and the second principal cone of a man in $C$.
\begin{lemma}
    \label{lem:two-rotations}
    The first principal cone of a simple man consists of a single swap that can be executed immediately from the reference state $\mu_D$ in the chosen direction.
    If $m\in C$, there are men $a,b$ such that $m,a,b$ are distinct and
    \begin{equation*}
        K_m(2) = \{\tau,\sigma\}, \quad \tau=(m,b), \quad \sigma=(m,a).
    \end{equation*}
    \ignorespaces
    The execution order is necessarily $\tau$ followed by $\sigma$.
\end{lemma}
\begin{proof}
    First let $m$ be a simple man. Since $s_m(1)=2$, the support $S_m(1)$ consists of $m$ and one other man $b$.
    Every rotation in this cone swaps the same two men. Executing two such swaps would return them to their initial partners, violating the non-revisiting property.
    Thus, the cone contains only one rotation, and $K_m(1)=\{\tau\}= \{(m,b)\}$.
    By downward closure of the cone, $\tau$ is a minimal element of the corresponding outward order and can therefore be executed immediately from $\mu_D$ in the chosen direction.
    That is, it can be eliminated in the forward direction, or its elimination can be undone in the backward direction.

    Now suppose that $m\in C$.
    Since $s_m(2)=3$, the support $S_m(2)$ consists of $m,b$ and one other man $a$.
    By \autoref{lem:three-person-cone}, the cone $K_m(2)$ contains at most two rotations.
    Since the second cone from the boundary exists, $K_m(2)$ contains at least two rotations.
    The first rotation is a swap, so the second is also a swap.
    Since the second rotation also involves $m$, it is $(m,a)$.
\end{proof}
For $m\in C$, we give a name to the man $a$ in \autoref{lem:two-rotations}.
\begin{definition}[Image of a simple man]
    \label{def:image}
    For $m\in C$, the man $a$ who swaps partners with $m$ in the second rotation $\sigma=(m,a)$ in \autoref{lem:two-rotations} is called the image of $m$; we write $\phi(m)=a$.
    Let $I=\phi(C)$ be the image of the map $\phi: C \to M$.
\end{definition}
Like $K_m(k),S_m(k),s_m(k)$, the classes $B,C,I$ and the map $\phi$ depend on the reference downset and outward direction; we suppress this dependence when both are fixed.
When considering both directions, we distinguish them by superscripts $+,-$ or explicitly in the text; when considering different reference downsets, we also specify the reference downset.
For $m\in C$, the support sizes at depths one and two attain their minimum values, whereas his image $a=\phi(m)$ has larger supports at both depths.
This correspondence allows the savings associated with images to compensate for the locally unfavorable cases when summed over all men.
\begin{lemma}
    \label{lem:image-properties}
    Let $a=\phi(m)$.
    \begin{enumerate}
        \item $K_a(1)=K_m(2)$.
        \item The map $\phi$ is injective.
        \item $s_a(1)=3, s_a(2)\geq4$.
    \end{enumerate}
\end{lemma}
\begin{proof}
    The cone $K_m(2)$ contains only two rotations. Since $a$ does not participate in the first but does participate in the second, his first rotation is the swap $(m,a)$.
    Both $K_m(2)$ and $K_a(1)$ are the smallest cone containing the swap $(m,a)$, so $K_a(1)=K_m(2)$.

    Next, suppose that $m,m'\in C$ and $\phi(m)=\phi(m')$.
    Set $a=\phi(m)=\phi(m')$.
    By what we have just shown, the first rotation of $a$ is both the swap $(m,a)$ and the swap $(m',a)$.
    Since the first rotation of $a$ is unique, its other participating man is also unique, giving $m=m'$.
    Thus, $\phi$ is injective.

    We have $s_a(1)=|S_a(1)|=|S_m(2)|=3$.
    It remains to show that $s_a(2)\geq4$.
    If $a$ has no second rotation, then $s_a(2)=\infty\geq4$, so assume that his second rotation exists.
    Suppose for a contradiction that $s_a(2)\leq3$.
    \autoref{lem:support-size} gives $s_a(2)\geq3$, so $s_a(2)=3$.
    But $|K_a(2)|\geq|K_a(1)|+1=3$,
    contradicting \autoref{lem:three-person-cone}.
    Hence $s_a(2)\geq4$.
\end{proof}
Since $\phi$ is injective and $I=\phi(C)$, we have $|I|=|C|$.
Every image $a\in I$ satisfies $s_a(1)=3$ and is therefore not simple, so $I\cap(B\cup C)=\emptyset$.
We will use the equality of the sizes of these two sets for each fixed $D$ and direction when averaging the entropy bound.

\subsection{Basic Credits}\label{subsec:basic-credits}
Fix a downset $D$ and a direction. We first determine the entropy savings available for each man.
\begin{lemma}
    \label{lem:basic-credits}
    Fix $m\in M$.
    \begin{enumerate}
        \item If $m$ is not simple, a saving of $\delta_1$ is available.
        \item If $m\in B$, a saving of $\delta_2$ is available.
        \item If $m\in I$, a saving of $\delta_1+\delta_2$ is available.
    \end{enumerate}
\end{lemma}
\begin{proof}
    If $m$ is not simple, \autoref{def:simple-b-c} gives $s_m(1)\geq3$.
    \autoref{lem:delta-d-savings} gives a saving from the tail-probability deficit at depth one, yielding $\EE_U[\log N_m(D,U)\mid D] \leq C_0-\delta_1$.
    If $m\in B$, \autoref{def:simple-b-c} gives $s_m(1)=2$ and $s_m(2)\geq4$.
    \autoref{lem:delta-d-savings} gives a saving from the tail-probability deficit at depth two, yielding $\EE_U[\log N_m(D,U)\mid D] \leq C_0-\delta_2$.
    Finally, if $m\in I$, \autoref{lem:image-properties} gives $s_m(1)=3$ and $s_m(2)\geq4$.
    \autoref{lem:delta-d-savings} gives savings from the tail-probability deficits at depths one and two, yielding $\EE_U[\log N_m(D,U)\mid D] \leq C_0-(\delta_1+\delta_2)$.
\end{proof}
For each class defined above, the following table summarizes the support sizes, the savings counted at each depth, and their total.
\begin{center}
    \begin{minipage}{\linewidth}
    \centering
    \captionsetup{hypcap=false}
    \captionof{table}{Support sizes and basic credits by class}
    \label{tab:basic-credits}
    \begin{tabular}{l|cc|ccc}
        Class                & $s_m(1)$ & $s_m(2)$ & Depth 1    & Depth 2    & Total               \\
        \hline
        $C$                  & $2$      & $3$      & $0$        & $0$        & $0$                 \\
        $B$                  & $2$      & $\geq4$  & $0$        & $\delta_2$ & $\delta_2$          \\
        $I$                  & $3$      & $\geq4$  & $\delta_1$ & $\delta_2$ & $\delta_1+\delta_2$ \\
        Other non-simple men & $\geq3$  & ---      & $\delta_1$ & $0$        & $\delta_1$
    \end{tabular}
    \end{minipage}
\end{center}
Here, ``other non-simple men'' means $M\setminus(B\cup C\cup I)$.
Since $B,C,I$ are pairwise disjoint, every man belongs to exactly one row of the table.
The savings in the table are guaranteed lower bounds for each class; a zero means that no basic credit is counted at that depth.
The dash means that no condition on $s_m(2)$ is used to assign the basic credit.
In the averages below, choose $D$ uniformly from all downsets corresponding to stable matchings and, independently, choose a man $m$ uniformly from $M$.
\begin{definition}
    The value in the total column of \autoref{tab:basic-credits} is called the basic credit and is denoted by $g_{\mathrm{basic}}^\pm(D,m)$, according to the direction.
    Define the average saving by
    $$g_{\mathrm{basic}} = \EE_{D,m}[g_{\mathrm{basic}}^+(D,m)+g_{\mathrm{basic}}^-(D,m)].$$
    \ignorespaces
\end{definition}
In the next lemma, fix $D$ and one direction. The sets $B,C,I$ refer to the classes in the direction specified by the sign in $g_{\mathrm{basic}}^\pm$.
\begin{lemma}
    \label{lem:basic-credits-sum}
    The sum of the basic credits over all men is
    $$\sum_{m\in M} g_{\mathrm{basic}}^\pm(D,m) = n\delta_1 - (\delta_1 - \delta_2)(|B|+|C|).$$
    \ignorespaces
\end{lemma}
\begin{proof}
    Summing the basic credits over the classes gives
    \begin{align*}
        \sum_{m\in M} g_{\mathrm{basic}}^\pm(D,m) & = \sum_{m\in B} \delta_2 + \sum_{m\in I} (\delta_1+\delta_2) + \sum_{m\in M\setminus(B\cup C\cup I)} \delta_1 \\
                                 & = \delta_2(|B|+|I|) + \delta_1(|I|+|M\setminus(B\cup C\cup I)|)                                               \\
                                 & = \delta_2(|B|+|I|) + \delta_1(n - |B| - |C|).
    \end{align*}
    \ignorespaces
    Since $\phi$ is injective and $I=\phi(C)$, we have $|I|=|C|$. Substituting this equality gives
    \begin{align*}
        \delta_2(|B|+|I|) + \delta_1(n - |B| - |C|) & = \delta_2(|B|+|C|) + \delta_1(n - |B| - |C|) \\
                                                    & = n\delta_1 - (\delta_1 - \delta_2)(|B|+|C|),
    \end{align*}
    which proves the claim.
\end{proof}
Using $|I|=|C|$ in the sum reduces the dependence on the class sizes to the total number $|B|+|C|$ of simple men.
\begin{lemma}
    The average basic credit $g_{\mathrm{basic}}$ gives the following bound on the number of stable matchings:
    \label{lem:basic-credits-stable-matching-bound}
    $$\log |\Omega| \leq n(C_0-g_{\mathrm{basic}}).$$
    \ignorespaces
\end{lemma}
\begin{proof}
    By \autoref{lem:basic-credits}, the savings $g_{\mathrm{basic}}^+(D,m)$ and $g_{\mathrm{basic}}^-(D,m)$ are available for each man $m$.
    By \autoref{lem:delta-d-savings}, for the same $D,m$, savings from distinct direction--depth pairs satisfying the stated conditions may be summed, and the total subtracted from $C_0$.
    Thus, a total saving of $g_{\mathrm{basic}}^+(D,m)+g_{\mathrm{basic}}^-(D,m)$ is available for each man $m$.
    Averaging over all men $m$ and downsets $D$ gives the bound $\log |\Omega| \leq n(C_0-g_{\mathrm{basic}})$ on the number of stable matchings.
\end{proof}

\subsection{A Bijection Between Pointed Matchings}\label{subsec:pointed-bijection}
Since $\delta_1>\delta_2$, \autoref{lem:basic-credits-sum} shows that an upper bound on the average number of simple directions gives a lower bound on the average basic credit.
For this purpose, we construct a correspondence between different stable matchings.
Whether a man is simple, and his membership in the classes $B,C,I$, depend on the downset $D$, the chosen man $m$, and the direction.
To average over $D$ and $m$, we introduce the following definition.
\begin{definition}[Pointed Matching]
    \label{def:pointed-matching}
    Represent a stable matching by its downset $D$. A pair $(D,m)$ consisting of this downset and a distinguished man $m$ is called a \emph{pointed matching}.
\end{definition}
The pair $(D,m)$ allows us to consider the classes of $m$ in the backward and forward directions simultaneously.
\begin{lemma}
    \label{lem:pointed-bijection}
    There is a bijection between pointed matchings $(D,m)$ that are simple in both directions and those for which $m$ belongs to $C$ in the backward direction.
    Likewise, there is a bijection between pointed matchings that are simple in both directions and those for which $m$ belongs to $C$ in the forward direction.
\end{lemma}
\begin{proof}
    Let $(D,m)$ be a pointed matching that is simple in both directions.
    Then each of the first principal cones $K^-_{m,D}(1)$ and $K^+_{m,D}(1)$ consists of a single swap.
    Denote these swaps by $\tau_1$ and $\tau_2$, respectively.
    From $D$, we can immediately eliminate $\tau_2$ or undo the elimination of $\tau_1$, without operating on any other rotation.
    Set $D'=D\cup\{\tau_2\}$. Then $D'$ is a downset, and we have the following legal forward transitions:
    \[
        D\setminus\{\tau_1\}
        \xrightarrow{\tau_1} D
        \xrightarrow{\tau_2} D'=D\cup\{\tau_2\}.
    \]
    The subscripts $1,2$ indicate the forward order of the two swaps in this sequence.
    At the middle state $D$, the man $m$ is simple in both directions. We show that at the right endpoint $D'$, he belongs to $C$ in the backward direction, and that undoing $\tau_2$ to return to $D$ gives the inverse map.
    From $D'$, we can undo the eliminations of $\tau_2$ and $\tau_1$ in that order without operating on any other rotation.
    This means that $K^-_{m,D'}(1)=\{\tau_2\}, K^-_{m,D'}(2)=\{\tau_2,\tau_1\}$.
    Both swaps involve $m$. If their other participant were the same, the two swaps would return the men to their previous partners, contradicting \autoref{lem:non-retrogression}.
    The participants other than $m$ are therefore distinct, and the union of the supports of the two swaps consists of exactly three men.
    Hence $s^-_{m,D'}(1)=2, s^-_{m,D'}(2)=3$, so $m$ belongs to $C$ in the backward direction.

    Conversely, let $(D',m)$ be a pointed matching for which $m$ belongs to $C$ in the backward direction. Then $K^-_{m,D'}(1)$ consists of one swap and $K^-_{m,D'}(2)$ consists of two swaps.
    Let $\tau_2$ be the unique swap in $K^-_{m,D'}(1)$, and let $\tau_1$ be the other swap in $K^-_{m,D'}(2)$.
    From $D'$, we can immediately undo the eliminations of $\tau_2$ and $\tau_1$ in that order without operating on any other rotation.
    Set $D=D'\setminus\{\tau_2\}$. Then $D$ is a downset, and we can eliminate $\tau_2$ or undo the elimination of $\tau_1$ without operating on any other rotation.
    This means that $K^-_{m,D}(1)=\{\tau_1\}, K^+_{m,D}(1)=\{\tau_2\}$.
    Thus, $s^-_{m,D}(1)=s^+_{m,D}(1)=2$, and both directions are simple.

    Adding $\tau_2$ to $D$ and removing $\tau_2$ from $D'$ are inverse operations, giving a bijection between pointed matchings that are simple in both directions and those for which $m$ belongs to $C$ in the backward direction.
    The analogous construction gives a bijection between pointed matchings that are simple in both directions and those for which $m$ belongs to $C$ in the forward direction.
\end{proof}
Consequently, the number of pointed matchings that are simple in both directions equals both the number for which $m$ belongs to $C$ in the backward direction and the number for which $m$ belongs to $C$ in the forward direction.
This equality of counts ranges over all pairs of downsets and men; the bijection changes the reference downset between corresponding pairs.
We now bound the average number of simple directions.
\begin{definition}
    For a pointed matching chosen uniformly from $\Omega\times M$, let $e_s$ be the sum of the probability that the backward direction is simple and the probability that the forward direction is simple.
    Under the same distribution, let $p_c$ be the probability that the man $m$ belongs to $C$ in the backward direction.
\end{definition}
\begin{lemma}
    \label{lem:es-bound}
    $e_s\leq1+p_c$, $e_s\leq2(1-p_c)$, and $e_s\leq\frac{4}{3}$.
\end{lemma}
\begin{proof}
    By \autoref{lem:pointed-bijection}, the number of pointed matchings that are simple in both directions equals the number for which $m$ belongs to $C$ in the backward direction.
    Using also the bijection for the forward direction in the same lemma, both the probability that $m$ belongs to $C$ in the forward direction and the probability that both directions are simple equal $p_c$.
    Thus,
    \begin{align*}
        e_s & = 2\PP[\text{both directions are simple}] + \PP[\text{only one direction is simple}] \\
            & = p_c + \PP[\text{At least one direction is simple}]                                 \\
            & \leq p_c + 1.
    \end{align*}
    \ignorespaces
    Since $|C|=|I|$ for each fixed $D$ and direction, the probability that $m$ belongs to $I$ in the backward direction is also $p_c$ when a pointed matching is chosen uniformly from $\Omega\times M$.
    No man in $I$ is simple, so the probability that the backward direction is simple is at most $1-p_c$.
    Likewise, the probability that the forward direction is simple is at most $1-p_c$.
    Hence $e_s\leq 2(1-p_c)$.
    We have shown both $e_s\leq p_c+1$ and $e_s\leq 2(1-p_c)$.
    Twice the first inequality plus the second gives
    \[
        3e_s\leq 2(p_c+1)+2(1-p_c)=4,
    \]
    and therefore $e_s\leq\frac{4}{3}$.
\end{proof}

\subsection{Averaging Basic Credits and Improving the Bound}\label{subsec:basic-bound}
We combine these results to bound the average saving from basic credits and improve the upper bound on the number of stable matchings.
\begin{lemma}
    \label{lem:basic-credits-average-bound}
    The average basic credit $g_{\mathrm{basic}}$ satisfies
    $$g_{\mathrm{basic}}=2\delta_1-(\delta_1-\delta_2)e_s\geq\frac{2}{3}(\delta_1+2\delta_2).$$
\end{lemma}
\begin{proof}
    By \autoref{lem:basic-credits-sum}, for fixed $D$ and direction, the sum of the basic credits over all men is
    $$\sum_{m\in M} g_{\mathrm{basic}}^\pm(D,m) = n\delta_1 - (\delta_1 - \delta_2)(|B|+|C|).$$
    \ignorespaces
    Write $B^+,C^+$ for the classes in the forward direction and $B^-,C^-$ for those in the backward direction.
    Adding the identities for both directions, dividing by $n$, and averaging over $D$ gives
    $$g_{\mathrm{basic}} = \EE_{D,m}[g_{\mathrm{basic}}^+(D,m)+g_{\mathrm{basic}}^-(D,m)] = 2\delta_1 - (\delta_1 - \delta_2)\EE_{D}[\frac{|B^+|+|C^+|+|B^-|+|C^-|}{n}].$$
    \ignorespaces
    For fixed $D$, the ratio $(|B^\pm|+|C^\pm|)/n$ is the probability that a uniformly chosen man is simple in that direction.
    Thus, the expectation on the right is $e_s$, the sum of the probabilities that the backward and forward directions are simple for a pointed matching chosen uniformly from $\Omega\times M$.
    That is, $g_{\mathrm{basic}} = 2\delta_1 - (\delta_1 - \delta_2)e_s$.
    By \autoref{def:delta-d},
    $$\delta_1 - \delta_2 = \frac{2\log2}{2\times3\times4} > 0.$$
    \ignorespaces
    Since $e_s\leq\frac{4}{3}$ by \autoref{lem:es-bound}, we obtain $g_{\mathrm{basic}} \geq 2\delta_1 - (\delta_1 - \delta_2)\frac{4}{3} = \frac{2}{3}(\delta_1+2\delta_2)$.
\end{proof}
We use this to improve the bound on $\log |\Omega|$.
\begin{theorem}
    \label{prop:basic-credits-bound}
    The number of stable matchings satisfies
    $$|\Omega| < 3.060^n.$$
    \ignorespaces
\end{theorem}
\begin{proof}
    \autoref{lem:basic-credits-stable-matching-bound} bounds the number of stable matchings by $\log |\Omega| \leq n(C_0-g_{\mathrm{basic}})$.
    \autoref{lem:basic-credits-average-bound} gives $g_{\mathrm{basic}}\geq\frac{2}{3}(\delta_1+2\delta_2)$.
    Combining these yields $\log |\Omega| \leq n(C_0-\frac{2}{3}(\delta_1+2\delta_2))$.
    \autoref{lem:numerical-thresholds} in \autoref{app:numerical-thresholds} gives $C_0-\frac{2}{3}(\delta_1+2\delta_2)<\log(3.060)$, so $\log |\Omega| < n \log(3.060)$ and hence $|\Omega| < 3.060^n$.
\end{proof}

\section{The \texorpdfstring{$2.999^n$}{2.999\string^n} Bound}\label{sec:charging}
In this section, we obtain additional savings from the tail-probability deficits at depths three and four, and from those at depths one and two that have not been counted toward the basic credits.
Since the required additional saving need not be available for each man individually, we redistribute the additional credits obtained for the men.
We first bound the available additional savings, then analyze the local structure in cases where these conditions alone do not guarantee the required amount.
We then identify suitable charge recipients, bound their loads, and average the resulting savings to obtain the $2.999^n$ bound.

\subsection{Required Credits and Available Savings}\label{subsec:credit-budgets}
By \autoref{lem:basic-credits-average-bound}, $g_{\mathrm{basic}}=2\delta_1-(\delta_1-\delta_2)e_s$. Since $\delta_1>\delta_2$, the average basic credit decreases as the average number $e_s$ of simple directions increases.

To improve this bound, we use further entropy savings that have not been counted toward the basic credits.
Since these savings are not necessarily distributed uniformly among men, we use a charging argument to establish a lower bound on their total.

Fix a downset $D$ and an outward direction.
We begin by identifying local support conditions that guarantee additional savings for individual men.
Recall that $s_m(k)=\infty$ is a convention indicating that the $k$th rotation in the chosen direction and its principal cone do not exist.
In this case, no candidate boundary lies $k$ or more steps away in that direction, yielding a tail-probability deficit that can also make a man self-sufficient under the conditions below.
\begin{definition}
    \label{def:self-sufficient-strong}
    \begin{enumerate}
        \item A man $m$ is called \emph{self-sufficient simple} if he is simple and satisfies at least one of $s_m(3)\geq5$ and $s_m(4)\geq6$.
        \item A man $m$ is called a \emph{self-sufficient image} if $m\in I$ and at least one of $s_m(2)\geq5$ and $s_m(3)\geq6$ holds.
        \item A man $m$ is called \emph{strong} if $4\leq s_m(1)<\infty$.
    \end{enumerate}
\end{definition}
The definition of \emph{strong} excludes a man $m$ who has no rotation in the chosen direction.
Each of these three types of men yields savings beyond the basic credits.
Set $t=\delta_4$.
\begin{lemma}
    \label{lem:additional-savings}
    Fix $m\in M$.
    \begin{enumerate}
        \item If $m$ is self-sufficient simple, a saving of at least $t$ is available in addition to his basic credit.
        \item If $m$ is a self-sufficient image, a saving of at least $4t$ is available in addition to his basic credit.
        \item If $m$ is strong, a saving of at least $10t$ is available in addition to his basic credit.
    \end{enumerate}
\end{lemma}
\begin{proof}
    If $m$ is self-sufficient simple, \autoref{def:self-sufficient-strong} gives $s_m(3)\geq5$ or $s_m(4)\geq6$.
    The basic credits are counted at depths one and two, so they do not overlap with the savings at depths three and four used here.
    If $s_m(3)\geq5$, \autoref{lem:delta-d-savings} gives a saving of $\delta_3$ from the tail-probability deficit at depth three.
    If $s_m(4)\geq6$, \autoref{lem:delta-d-savings} gives a saving of $\delta_4$ from the tail-probability deficit at depth four.
    Since $\delta_3-\delta_4=\frac{2\log\frac{4}{3}}{4\times5\times6}>0$, either case gives a saving of at least $t$.

    If $m$ is a self-sufficient image, \autoref{def:self-sufficient-strong} gives $s_m(2)\geq5$ or $s_m(3)\geq6$.
    If $s_m(2)\geq5$, \autoref{lem:delta-d-savings} gives a saving of at least $(1+\frac{3}{6})\delta_2$ at depth two.
    Subtracting the $\delta_2$ already counted toward the basic credit leaves an additional saving of at least $\frac{1}{2}\delta_2$.
    By nesting of the supports and the convention of assigning size $\infty$ when a rotation does not exist, we have $s_m(3)\geq s_m(2)\geq5$ in this case.
    \autoref{lem:delta-d-savings} therefore gives a saving of $\delta_3$ at depth three.
    The total additional saving is thus at least $\frac{1}{2}\delta_2+\delta_3$.
    If $s_m(3)\geq6$, \autoref{lem:delta-d-savings} gives a saving of $\delta_3+\frac{4}{7}\delta_3=\frac{11}{7}\delta_3$ from the tail-probability deficit at depth three.
    The same nesting and convention give $s_m(4)\geq s_m(3)\geq6$, so \autoref{lem:delta-d-savings} gives a saving of $\delta_4$ at depth four.
    Since the savings at depths three and four have not been counted toward the basic credit, the total additional saving is at least $\frac{11}{7}\delta_3+\delta_4$.
    \autoref{lem:delta-inequalities} in \autoref{app:credit-budgets} gives $\frac{1}{2}\delta_2+\delta_3>4t$ and $\frac{11}{7}\delta_3+\delta_4>4t$, so an additional saving of at least $4t$ is available in either case.

    Finally, if $m$ is strong, \autoref{def:self-sufficient-strong} gives $4\leq s_m(1)<\infty$.
    Since $s_m(1)\geq4$, the man $m$ is neither simple nor an image, so his basic credit consists only of $\delta_1$ at depth one.
    \autoref{lem:delta-d-savings} gives a saving of at least $(1+\frac{2}{5})\delta_1$ at depth one, leaving an additional saving of at least $\frac{2}{5}\delta_1$ after subtracting the basic credit.
    The same nesting and convention give $s_m(2)\geq s_m(1)\geq4$, so \autoref{lem:delta-d-savings} gives a saving of $\delta_2$ at depth two. This has not been counted toward the basic credit.
    The total additional saving is thus at least $\frac{2}{5}\delta_1+\delta_2$.
    \autoref{lem:delta-inequalities} in \autoref{app:credit-budgets} gives $\frac{2}{5}\delta_1+\delta_2>10t$, so an additional saving of at least $10t$ is available.

    In each case, the basic and additional amounts counted at a given direction and depth together satisfy the conditions of \autoref{lem:delta-d-savings}. Savings from distinct direction--depth pairs may be added simultaneously.
\end{proof}
The following table lists the amounts we use for redistribution from the additional savings guaranteed by each man's own tail-probability deficits.
We call these amounts additional credits and count them in addition to the basic credits.
\begin{center}
    \begin{minipage}{\linewidth}
    \centering
    \captionsetup{hypcap=false}
    \captionof{table}{Additional credits by type of man}
    \label{tab:additional-credits}
    \begin{tabular}{l|c}
        Type of man            & Additional credit \\
        \hline
        self-sufficient simple & $t$               \\
        self-sufficient image  & $4t$              \\
        strong                 & $10t$
    \end{tabular}
    \end{minipage}
\end{center}
Each man belongs to at most one row of the table.
Denote the additional credit specified in \autoref{tab:additional-credits} by $g_{\mathrm{add}}^\pm(D,m)$, with the sign indicating the direction.
For a man who belongs to none of the rows of the table, set $g_{\mathrm{add}}^\pm(D,m)=0$.
For the fixed $D$ and outward direction, our target is the following inequality, to be proved in \autoref{lem:additional-credits-sum}:
\[
    \sum_{m\in M}g_{\mathrm{add}}^\pm(D,m)
    \ge t(|B|+|C|)+4t|I|.
\]
The left-hand side is the total additional credit guaranteed by each man's own tail-probability deficits; the right-hand side is the total demand of $t$ for every simple man and $4t$ for every image.
Summing this demand over both directions and averaging over downsets and men gives $te_s+8tp_c$, which equals $4t$ at the extremal value $e_s=4/3$, where the inequalities in \autoref{lem:es-bound} force $p_c=1/3$.
Charging is an accounting device for proving this lower bound on the total: each man's entropy bound remains unchanged and continues to use his own credits.
Self-sufficient simple men and images cover their own required amounts with their additional credits; each remaining request is assigned to a strong man.

The local structure analysis in \autoref{subsec:four-person-structure} will be used in \autoref{subsec:choosing-recipients} to establish the existence of charge recipients with the following properties.
For each simple man $m$ who is not self-sufficient, \autoref{lem:charging-simple} will give a strong man $h$ satisfying
\[
    4\le s_h(1)\le5,\qquad m\in S_h(1).
\]
For each image $a$ who is not self-sufficient, \autoref{lem:charging-image} will give a strong man $h$ satisfying
\[
    4\le s_h(1)\le5,\qquad
    \phi^{-1}(a)\in S_h(1),\qquad I\cap S_h(1)\ne\emptyset.
\]
Here $\phi^{-1}(a)\in C$ is the unique preimage of $a$, well-defined by the injectivity of $\phi$.

These conditions place simple requesters and the preimages of image requesters in the recipient's first principal-cone support, which contains at most five men.
The extra condition $I\cap S_h(1)\ne\emptyset$ for images is also needed for the load bound: in \autoref{subsec:load-total-credit}, we will use these conditions to show that each strong man's load is at most his additional credit of $10t$.
The final subsection will average the resulting inequality for the total and combine it with the basic credits to obtain the $2.999^n$ bound.

\subsection{Local Structure of Clean Four-Person Cones}\label{subsec:four-person-structure}
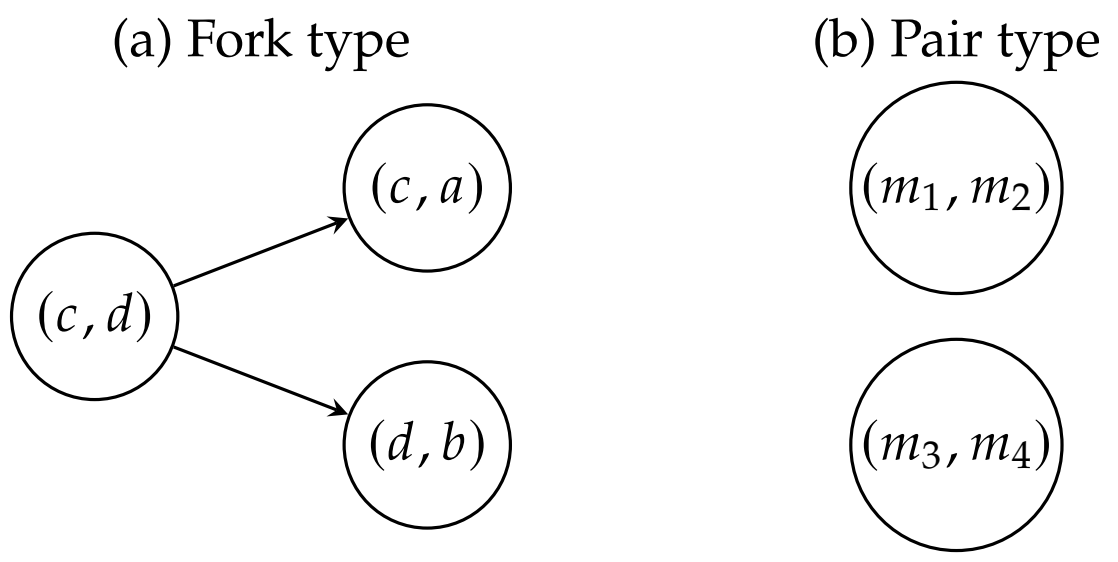
\begin{figure}[htbp]
    \centering
    \begin{tikzpicture}[x=1cm,y=1cm,>=stealth,
            every node/.style={font=\small},
            rotation/.style={circle,draw,minimum size=11mm,inner sep=1pt,line width=0.6pt}]
        \node at (1.1,1.8) {(a) Fork type};
        \node[rotation] (rho) at (0,0) {$(c,d)$};
        \node[rotation] (tau1) at (2.2,0.85) {$(c,a)$};
        \node[rotation] (tau2) at (2.2,-0.85) {$(d,b)$};
        \draw[->,line width=0.6pt] (rho) -- (tau1);
        \draw[->,line width=0.6pt] (rho) -- (tau2);
        \node at (5.7,1.8) {(b) Pair type};
        \node[rotation] at (5.7,0.85) {$(m_1,m_2)$};
        \node[rotation] at (5.7,-0.85) {$(m_3,m_4)$};
    \end{tikzpicture}
    \caption{Local structures of clean four-person cones, up to relabeling. Circles represent rotations, with the chosen outward order proceeding from left to right. The fork branches from its unique minimal element, the swap $(c,d)$, to $(c,a)$ and $(d,b)$, where $a=\phi(c)$ and $b=\phi(d)$. The pair has two incomparable minimal elements, the swaps $(m_1,m_2)$ and $(m_3,m_4)$. Later rotations are omitted.}
    \label{fig:fork-pair}
\end{figure}

We study the local structure of cones whose supports consist of exactly four men and contain no strong man, for use in choosing charge recipients in the next subsection.
\begin{definition}
    \label{def:four-person-cone}
    For a fixed downset $D$ and direction, a \emph{clean four-person cone} is a cone $K$ whose support consists of four men and for which $S(K)$ contains no \emph{strong} man.
\end{definition}
\begin{lemma}
    \label{lem:four-person-cone-swap}
    For a clean four-person cone $K$, the set of minimal elements of $K$ consists of either
    \begin{itemize}
        \item a single swap, or
        \item two swaps with disjoint sets of participating men.
    \end{itemize}
\end{lemma}
\begin{proof}
    Throughout this proof, we consider only rotations in $K$.
    By downward closure, the minimal elements of $K$ are precisely the rotations that can be executed immediately from the reference state in the chosen direction.
    First suppose that a 3-cycle $\rho$ can be executed immediately.
    Let $v$ be the man in $S(K)$ who does not participate in $\rho$.
    Since $v\in S(K)$, his first rotation exists; denote it by $\tau$.
    Downward closure gives $\tau\in K$ and $K_v(1)\subseteq K$.
    Choose any man $u$ other than $v$ who participates in $\tau$.
    The man $u$ also participates in $\rho$, so $\rho$ and $\tau$ share $u$ and are comparable in the corresponding outward order.
    Since $\rho\ne\tau$ and $\rho$ is minimal in this outward order, $\rho$ precedes $\tau$.
    Thus, $\rho\in K_v(1)$.
    It follows that $S_v(1)=S(K)$, making $v$ strong and contradicting the assumption that $K$ is a clean four-person cone.
    An immediately executable rotation of length four would likewise make all four men strong, again contradicting the assumption that $K$ is a clean four-person cone.
    Hence every immediately executable rotation is a swap.
    If two immediately executable swaps shared a man, the rotations would be comparable, contradicting their immediate executability.
    Since $K$ is finite and nonempty, it has a minimal element. By downward closure, this rotation can be executed immediately from the reference state in the chosen direction, so at least one such rotation exists.
    Each swap involves two men, and distinct immediately executable swaps have disjoint sets of participants. Since $|S(K)|=4$, there are at most two such swaps.
    Thus, the minimal elements of $K$ consist of one or two swaps, with disjoint sets of participants in the latter case.
\end{proof}
\begin{definition}
    \label{def:four-person-cone-types}
    A clean four-person cone $K$ is of fork type if it has one minimal element, and of pair type if it has two.
\end{definition}
\autoref{fig:fork-pair} shows the local structures of the two types. The branching in the fork type is established in the proof of the next lemma.
We now examine the rotations within $K$ to determine which of its four men are simple or images, and what can be inferred about the supports of their principal cones.
In the next subsection, we use this information to analyze the first principal cones of candidate charge recipients, showing that they are strong and verifying the required support inclusions.
\begin{lemma}[Fork-type structure]
    \label{lem:fork-cone-properties}
    Let $K$ be a clean four-person cone of fork type, with unique minimal element $\rho=(c,d)$, a swap.
    \begin{enumerate}
        \item[(i)] $c,d\in C$ and $S(K)=\{a,b,c,d\}$, where $a=\phi(c)$ and $b=\phi(d)$.
        \item[(ii)] For every $v\in S(K)$, there exists a positive integer $k$ such that $K_v(k)\subseteq K$ and $s_v(k)=3$.
    \end{enumerate}
\end{lemma}
\begin{proof}
    The cone $K$ is finite with unique minimal element $\rho$, so $\rho$ precedes every other rotation of $K$ in the corresponding outward order.
    Since $\rho$ is also minimal in the outward order, $K_c(1)=K_d(1)=\{\rho\}$, so both $c,d$ are simple.
    Let $a$ and $b$ be the men in $S(K)$ other than $c,d$.
    Let $\tau_1$ be the first rotation of $a$.
    Downward closure and the precedence of $\rho$ give $\rho\in K_a(1)\subseteq K$, so $S_a(1)$ contains $c,d,a$.
    Since the first rotation of $a$ exists and $K$ is clean, we also have $s_a(1)\leq3$.
    Hence $S_a(1)=\{c,d,a\}$.
    Executing the entire principal cone $K_a(1)$ along a linear extension begins with the swap $\rho$.
    By \autoref{lem:three-person-cone}, there are at most two rotations. Since $\rho$ and $\tau_1$ are distinct, $K_a(1)=\{\rho,\tau_1\}$.
    The same lemma implies that $\tau_1$ is also a swap, involving $a$ and one of $c,d$.
    Similarly, if $\tau_2$ is the first rotation of $b$, then $S_b(1)=\{c,d,b\}$ and $K_b(1)=\{\rho,\tau_2\}$, where $\tau_2$ swaps $b$ with one of $c,d$.
    If $\tau_1,\tau_2$ share a man, they are comparable.
    If $\tau_1$ precedes $\tau_2$, then $\{\rho,\tau_1,\tau_2\}\subseteq K_b(1)$.
    This gives $s_b(1)=4$, contradicting the assumption that $K$ is a clean four-person cone.
    The case where $\tau_2$ precedes $\tau_1$ is analogous.
    Thus, $\tau_1$ and $\tau_2$ have disjoint sets of participating men.
    By symmetry, assume that $\tau_1$ is the swap $(c,a)$ and $\tau_2$ is the swap $(d,b)$.
    Since $K_a(1)=\{\rho,\tau_1\}$, the rotation $\tau_1$ is the second rotation of $c$. Similarly, $\tau_2$ is the second rotation of $d$.
    Therefore,
    \begin{align*}
        K_c(2) & = K_a(1) = \{\rho,\tau_1\}, \quad s_c(2) = s_a(1) = 3, \\
        K_d(2) & = K_b(1) = \{\rho,\tau_2\}, \quad s_d(2) = s_b(1) = 3,
    \end{align*}
    so for every $v\in S(K)$, there exists a positive integer $k$ such that $K_v(k)\subseteq K$ and $s_v(k)=3$.
    Also, $c,d$ are simple with $s_c(2)=s_d(2)=3$, so $c,d\in C$; by the definition of an image, $\phi(c)=a$ and $\phi(d)=b$. Hence $S(K)=\{c,d,\phi(c),\phi(d)\}$.
\end{proof}
\begin{lemma}[Pair-type structure]
    \label{lem:pair-cone-properties}
    Let $K$ be a clean four-person cone of pair type.
    \begin{enumerate}
        \item[(i)] All four men are simple: $S(K)\subseteq B\cup C$.
        \item[(ii)] If there exists $m\in S(K)$ with $|C_m\cap K|\geq3$, every $v\in S(K)$ satisfies $K_v(2)\subseteq K$ and $s_v(2)=4$.
    \end{enumerate}
\end{lemma}
\begin{proof}
    By \autoref{lem:four-person-cone-swap}, the two minimal elements of $K$ are swaps on disjoint pairs of men.
    Denote them by $\rho_1=(m_1,m_2)$ and $\rho_2=(m_3,m_4)$.
    By downward closure, each swap is also minimal in the outward order and is the only rotation in each participant's first principal cone. Thus, all four men are simple, giving $S(K)\subseteq B\cup C$.
    Suppose that there exists $m\in S(K)$ with $|C_m\cap K|\geq3$.
    By symmetry, we may assume $m=m_1$ without loss of generality.
    Let $\tau$ and $\sigma$ be the second and third rotations of $m_1$, respectively.
    We first show that the second rotation of every $v\in S(K)$ belongs to $K$.
    For $v=m_1$, the assumption says that his first three rotations belong to $K$, so his second rotation does as well.
    Next, consider $v=m_2$.
    Suppose that the second rotation of $m_2$ does not belong to $K$.
    Let $D'$ be the downset obtained by executing $\rho_1$ from $D$.
    Since $\rho_1$ is minimal in $K$, the set $K\setminus\{\rho_1\}$ is a cone in the same direction from the new reference downset $D'$.
    This cone does not involve $m_2$ and contains $\rho_2,\tau,\sigma$, so its support is $\{m_1,m_3,m_4\}$.
    It is therefore a cone with support size three containing at least three rotations, contradicting \autoref{lem:three-person-cone}.
    Next, consider $v=m_3$.
    Suppose that the second rotation of $m_3$ does not belong to $K$.
    Let $D'$ be the downset obtained by executing $\rho_2$ from $D$.
    Similarly, $K\setminus\{\rho_2\}$ is a cone in the same direction from the new reference downset $D'$ and contains $\rho_1,\tau,\sigma$.
    Since this cone does not involve $m_3$, its support is contained in $\{m_1,m_2,m_4\}$.
    If $m_4$ were also absent, every rotation would swap $m_1,m_2$. A repeated swap would return them to former partners, violating the non-revisiting property.
    The support is therefore $\{m_1,m_2,m_4\}$. Since the cone contains at least three rotations, this contradicts \autoref{lem:three-person-cone}.
    The case $v=m_4$ is analogous.
    Thus, the second rotation of every $v\in S(K)$ belongs to $K$.
    This implies $K_v(2)\subseteq K$.
    Let $w$ be the other participant in the first swap of $v$.
    From this swap until just before his second rotation, $v$ is matched to the partner of $w$ in the reference state $\mu_D$.
    If the second rotation of $v$ were also the swap $(v,w)$, it would return $w$ to his partner in the reference state, violating the non-revisiting property.
    The second rotation of $v$ therefore involves a man from the other pair. By downward closure, that man's first rotation also belongs to $K_v(2)$, so $K_v(2)$ contains both $\rho_1$ and $\rho_2$.
    Hence $s_v(2)=4$ for every $v\in S(K)$.
\end{proof}

The next lemma rules out an image participating in three rotations within a clean four-person cone.
In the image charging argument of the next subsection, this bound ensures that, in the clean case, the third principal cone of an image who is not self-sufficient contains a fifth man.
\begin{lemma}[Image rotation bound]
    \label{lem:four-person-cone-image}
    Let $K$ be a clean four-person cone.
    Every man $v\in I\cap S(K)$ satisfies $|C_v\cap K|\leq 2$.
\end{lemma}
\begin{proof}
    By \autoref{lem:pair-cone-properties}(i), if $K$ were of pair type, $v$ would be simple, contradicting $v\in I$. Thus, $K$ is of fork type.
    Write $\rho=(c,d)$ for the swap that is the unique minimal element of $K$, and set $a=\phi(c)$, $b=\phi(d)$. By \autoref{lem:fork-cone-properties}(i), $S(K)=\{a,b,c,d\}$, and since $c,d\in C$ are simple, the only images in this support are $a,b$.
    As established in the proof of that lemma, $K$ contains the three swaps $\rho,\tau_1=(c,a),\tau_2=(d,b)$.
    Without loss of generality, assume $v=a$.
    Since $K_a(1)\cup K_b(1)=\{\rho,\tau_1,\tau_2\}$ is a cone, choose a linear extension of $K$ that begins by executing $\rho,\tau_1,\tau_2$ in this order.
    Write $w_x=\mu_D(x)$ ($x\in\{a,b,c,d\}$) for the partners in the reference state. Immediately after these three rotations, the current partners and the partner histories are as follows.
    \begin{center}
        \begin{tabular}{c|c|c|c}
            Man & Current partner & \shortstack{Partners encountered in order\\(including the initial state)} & \shortstack{Partners not yet\\encountered among the four} \\
            \hline
            $c$ & $w_a$           & $w_c,w_d,w_a$                                                             & $w_b$                                                     \\
            $d$ & $w_b$           & $w_d,w_c,w_b$                                                             & $w_a$                                                     \\
            $a$ & $w_d$           & $w_a,w_d$                                                                 & $w_c,w_b$                                                 \\
            $b$ & $w_c$           & $w_b,w_c$                                                                 & $w_d,w_a$
        \end{tabular}
    \end{center}
    By the non-revisiting property, the only partner that $c$ can next obtain by executing rotations in $K$ is $w_b$; for $d$, the only such partner is $w_a$.
    These are currently the partners of $d,c$, respectively. Since a rotation is a single cyclic permutation, the only possible rotation involving $c$ or $d$ in this state is the swap $(c,d)$.
    The only possible rotation involving neither $c$ nor $d$ is the swap $(a,b)$ of the remaining two men.
    After the swap $(c,d)$, both $c,d$ have encountered all four partners, so the non-revisiting property prevents them from participating in any subsequent rotation of $K$.
    If the swap $(a,b)$ is executed first, the partners and histories of $c,d$ remain unchanged, so the only possible rotation involving either of them is still the swap $(c,d)$.
    Moreover, the swap $(c,d)$ does not change the partners of $a,b$. Thus, executing the swap $(a,b)$ twice would return them to former partners, violating the non-revisiting property.
    Hence, in either order, $a$ can participate in at most one additional rotation, namely the swap $(a,b)$.
    Among the first three rotations, $a$ participates only in $\tau_1$, so $|C_a\cap K|\leq2$.
\end{proof}

\subsection{Choosing Charge Recipients}\label{subsec:choosing-recipients}
We now show that each simple man or image who is not self-sufficient can assign his required credit to a suitable strong man.

Fix a downset $D$ and an outward direction.
We say that a man's required amount is charged to a strong man $h$ when the request for that amount is assigned to $h$.
Since $\phi: C\to I$ is a bijection, its inverse $\phi^{-1}: I\to C$ is well-defined. We call $\phi^{-1}(a)$ the preimage of $a$.

To show that a candidate charge recipient is strong, we compare his first rotation with a principal cone of another man participating in that rotation. We first establish a general containment property for use in the two charging arguments below.
\begin{lemma}[First-rotation containment]
    \label{lem:first-rotation-containment}
    Suppose that the first rotation of a man $h$ exists, and denote it by $\tau$. Let $v\in M(\tau)$ and let $k$ be a positive integer such that the $k$th rotation of $v$ exists and $h\notin S_v(k)$. Then
    \[
        K_v(k)\subseteq K_h(1),\qquad s_h(1)\geq s_v(k)+1.
    \]
\end{lemma}
\begin{proof}
    Let $\rho$ be the $k$th rotation of $v$.
    The rotations $\rho$ and $\tau$ share the man $v$, so they are comparable in the corresponding outward order.
    If $\tau$ preceded or equaled $\rho$ in this order, the definition of a principal cone would give $\tau\in K_v(k)$. Since $h$ participates in $\tau$, this would imply $h\in S_v(k)$, a contradiction.
    Thus, $\rho$ strictly precedes $\tau$ in this outward order, giving $K_v(k)\subseteq K_h(1)$.
    Hence $S_v(k)\subseteq S_h(1)$, and since $h\in S_h(1)\setminus S_v(k)$, we obtain $s_h(1)\geq s_v(k)+1$.
\end{proof}

Both charging arguments begin with a cone $K$ whose support has size four. If $K$ is not clean, a strong man in $S(K)$ serves as a charge recipient.
In the clean case, we use the respective hypotheses to choose a man $h\notin S(K)$ whose first rotation involves some $v\in S(K)$.
The structural lemmas in \autoref{subsec:four-person-structure} then provide a principal cone $K_v(k)\subseteq K$ with $s_v(k)\geq3$, so \autoref{lem:first-rotation-containment} gives $s_h(1)\geq4$.
The remaining support conditions are verified separately in each argument.

We first consider where to charge a simple man $m$ who is not self-sufficient.
We require the recipient $h$ to have a first principal-cone support of size at most five containing the requester $m$, so that we can later bound the total amount charged to $h$.
\begin{lemma}
    \label{lem:charging-simple}
    For every simple man $m$ who is not self-sufficient, there exists a strong man $h$ satisfying
    $$4\le s_h(1)\le 5,\quad m\in S_h(1).$$
    \ignorespaces
\end{lemma}
\begin{proof}
    Let $m$ be any simple man who is not self-sufficient.
    By \autoref{def:self-sufficient-strong}, $m$ is simple and satisfies $s_m(3)\le 4$ and $s_m(4)\le 5$.
    \autoref{lem:support-size} gives $s_m(3)\ge 4$ and $s_m(4)\ge 5$, so $s_m(3)=4$ and $s_m(4)=5$.
    Let $K=K_m(3)$ be the third principal cone of $m$.
    The support of $K$ has size four.
    \par\smallskip
    \noindent\textbf{Case 1: $K$ is not clean.}
    Then there exists a strong man $h\in S(K)$.
    The first rotation of $h$ belongs to $K$, so $S_h(1)\subseteq S(K)$. Since $h$ is strong, $s_h(1)\geq 4$.
    Thus, $S_h(1)=S(K),s_h(1)=4$, and $m\in S(K)= S_h(1)$.
    \par\smallskip
    \noindent\textbf{Case 2: $K$ is clean.}
    By \autoref{lem:four-person-cone-swap} and \autoref{def:four-person-cone-types}, $K$ is of fork type or pair type.
    Let $h$ be the unique man in $S_m(4)\setminus S_m(3)$.
    Let $\tau$ be the first rotation of $h$, and choose a man $v$ other than $h$ who participates in $\tau$.
    Since $\tau\in K_m(4)$ and $S_m(4)=S(K)\cup\{h\}$, we have $v\in S(K)$.
    Downward closure also gives $K_h(1)\subseteq K_m(4)$, so $S_h(1)\subseteq S_m(4)$.
    Hence $s_h(1)\le s_m(4)=5$.
    \par
    \noindent\emph{Case 2(a): Fork type.}
    By \autoref{lem:fork-cone-properties}(ii), there exists a positive integer $k$ such that $K_v(k)\subseteq K$ and $s_v(k)=3$.
    Since $K_v(k)\subseteq K$ and $h\notin S(K)$, we have $h\notin S_v(k)$, so \autoref{lem:first-rotation-containment} gives $K_v(k)\subseteq K_h(1)$ and $s_h(1)\ge4$.
    By \autoref{lem:fork-cone-properties}(i) and the fact that images are not simple, the only simple men in $S(K)$ are the two participants in the swap $\rho$ that is the unique minimal element of $K$. Thus, $m$ participates in $\rho$.
    The nonempty downset $K_v(k)\subseteq K$ contains $\rho$, giving $m\in S_v(k)\subseteq S_h(1)$.
    \par
    \noindent\emph{Case 2(b): Pair type.}
    Since the third rotation of $m$ belongs to $K$, \autoref{lem:pair-cone-properties}(ii) gives $K_v(2)\subseteq K$ and $s_v(2)=4$.
    Since $h\notin S(K)$ implies $h\notin S_v(2)$, \autoref{lem:first-rotation-containment} gives $s_h(1)\ge s_v(2)+1=5$.
    Together with $S_h(1)\subseteq S_m(4)$ and $s_m(4)=5$, this yields $s_h(1)=5$ and $S_h(1)=S_m(4)$, so in particular $m\in S_h(1)$.
\end{proof}

We next consider where to charge an image $a$ who is not self-sufficient.
We will count image requests sent to a strong man $h$ by their distinct preimages in $S_h(1)$.
We also require this support to contain an image, ensuring that besides $h$ it contains another non-simple man and thus leaving fewer positions for simple men.
\begin{lemma}
    \label{lem:charging-image}
    For every image $a$ who is not self-sufficient, there exists a strong man $h$ satisfying
    $$4\le s_h(1)\le 5,\quad \phi^{-1}(a)\in S_h(1),\quad I\cap S_h(1)\neq\emptyset.$$
    \ignorespaces
\end{lemma}
\begin{proof}
    Let $a$ be any image who is not self-sufficient.
    By \autoref{def:self-sufficient-strong}, $a\in I$ and $s_a(2)\le 4$ and $s_a(3)\le 5$.
    \autoref{lem:image-properties} also gives $s_a(2)\ge 4$, so $s_a(2)=4$.
    Let $K=K_a(2)$ be the second principal cone of $a$.
    By \autoref{lem:image-properties}, $K_{\phi^{-1}(a)}(2)=K_a(1)\subseteq K$, so the preimage $\phi^{-1}(a)$ belongs to the support of $K$.
    The support of $K$ has size four.
    \par\smallskip
    \noindent\textbf{Case 1: $K$ is not clean.}
    Then there exists a strong man $h\in S(K)$.
    The first rotation of $h$ belongs to $K$, so $S_h(1)\subseteq S(K)$. Since $h$ is strong, $s_h(1)\geq 4$.
    Thus, $S_h(1)=S(K)$, so $s_h(1)=4$ and $\phi^{-1}(a)\in S(K)=S_h(1)$.
    Moreover, since $a\in I$ and $a\in S(K)=S_h(1)$, we have $I\cap S_h(1)\neq\emptyset$.
    \par\smallskip
    \noindent\textbf{Case 2: $K$ is clean.}
    If it were of pair type, \autoref{lem:pair-cone-properties}(i) would make every man in $S(K)$ simple, contradicting the fact that $a$ is an image.
    Hence $K$ is of fork type.
    \autoref{lem:support-size} gives $s_a(3)\ge 4$, so $s_a(3)$ is either four or five.
    First consider the case $s_a(3)=4$.
    Nesting of the supports gives $S_a(2)\subseteq S_a(3)$. Since both have size four, $S_a(3)=S(K)$.
    This support contains no strong man, so $K_a(3)$ is also clean.
    Thus, $K'=K_a(3)$ is a clean four-person cone containing three rotations in which the image $a$ participates, contradicting \autoref{lem:four-person-cone-image}.
    Therefore, $s_a(3)=5$.
    Let $h$ be the unique man in $S_a(3)\setminus S_a(2)$.
    Let $\tau$ be the first rotation of $h$, and choose a man $v$ other than $h$ who participates in $\tau$.
    Since $\tau\in K_a(3)$ and $S_a(3)=S(K)\cup\{h\}$, we have $v\in S(K)$.
    Downward closure also gives $K_h(1)\subseteq K_a(3)$, so $S_h(1)\subseteq S_a(3)$.
    Hence $s_h(1)\leq s_a(3)=5$.
    By \autoref{lem:fork-cone-properties}(ii), there exists a positive integer $k$ such that $K_v(k)\subseteq K$ and $s_v(k)=3$.
    Since $K_v(k)\subseteq K$ and $h\notin S(K)$, we have $h\notin S_v(k)$, so \autoref{lem:first-rotation-containment} gives $K_v(k)\subseteq K_h(1)$ and $s_h(1)\ge4$.
    Write $\rho=(c,d)$ for the swap that is the unique minimal element of $K$. By \autoref{lem:fork-cone-properties}(i), we have $c,d\in C$ and $S(K)=\{c,d,\phi(c),\phi(d)\}$.
    The nonempty downset $K_v(k)\subseteq K$ contains $\rho$, so $\{c,d\}\subseteq S_v(k)$. Since $S_v(k)\subseteq S(K)$ and $|S_v(k)|=3$, this support consists of $c,d$ and one image.
    Since $a\in I\cap S(K)$, its preimage belongs to $\{c,d\}$. The inclusion $S_v(k)\subseteq S_h(1)$ therefore gives both $\phi^{-1}(a)\in S_h(1)$ and $I\cap S_h(1)\neq\emptyset$.
\end{proof}

Thus, each simple man who is not self-sufficient can charge his required amount $t$ to a strong man provided by \autoref{lem:charging-simple}, and each image who is not self-sufficient can charge his required amount $4t$ to a strong man provided by \autoref{lem:charging-image}.
We now establish properties of this charging scheme.

\subsection{Bounding the Load and Total Savings}\label{subsec:load-total-credit}
Using the charging scheme defined in \autoref{subsec:choosing-recipients}, we show that the total amount charged to any strong man $h$ is at most $10t$.

Continue to fix a downset $D$ and an outward direction.
We also fix the assignment of charges as follows.

\begin{itemize}
    \item For each simple man $m$ who is not self-sufficient, charge $t$ to a strong man $h$ provided by \autoref{lem:charging-simple}.
    \item For each image $a$ who is not self-sufficient, charge $4t$ to a strong man $h$ provided by \autoref{lem:charging-image}.
\end{itemize}
There may be several strong men $h$ satisfying the required conditions. For each requester, choose one arbitrarily and fix this choice.

\begin{definition}
    Let $h$ be a strong man.
    Let $n_s(h)$ and $n_i(h)$ denote the numbers of simple men and images, respectively, whose requests are charged to $h$.
    Write $\ell(h)$ for the total amount charged to $h$.
    That is,
    $$\ell(h)=t\,n_s(h)+4t\,n_i(h).$$
\end{definition}

The recipient conditions established in the preceding subsection let us bound $n_s(h)$ and $n_i(h)$ separately using the simple men in $S_h(1)$.
Each simple request corresponds to its requester in $S_h(1)$, while image requests correspond to distinct preimages in $C\cap S_h(1)$ by the injectivity of $\phi^{-1}$.

If $h$ receives an image request, then $s_h(1)\le5$, and $S_h(1)$ contains both $h$ and an image, which are distinct non-simple men.
Hence at most three men in $S_h(1)$ are simple. The following parity property reduces this bound to two.
\begin{lemma}
    \label{lem:even-simple}
    For every cone $K$, the number of simple men in $S(K)$ is even.
\end{lemma}
\begin{proof}
    Suppose that $\rho$ is a minimal element of $K$ that is a swap.
    By downward closure of $K$, $\rho$ is also minimal in the full outward order.
    Thus, each participant in $\rho$ has first principal cone $\{\rho\}$, so both are simple and belong to $S(K)$.

    Conversely, the first principal cone of a simple man $m$ in $S(K)$ consists of a single swap $\rho$.
    Since $K$ contains a rotation in which $m$ participates, the chain of his participating rotations and downward closure of $K$ imply that his first rotation $\rho$ also belongs to $K$.
    Moreover, $K_m(1)=\{\rho\}$ implies that $\rho$ is minimal in the full outward order and hence also in $K$.

    If two swaps that are distinct minimal elements of $K$ shared a man, the chain of his participating rotations would make them comparable, contradicting minimality. Thus, the sets of men participating in these swaps are pairwise disjoint.
    It follows that the number of simple men in $S(K)$ is twice the number of minimal elements of $K$ that are swaps.
\end{proof}
\begin{lemma}
    \label{lem:load-bound}
    For every strong man $h$, $\ell(h)\le 10t$.
\end{lemma}
\begin{proof}
    If both $n_s(h)$ and $n_i(h)$ are zero, then $\ell(h)=0$. We therefore assume that at least one is positive.

    First consider the case $n_i(h)=0$.
    In this case, only simple men who are not self-sufficient charge requests to $h$.
    By \autoref{lem:charging-simple}, every simple man $m$ whose request is charged to $h$ belongs to $S_h(1)$.
    We also have $s_h(1)\le 5$.
    Since $h\in S_h(1)$ and $h$ is not simple, $S_h(1)$ contains at most four simple men.
    Hence $n_s(h)\le 4$ and $\ell(h)=n_s(h)\cdot t\le 4t<10t$.

    Next consider the case $n_i(h)>0$.
    By \autoref{lem:charging-image}, every image $a$ whose request is charged to $h$ satisfies $\phi^{-1}(a)\in S_h(1)$.
    The same lemma gives $s_h(1)\le 5$.
    We also have $I\cap S_h(1)\neq\emptyset$.
    Thus, $S_h(1)$ contains at least one image.
    Moreover, $h\in S_h(1)$.
    Since $I$, the set of simple men, and the set of strong men are pairwise disjoint, $S_h(1)$ contains at most three simple men.
    By \autoref{lem:even-simple}, the number of simple men in $S_h(1)$ is even, so there are at most two.
    Since $\phi^{-1}:I\to C$ is injective and every man in $C$ is simple, at most two images have their preimages in $S_h(1)$.
    Consequently, $n_i(h)\le 2$.
    Also, \autoref{lem:charging-simple} implies that every simple man $m$ whose request is charged to $h$ belongs to $S_h(1)$, so $n_s(h)\le 2$.
    Combining these bounds yields $\ell(h)=n_s(h)\cdot t+n_i(h)\cdot 4t\le 2t+2\cdot 4t=10t$.
\end{proof}

We have shown that the total amount charged to any strong man $h$ is at most $10t$.
Thus, the additional credits of strong men suffice to cover the total amount required by simple men and images that are not self-sufficient.
We use this fact to prove the following lower bound on the sum of the additional credits.

Below, $B,C,I$ denote the classes for the fixed $D$ and the direction corresponding to the sign in $g_{\mathrm{add}}^\pm$.
\begin{lemma}
    \label{lem:additional-credits-sum}
    The sum of the additional credits over all men satisfies
    \[
        \sum_{m\in M}g_{\mathrm{add}}^\pm(D,m)
        \ge t(|B|+|C|)+4t|I|.
    \]
    \ignorespaces
\end{lemma}
\begin{proof}
    Each self-sufficient simple man can cover his required amount $t$ with his own additional credit, and each self-sufficient image can likewise cover his required amount $4t$.
    The request of every other simple man or image is charged to a single strong man.
    By \autoref{lem:load-bound}, the total amount requested from each strong man $h$ satisfies $\ell(h)\le10t=g_{\mathrm{add}}^\pm(D,h)$, so his additional credit suffices to cover it.
    Hence the total additional credit is at least the total amount required by all simple men and images.
    The set of simple men is $B\cup C$ and is disjoint from $I$, so this total required amount is $t(|B|+|C|)+4t|I|$.
\end{proof}

\subsection{Averaging Credits and the Main Theorem}
Following the approach of \autoref{sec:histories-balance}, we average the sum of the basic and additional credits to improve the upper bound on the number of stable matchings.
In each man's entropy bound, we continue to use the additional credit $g_{\mathrm{add}}^\pm(D,m)$ obtained from his own tail-probability deficits.
The redistribution in the preceding subsection shows that the total of these credits is at least the total amount required by simple men and images; we use this lower bound on the total to bound the average additional credit.

We first define the average additional credit.
\begin{definition}
    Choose $(\mu,m)$ uniformly from $\Omega\times M$ and set $D=D(\mu)$.
    Define the average additional credit by
    $$g_{\mathrm{add}} = \EE_{D,m}[g_{\mathrm{add}}^+(D,m)+g_{\mathrm{add}}^-(D,m)].$$
\end{definition}
As discussed in \autoref{subsec:credit-budgets}, this quantity can be added to $g_{\mathrm{basic}}$.
\begin{lemma}
    \label{lem:basic-additional-credits}
    $$\log |\Omega| \leq n(C_0-g_{\mathrm{basic}}-g_{\mathrm{add}}).$$
\end{lemma}
\begin{proof}
    As discussed in \autoref{subsec:credit-budgets}, the additional credits exclude the amounts already counted as basic credits. Thus, for each man $m$ and each direction, we may count $g_{\mathrm{basic}}^\pm(D,m)+g_{\mathrm{add}}^\pm(D,m)$.
    By \autoref{lem:delta-d-savings}, for each fixed $D,m$, the savings from the two directions can be added, giving
    $$\EE_U[\log N_m(D,U)\mid D]\leq C_0 -g_{\mathrm{basic}}^+(D,m)-g_{\mathrm{basic}}^-(D,m) -g_{\mathrm{add}}^+(D,m)-g_{\mathrm{add}}^-(D,m).$$
    \ignorespaces
    Averaging over $D$ and summing over all men, the definitions of $g_{\mathrm{basic}}$ and $g_{\mathrm{add}}$ give
    $$\sum_{m\in M}\EE_{D,U}[\log N_m(D,U)]\leq n(C_0-g_{\mathrm{basic}}-g_{\mathrm{add}}).$$
    \ignorespaces
    Applying \autoref{lem:entropy-bound-candidates} proves the claim.
\end{proof}

\begin{lemma}
    \label{lem:additional-credits-average}
    The average additional credit satisfies
    $$g_{\mathrm{add}} \geq te_s + 8tp_c.$$
\end{lemma}
\begin{proof}
    By \autoref{lem:additional-credits-sum}, for a fixed downset $D$ and outward direction, the sum of the additional credits is at least $t(|B|+|C|)+4t|I|$.
    Since $|C|=|I|$, this sum is at least $t(|B|+|C|)+4t|C|$.
    To distinguish the outward directions, write $B^+$ and $C^+$ for $B$ and $C$ in the forward direction, and $B^-$ and $C^-$ for $B$ and $C$ in the backward direction.
    Then the average additional credit satisfies
    \begin{align*}
        g_{\mathrm{add}} & = \EE_{D,m}[g_{\mathrm{add}}^+(D,m)+g_{\mathrm{add}}^-(D,m)]                         \\
                         & \geq t\EE_{D}[\frac{|B^+|+|C^+|+|B^-|+|C^-|}{n}] + 4t\EE_{D}[\frac{|C^+|+|C^-|}{n}].
    \end{align*}
    Here, $\EE_{D}[\frac{|B^+|+|C^+|+|B^-|+|C^-|}{n}]$ is the sum of the probabilities that the backward and forward directions are simple for a pointed matching chosen uniformly from $\Omega\times M$, and thus equals $e_s$.
    Similarly, $\EE_{D}[\frac{|C^+|+|C^-|}{n}]$ is the sum of the probabilities that $m$ belongs to $C$ in the backward and forward directions for a pointed matching chosen uniformly from $\Omega\times M$. By \autoref{lem:pointed-bijection} and the definition of $p_c$, this sum equals $2p_c$.
    Substituting these identities gives $g_{\mathrm{add}}\geq te_s+8tp_c$.
\end{proof}
We next bound the sum of the average basic and additional credits.
The additional credit alone need not admit a uniform lower bound of $4t$.
However, as the average number of simple directions decreases, the basic credit increases sufficiently to compensate for a possible decrease in the guaranteed additional credit.
The following lemma uses this tradeoff to bound their sum.
\begin{lemma}
    \label{lem:credits-average}
    The sum of the average basic and additional credits satisfies
    $$g_{\mathrm{basic}} + g_{\mathrm{add}} \geq \frac{2}{3}(\delta_1 + 2\delta_2) + 4\delta_4.$$
\end{lemma}
\begin{proof}
    By \autoref{lem:basic-credits-average-bound}, $g_{\mathrm{basic}} = 2\delta_1 - (\delta_1 - \delta_2)e_s$.
    Also, \autoref{lem:additional-credits-average} gives $g_{\mathrm{add}} \geq te_s + 8tp_c$.
    Thus, $g_{\mathrm{basic}} + g_{\mathrm{add}} \geq 2\delta_1 - (\delta_1 - \delta_2-t)e_s + 8tp_c$.
    Since $p_c \geq e_s - 1$ by \autoref{lem:es-bound}, we obtain
    \begin{align*}
        g_{\mathrm{basic}} + g_{\mathrm{add}} & \geq 2\delta_1 - (\delta_1 - \delta_2-t)e_s + 8t(e_s - 1) \\
                                              & = 2\delta_1 - 8t - (\delta_1 - \delta_2 - 9t)e_s.
    \end{align*}
    \ignorespaces
    \autoref{lem:delta-inequalities} in \autoref{app:credit-budgets} gives $\delta_1 - \delta_2 - 9t > 0$.
    Since $t=\delta_4$ and $e_s \leq 4/3$ by \autoref{lem:es-bound}, we obtain the following bound.
    \begin{align*}
        g_{\mathrm{basic}} + g_{\mathrm{add}} & \geq 2\delta_1 - 8t - \frac{4}{3}(\delta_1 - \delta_2 - 9t) \\
                                              & = \frac{2}{3}(\delta_1 + 2\delta_2) + 4\delta_4.
    \end{align*}
\end{proof}

\begin{theorem}
    \label{thm:main}
    For a stable matching instance with $n\ge1$ men and $n$ women, the number of stable matchings $|\Omega|$ satisfies
    $$|\Omega| < 2.999^n.$$
    \ignorespaces
\end{theorem}
\begin{proof}
    \autoref{lem:basic-additional-credits} and \autoref{lem:credits-average} give
    \begin{align*}
        \log |\Omega| & \leq n(C_0 - g_{\mathrm{basic}} - g_{\mathrm{add}})          \\
                      & \leq n(C_0 - \frac{2}{3}(\delta_1 + 2\delta_2) - 4\delta_4).
    \end{align*}
    \ignorespaces
    By \autoref{lem:numerical-thresholds} in \autoref{app:numerical-thresholds}, we have
    $$C_0 - \frac{2}{3}(\delta_1 + 2\delta_2) - 4\delta_4 < \log(2.999),$$
    so $\log |\Omega| < n\log(2.999)$, proving the claim.
\end{proof}

\section{Conclusion}\label{sec:conclusion}

In this paper, we obtained a new upper bound of $2.999^n$ on the number of stable matchings with $n$ men and $n$ women. In \autoref{sec:cones-entropy}, we introduce the notions of cones and supports, and prove that the support size of the $k$-th principal cone satisfies $s_m(k) \geq k+1$. This establishes a $3.332^n$ bound. In \autoref{sec:histories-balance}, we utilize the cone of up to three people and control how often unfavorable conditions (such as $s_m(1) = 2$ and $s_m(2) = 3$) can occur. This establishes a $3.060^n$ bound. In \autoref{sec:charging}, we extend the discussion to four people using the charging argument. This establishes the main $2.999^n$ bound, which is less than $3^n$.

However, a gap still remains between the known lower bound of $\frac{2.28^n}{\text{poly}(n)}$ and our main result of $2.999^n$. We strongly believe we can obtain a better upper bound by extending our proof method, possibly with computer-assisted case analysis. For example, computing local structures with cones of five people or more, or refining the case analysis to exploit configurations with larger supports, such as the $s_m(1) \geq 5$ or $s_m(1) \geq 6$ cases, may be possible next options. Finally, we conclude this paper by proposing the following conjecture:

\begin{conjecture}
    Every stable matching instance with $n$ men and $n$ women admits fewer than $2.3^n$ stable matchings.
\end{conjecture}
\section*{AI Disclosure}

The authors extensively used OpenAI's large language model, GPT-6 Astra, during the proof exploration phase. Without using AI, the authors initially obtained a weaker upper bound of $3.82^n$. This is still better than the $4.01^n + O(1)$ bound obtained by correcting the numerical calculation in \cite{palmer2022most}. We obtained this $3.82^n$ bound by extending the \cite{palmer2022most} framework (with $2n$ chains) from two to three auxiliary participants, on each side of the boundary. At this point, we did not introduce the new notions of cones and supports; however, we obtained structural properties that imply $s_m(k) \geq k+1$ (\autoref{lem:support-size}) for $k \in \{1, 2, 3\}$. Then, we provided the proof draft of the $3.82^n$ bound to GPT-6 Astra, and through an iterative process with the authors, GPT-6 Astra generalized the proof of $s_m(k) \geq k+1$ for all $k$, and we established the final $2.999^n$ bound.

This paper is entirely written by human authors. However, we used GPT-5.6 Sol, GPT-6 Sol, and GPT-6 Astra to polish the presentation of this paper. Finally, the authors take full responsibility for the correctness and the contents of the paper.

\printbibliography
\appendix

\newpage

\section{Inequalities Certifying the Credit Budgets}\label{app:credit-budgets}
We prove inequalities for the savings $\delta_1,\delta_2,\delta_3,\delta_4$ obtained from tail-probability deficits.
By \autoref{def:delta-d}, $\delta_d$ is given by
$$\delta_d = \sum_{j\geq d} \frac{2\log \frac{j+1}{j}}{(j+1)(j+2)(j+3)}.$$
Also, $t=\delta_4$ by \autoref{subsec:credit-budgets}.
Our targets are the four inequalities in \autoref{lem:delta-inequalities}: the first two certify the additional credit of $4t$ for self-sufficient images, and the third certifies the additional credit of $10t$ for strong men in \autoref{lem:additional-savings}.
The last inequality, $\delta_1-\delta_2-9t>0$, gives the sign of the coefficient needed to use $e_s\leq4/3$ when bounding the sum of the average basic and additional credits in \autoref{lem:credits-average}.

We define the difference between $\delta_i$ and $\delta_{i+1}$.
\begin{definition}
    $$\alpha_i=\delta_i-\delta_{i+1}\qquad(i=1,2,3).$$
\end{definition}
\begin{lemma}
    \label{lem:alpha-lower-bound}
    $$\alpha_1\geq \frac{1}{18},\quad \alpha_2\geq \frac{1}{75},\quad \alpha_3\geq \frac{1}{210}$$
\end{lemma}
\begin{proof}
    By definition,
    $$\alpha_i=\sum_{j\geq i} \frac{2\log \frac{j+1}{j}}{(j+1)(j+2)(j+3)}-\sum_{j\geq i+1} \frac{2\log \frac{j+1}{j}}{(j+1)(j+2)(j+3)}=\frac{2\log (1+\frac{1}{i})}{(i+1)(i+2)(i+3)}$$
    \ignorespaces
    The convexity of $1/x$ gives
    $$\log(1+\frac{1}{i})=\int_i^{i+1}\frac{1}{x}\,dx\geq\frac{1}{i+1/2}=\frac{2}{2i+1}$$
    and hence
    $$\alpha_i\geq \frac{4}{(2i+1)(i+1)(i+2)(i+3)}$$
    \ignorespaces
    Substituting $i=1,2,3$ yields
    $$\alpha_1\geq \frac{4}{3\cdot 2\cdot 3\cdot 4}=\frac{1}{18},\quad \alpha_2\geq \frac{4}{5\cdot 3\cdot 4\cdot 5}=\frac{1}{75},\quad \alpha_3\geq \frac{4}{7\cdot 4\cdot 5\cdot 6}=\frac{1}{210}$$
    \ignorespaces
\end{proof}

\begin{lemma}
    \label{lem:delta4-upper-bound}
    $$\delta_4<\frac{1}{192}$$
\end{lemma}
\begin{proof}
    For $j\geq 4$, the alternating series expansion of the logarithm gives
    \begin{align*}
        \log(1+\frac{1}{j}) & \leq \frac{1}{j}-\frac{1}{2j^2}+\frac{1}{3j^3}  \\
                            & \leq \frac{1}{j}-\frac{1}{2j^2}+\frac{1}{12j^2} \\
                            & = \frac{1}{j}-\frac{5}{12j^2}
    \end{align*}

    It follows that
    \begin{align*}
        \delta_4 & = \sum_{j\geq 4} \frac{2\log \frac{j+1}{j}}{(j+1)(j+2)(j+3)}                                  \\
                 & \leq \sum_{j\geq 4} \frac{2}{j(j+1)(j+2)(j+3)} - \sum_{j\geq 4} \frac{5}{6j^2(j+1)(j+2)(j+3)}
    \end{align*}
    \ignorespaces
    Telescoping the first sum gives
    \begin{align*}
        \sum_{j\geq 4} \frac{2}{j(j+1)(j+2)(j+3)} & = \frac{2}{3}\sum_{j\geq 4} \left(\frac{1}{j(j+1)(j+2)}-\frac{1}{(j+1)(j+2)(j+3)}\right) \\
                                                  & = \frac{2}{3}\cdot\frac{1}{4\cdot 5\cdot 6}                                              \\
                                                  & = \frac{1}{180}
    \end{align*}
    For the second sum,
    \begin{align*}
        \sum_{j\geq 4} \frac{5}{6j^2(j+1)(j+2)(j+3)} & = \frac{5}{6}\left(\frac{1}{4^2\cdot 5\cdot 6\cdot 7} + \sum_{j\geq 5} \frac{1}{j^2(j+1)(j+2)(j+3)}\right) \\
                                                     & \geq \frac{5}{6}\left(\frac{1}{3360} + \sum_{j\geq 5} \frac{1}{j(j+1)(j+2)(j+3)(j+4)}\right)
    \end{align*}
    Telescoping the sum $\sum_{j\geq 5} \frac{1}{j(j+1)(j+2)(j+3)(j+4)}$ gives
    \begin{align*}
         & \sum_{j\geq 5} \frac{1}{j(j+1)(j+2)(j+3)(j+4)}                                                     \\
         & = \frac{1}{4}\sum_{j\geq 5} \left(\frac{1}{j(j+1)(j+2)(j+3)}-\frac{1}{(j+1)(j+2)(j+3)(j+4)}\right) \\
         & = \frac{1}{4}\cdot\frac{1}{5\cdot 6\cdot 7\cdot 8}                                                 \\
         & = \frac{1}{6720}
    \end{align*}
    Combining these estimates yields
    $$\delta_4 \leq \frac{1}{180}-\frac{5}{6}\left(\frac{1}{3360}+\frac{1}{6720}\right) = \frac{209}{40320} <\frac{1}{192}$$
\end{proof}

\begin{lemma}
    \label{lem:delta-inequalities}
    $$\frac{1}{2}\delta_2 + \delta_3 > 4t, \quad \frac{11}{7}\delta_3 + \delta_4 > 4t, \quad \frac{2}{5}\delta_1 + \delta_2 > 10t, \quad \delta_1 - \delta_2 - 9t > 0$$
\end{lemma}
\begin{proof}
    By the definition of $\alpha_i$, $\delta_1 = \alpha_1 + \alpha_2 + \alpha_3 + t, \delta_2 = \alpha_2 + \alpha_3 + t, \delta_3 = \alpha_3 + t$.
    Using \autoref{lem:alpha-lower-bound} and \autoref{lem:delta4-upper-bound}, we obtain
    \begin{align*}
        \frac{1}{2}\delta_2+\delta_3-4t
         & = \frac{1}{2}\alpha_2+\frac{3}{2}\alpha_3-\frac{5}{2}t
        > \frac{1}{150}+\frac{1}{140}-\frac{5}{384}=\frac{53}{67200}>0,                \\
        \frac{11}{7}\delta_3+\delta_4-4t
         & = \frac{11}{7}\alpha_3-\frac{10}{7}t
        > \frac{11}{1470}-\frac{10}{1344}=\frac{1}{23520}>0,                           \\
        \frac{2}{5}\delta_1+\delta_2-10t
         & = \frac{2}{5}\alpha_1+\frac{7}{5}\alpha_2+\frac{7}{5}\alpha_3-\frac{43}{5}t
        > \frac{1}{45}+\frac{7}{375}+\frac{1}{150}-\frac{43}{960}=\frac{199}{72000}>0, \\
        \delta_1-\delta_2-9t
         & = \alpha_1-9t
        > \frac{1}{18}-\frac{9}{192}=\frac{5}{576}>0.
    \end{align*}

\end{proof}

\section{Numerical Certificates for the Three Upper Bounds}\label{app:numerical-thresholds}
\label{app:eight-term-certificate}
\label{subsec:baseline-basic-thresholds}
\label{subsec:eight-term-certificate}
Our goal is \autoref{lem:numerical-thresholds}, which certifies the numerical inequalities that turn the entropy estimates in the main text into the three bounds: $3.332^n$ in \autoref{sec:cones-entropy}, $3.060^n$ in \autoref{prop:basic-credits-bound}, and $2.999^n$ in \autoref{thm:main}.
We split the series into finite sums and remainders and use rational bounds on logarithms, so that all three comparisons can be checked by finite rational arithmetic.
By \autoref{subsec:geometric-benchmark}, $C_0$ is given by
$$C_0 = \sum_{j\geq 1} \frac{2\log j}{(j+1)(j+2)}$$
Also, by \autoref{def:delta-d}, $\delta_d$ is given by
$$\delta_d = \sum_{j\geq d} \frac{2\log \frac{j+1}{j}}{(j+1)(j+2)(j+3)}.$$

\begin{lemma}
    \label{lem:C0-series}
    $$C_0 = \sum_{j\geq 1} \frac{2\log (1+\frac{1}{j})}{j+2}$$
\end{lemma}
\begin{proof}
    Partial fraction decomposition gives
    $$\frac{2\log j}{(j+1)(j+2)} = \frac{2\log j}{j+1}-\frac{2\log j}{j+2}$$
    \ignorespaces
    Thus, for every positive integer $K$,
    \begin{align*}
        \sum_{j=1}^{K}\frac{2\log j}{(j+1)(j+2)}
         & = \sum_{j=1}^{K}\frac{2\log j}{j+1}-\sum_{j=1}^{K}\frac{2\log j}{j+2}                           \\
         & = \sum_{j=1}^{K}\frac{2\log(j+1)}{j+2}-\frac{2\log(K+1)}{K+2}-\sum_{j=1}^{K}\frac{2\log j}{j+2} \\
         & = \sum_{j=1}^{K}\frac{2\log(1+\frac{1}{j})}{j+2}-\frac{2\log(K+1)}{K+2}
    \end{align*}
    As $K\to\infty$, the left-hand side converges to $C_0$ and the boundary term $2\log(K+1)/(K+2)$ tends to zero, proving the claim.
\end{proof}
We now split the series for $C_0$ into a finite sum and a remainder.
\begin{lemma}
    \label{lem:C0-upper-bound}
    For every positive integer $N$,
    $$C_0 < \sum_{j=1}^{N} \frac{2\log (1+\frac{1}{j})}{j+2} + \frac{1}{2(N+1)} + \frac{3}{2(N+2)}$$
\end{lemma}
\begin{proof}
    By \autoref{lem:C0-series}, $C_0 = \sum_{j\geq 1} \frac{2\log (1+\frac{1}{j})}{j+2}$.
    For every positive integer $j$, $\log(1+\frac{1}{j})=\int_j^{j+1}\frac{1}{x}\,dx$. The trapezoidal rule and the strict convexity of $1/x$ give $\log(1+\frac{1}{j}) < \frac{1}{2}(\frac{1}{j}+\frac{1}{j+1})$.
    It follows that
    \begin{align*}
        \sum_{j\geq N+1} \frac{2\log (1+\frac{1}{j})}{j+2} & < \sum_{j\geq N+1} \frac{1}{j(j+2)}+\sum_{j\geq N+1} \frac{1}{(j+1)(j+2)}                              \\
                                                           & = \sum_{j\geq N+1}\frac{1}{2}(\frac{1}{j}-\frac{1}{j+2})+\sum_{j\geq N+1}(\frac{1}{j+1}-\frac{1}{j+2}) \\
                                                           & = \frac{1}{2}(\frac{1}{N+1}+\frac{1}{N+2})+\frac{1}{N+2}                                               \\
                                                           & = \frac{1}{2(N+1)}+\frac{3}{2(N+2)}
    \end{align*}
    \ignorespaces
    Therefore,
    $$C_0 = \sum_{j=1}^{N} \frac{2\log (1+\frac{1}{j})}{j+2} + \sum_{j\geq N+1} \frac{2\log (1+\frac{1}{j})}{j+2} < \sum_{j=1}^{N} \frac{2\log (1+\frac{1}{j})}{j+2} + \frac{1}{2(N+1)} + \frac{3}{2(N+2)}$$
\end{proof}
\begin{lemma}
    \label{lem:delta2-lower-bound}
    For every integer $N\geq 2$,
    $$\delta_2 > \sum_{j=2}^{N} \frac{2\log (1+\frac{1}{j})}{(j+1)(j+2)(j+3)} + \frac{2}{3(N+2)(N+3)(N+4)}$$
\end{lemma}
\begin{proof}
    By definition, $\delta_2 = \sum_{j\geq 2} \frac{2\log (1+\frac{1}{j})}{(j+1)(j+2)(j+3)}$.
    For every positive integer $j$, $\log(1+\frac{1}{j})=\int_j^{j+1}\frac{1}{x}\,dx$. Since $1/x$ is strictly decreasing, $\log(1+\frac{1}{j}) > \frac{1}{j+1}$.
    It follows that
    \begin{align*}
        \sum_{j\geq N+1} \frac{2\log (1+\frac{1}{j})}{(j+1)(j+2)(j+3)} & > \sum_{j\geq N+1} \frac{2}{(j+1)^2(j+2)(j+3)}                                      \\
                                                                       & > \sum_{j\geq N+1} \frac{2}{(j+1)(j+2)(j+3)(j+4)}                                   \\
                                                                       & = \sum_{j\geq N+1} \frac{2}{3}(\frac{1}{(j+1)(j+2)(j+3)}-\frac{1}{(j+2)(j+3)(j+4)}) \\
                                                                       & = \frac{2}{3(N+2)(N+3)(N+4)}
    \end{align*}
    \ignorespaces
    Therefore,
    \begin{align*}
        \delta_2 & = \sum_{j=2}^{N} \frac{2\log (1+\frac{1}{j})}{(j+1)(j+2)(j+3)} + \sum_{j\geq N+1} \frac{2\log (1+\frac{1}{j})}{(j+1)(j+2)(j+3)} \\
                 & > \sum_{j=2}^{N} \frac{2\log (1+\frac{1}{j})}{(j+1)(j+2)(j+3)} + \frac{2}{3(N+2)(N+3)(N+4)}
    \end{align*}
\end{proof}

\begin{lemma}
    \label{lem:log-approximation}
    For every positive integer $p$ and every $z>1$, let $u=\frac{z-1}{z+1}$.
    $$2\sum_{r=0}^{p-1}\frac{u^{2r+1}}{2r+1} < \log z < 2\sum_{r=0}^{p-1}\frac{u^{2r+1}}{2r+1} + \frac{2u^{2p+1}}{(2p+1)(1-u^2)}$$
\end{lemma}
\begin{proof}
    For $z>1$, set $u=\frac{z-1}{z+1}$. Thus $z=\frac{1+u}{1-u}$ and $\log z = \log(1+u)-\log(1-u)$.
    Hence $\log z=2\int_0^u \frac{1}{1-x^2}\,dx$.
    For every positive integer $p$,
    $$\frac{1}{1-x^2} = \sum_{r=0}^{p-1} x^{2r} + \frac{x^{2p}}{1-x^2}$$
    Integrating gives
    $$\log z = 2\sum_{r=0}^{p-1}\frac{u^{2r+1}}{2r+1} + 2\int_0^u \frac{x^{2p}}{1-x^2}\,dx$$
    For $0<x<u<1$,
    $$\frac{x^{2p}}{1-x^2} < \frac{x^{2p}}{1-u^2}$$
    and hence
    $$\int_0^u \frac{x^{2p}}{1-x^2}\,dx < \frac{u^{2p+1}}{(2p+1)(1-u^2)}$$
    This proves the upper bound.
    For the lower bound, note that, for $0<x<u<1$,
    $$\frac{x^{2p}}{1-x^2} > 0$$
    \ignorespaces
\end{proof}

\begin{lemma}
    \label{lem:numerical-thresholds}
    $$C_0 < \log(3.332),\quad C_0-\frac{2}{3}(\delta_1+2\delta_2) < \log(3.060),\quad C_0-\frac{2}{3}(\delta_1+2\delta_2)-4\delta_4 < \log(2.999).$$
\end{lemma}
\begin{proof}
    First, we relate $\delta_1$ and $\delta_4$ to $\delta_2$:
    \begin{align*}
        \delta_1-\delta_2
         & = \frac{2\log2}{2\cdot3\cdot4}=\frac{\log2}{12},             \\
        \delta_2-\delta_4
         & = \frac{\log(3/2)}{30}+\frac{\log(4/3)}{60}=\frac{\log3}{60}
    \end{align*}
    Thus, the three quantities to be bounded can be written as
    \[
        C_0,\qquad C_0-2\delta_2-\frac{\log2}{18},\qquad
        C_0-6\delta_2-\frac{\log2}{18}+\frac{\log3}{15}
    \]
    \ignorespaces

    In what follows, let $N=24$ and $p=6$. For $z>1$, set $u(z)=(z-1)/(z+1)$ and define
    \[
        L(z)=2\sum_{r=0}^{5}\frac{u(z)^{2r+1}}{2r+1},\qquad
        H(z)=L(z)+\frac{2u(z)^{13}}{13(1-u(z)^2)}
    \]
    By \autoref{lem:log-approximation}, $L(z)<\log z<H(z)$.
    In particular, for $z=1+1/j$, $u(z)=1/(2j+1)$.

    Denote the bounds on the remainders in \autoref{lem:C0-upper-bound} and \autoref{lem:delta2-lower-bound} by
    \[
        A=\frac1{50}+\frac3{52}=\frac{101}{1300},\qquad
        B=\frac2{3\cdot26\cdot27\cdot28}=\frac1{29484}
    \]
    For $k\in\{0,2,6\}$, these lemmas give
    \begin{align*}
        C_0-k\delta_2
         & < \frac23\log2
        +\sum_{j=2}^{24}\frac2{j+2}
        \left(1-\frac{k}{(j+1)(j+3)}\right)\log(1+1/j)+A-kB.
    \end{align*}
    For $j\ge2$, $(j+1)(j+3)\ge15>k$, so the coefficient of each logarithm is positive.
    Define the rational number
    \[
        Q_k=\frac23 H(2)
        +\sum_{j=2}^{24}\frac2{j+2}
        \left(1-\frac{k}{(j+1)(j+3)}\right)H(1+1/j)+A-kB
    \]
    Then $C_0-k\delta_2<Q_k$.

    Using $\log3=\log2+\log(3/2)$, define
    \[
        R_0=Q_0,\qquad
        R_2=Q_2-\frac{L(2)}{18},\qquad
        R_6=Q_6-\frac{L(2)}{18}+\frac{H(2)+H(3/2)}{15}
    \]
    Then
    \begin{align*}
        C_0                                       & < R_0, \\
        C_0-\frac23(\delta_1+2\delta_2)           & < R_2, \\
        C_0-\frac23(\delta_1+2\delta_2)-4\delta_4 & < R_6
    \end{align*}
    On the other hand, for $b\in\{3.332,3.060,2.999\}$,
    \[
        D(b)=L(2)+L(b/2)<\log2+\log(b/2)=\log b
    \]
    \ignorespaces

    The quantities $R_k$ and $D(b)$ defined above are all rational. Evaluating the finite sums using rational arithmetic gives the following bounds.
    The terminating decimals in the table represent exact rational numbers. The upper bounds on $R_k$ are rounded upward, and the lower bounds on $D(b)$ are rounded downward.
    \begin{center}
        \begin{tabular}{cc|cc}
            $k$ & $b$     & Upper bound on $R_k$ & Lower bound on $D(b)$ \\
            \hline
            $0$ & $3.332$ & $1.203571720327$     & $1.203572615104$      \\
            $2$ & $3.060$ & $1.118295401900$     & $1.118414808924$      \\
            $6$ & $2.999$ & $1.097999925938$     & $1.098278792837$
        \end{tabular}
    \end{center}
    In each row, the upper bound on $R_k$ is smaller than the lower bound on $D(b)$. Thus $R_k<D(b)<\log b$, proving all three claims.
\end{proof}

\section{Correcting the Numerical Evaluation in Previous Work}
\label{app:previous-numerical-evaluation}

Palmer and P{\'a}lv{\"o}lgyi~\cite[proof of Claim~28, p.~226]{palmer2022most} state the following numerical bound:
\begin{equation*}
    \begin{aligned}
        o(1)+\frac{\log2}{12}+\frac{23\log3}{630}
        +\sum_{k\ge4}
        \frac{2k(k+7)+72}{(k+3)(k+5)(k+6)(k+7)}\log k
        \le 0.6331+o(1).
    \end{aligned}
\end{equation*}
In this appendix, we show that this numerical inequality does not hold and reevaluate the same series.
To this end, set
$$a_2=\frac1{12},\qquad a_3=\frac{23}{630},\qquad a_k=\frac{2k(k+7)+72}{(k+3)(k+5)(k+6)(k+7)}\quad(k\ge4)$$
and define
$$\sigma_m=\sum_{k=2}^{m}a_k\log k,\qquad \kappa_{\mathrm{PP}}=\sum_{k=2}^{\infty}a_k\log k$$
\ignorespaces
We omit the $k=1$ term because $\log1=0$.

\begin{lemma}
    \label{lem:pp-numerical-bounds}
    The following inequalities hold.
    \begin{align*}
        \sigma_{201}   & >0.6331,                                   \\
        0.693972343523 & <\kappa_{\mathrm{PP}}<0.693972362645,      \\
        4.006606753860 & <e^{2\kappa_{\mathrm{PP}}}<4.006606907079.
    \end{align*}
\end{lemma}

The first bound in \autoref{lem:pp-numerical-bounds} shows that the partial sum through $k=201$ already exceeds $0.6331$. Since every term is positive,
$$\kappa_{\mathrm{PP}}>\sigma_{201}>0.6331$$
and the positive constant gap cannot be absorbed into an $o(1)$ term.
Thus, the numerical inequality quoted at the beginning of this appendix is incorrect.
This identifies an error in a numerical inequality in the proof of Claim~28; it does not provide a counterexample to the claim itself.

Since the original paper sums the bounds over $2n$ coordinates, the corresponding exponential base is $e^{2\kappa_{\mathrm{PP}}}$. The remaining bounds in the lemma show that this base is approximately $4.006607$, so the derivation of $3.55$ from the quoted numerical inequality is invalid.

\begin{proof}
    We certify the three bounds in the lemma using rational arithmetic, treating the partial sums, the remainder, and the exponential function in that order.

    \paragraph{Rational bounds on partial sums.}
    We apply \autoref{lem:log-approximation} in \autoref{app:numerical-thresholds} with $p=20$.
    For $1<v\le2$, set $u=(v-1)/(v+1)$ and define
    $$L_{20}(v)=2\sum_{r=0}^{19}\frac{u^{2r+1}}{2r+1},\qquad H_{20}(v)=L_{20}(v)+\frac{2u^{41}}{41(1-u^2)}$$
    For any rational $z>1$, write $z=2^e v$, where $e\ge0$ is an integer and $1<v\le2$, and set
    $$\ell(z)=eL_{20}(2)+L_{20}(v),\qquad h(z)=eH_{20}(2)+H_{20}(v)$$
    Then $\ell(z)<\log z<h(z)$.
    Thus, the rational numbers
    $$\sigma_m^-=\sum_{k=2}^{m}a_k\ell(k),\qquad \sigma_m^+=\sum_{k=2}^{m}a_k h(k)$$
    satisfy $\sigma_m^-<\sigma_m<\sigma_m^+$.
    Evaluating the finite sums using rational arithmetic gives
    $$0.633311143065<\sigma_{201}^-,\qquad \sigma_{201}^+<0.633311143066$$
    In particular, $\sigma_{201}>0.6331$, proving the first bound in the lemma.

    \paragraph{Bounds on the remainder.}
    To evaluate the entire infinite series, for an integer $N\ge4$, we bound the remainder
    $$\rho_N=\sum_{k>N}a_k\log k$$
    \ignorespaces
    For real $t\ge4$, define
    $$a(t)=\frac{2t(t+7)+72}{(t+3)(t+5)(t+6)(t+7)}$$
    \ignorespaces
    Putting the fractions over common denominators gives
    \begin{align*}
        \frac2{t^2}-a(t)
         & =\frac{28t^3+250t^2+1062t+1260}
            {t^2(t+3)(t+5)(t+6)(t+7)}>0,            \\
        a(t)-\frac2{(t+7)^2}
         & =\frac{44t+324}{(t+3)(t+5)(t+6)(t+7)^2}>0
    \end{align*}
    \ignorespaces
    Also, using $(1+x)^{-2}\ge1-2x$ for $x\ge0$,
    $$\frac2{(t+7)^2}=\frac2{t^2}\left(1+\frac7t\right)^{-2}\ge\frac2{t^2}-\frac{28}{t^3}$$
    \ignorespaces
    Therefore,
    $$\frac2{t^2}-\frac{28}{t^3}\le a(t)\le\frac2{t^2}.$$
    In particular, the series defining $\kappa_{\mathrm{PP}}$ converges.

    Since $\log t/t^2$ and $\log t/t^3$ are positive and decreasing for $t\ge4$, comparison of sums with integrals gives
    $$2\int_{N+1}^{\infty}\frac{\log t}{t^2}\,dt -28\int_N^{\infty}\frac{\log t}{t^3}\,dt\le\rho_N\le2\int_N^{\infty}\frac{\log t}{t^2}\,dt.$$
    To obtain the lower bound, we bound the sum of the positive terms from below by an integral and the sum being subtracted from above by an integral.
    Integration by parts gives
    $$\int_b^{\infty}\frac{\log t}{t^2}\,dt=\frac{\log b+1}{b},\qquad\int_b^{\infty}\frac{\log t}{t^3}\,dt=\frac{\frac12\log b+\frac14}{b^2}$$
    We substitute these identities and then bound the logarithms using $\ell,h$.
    Define
    \begin{align*}
        \tau_N^- & =\frac{2(\ell(N+1)+1)}{N+1} -\frac{28}{N^2}\left(\frac{h(N)}{2}+\frac{1}{4}\right), \\
        \tau_N^+ & =\frac{2(h(N)+1)}{N}
    \end{align*}
    Then
    $$\tau_N^-<\rho_N<\tau_N^+$$
    \ignorespaces
    These lower and upper bounds are also rational.

    \paragraph{A certified interval for the infinite sum.}
    Setting $N=100000$, the bounds above give
    $$\sigma_N^-+\tau_N^-<\kappa_{\mathrm{PP}}<\sigma_N^++\tau_N^+$$
    \ignorespaces
    Set
    $$\kappa_-=\sigma_N^-+\tau_N^-,\qquad\kappa_+=\sigma_N^++\tau_N^+$$
    \ignorespaces
    We round the displayed decimal lower bounds downward and the upper bounds upward. For the evaluations of sums and exponentials, we use the rational bounds before rounding.
    Evaluating the finite sums using rational arithmetic gives
    \begin{align*}
        0.693722104135 & <\sigma_N^-<\sigma_N<\sigma_N^+<0.693722104136, \\
        0.000250239388 & <\tau_N^-<\rho_N<\tau_N^+<0.000250258510.
    \end{align*}
    Evaluating the rational bounds $\kappa_-,\kappa_+$ before rounding then yields the second bound in the lemma.

    \paragraph{The corresponding exponential base.}
    Finally, we certify the third bound in the lemma using rational arithmetic.
    For $0<z\le2$, set
    $$E_{40}(z)=\sum_{j=0}^{40}\frac{z^j}{j!},\qquad\varepsilon_{40}(z)=\frac{z^{41}}{41!}\frac1{1-z/42}$$
    \ignorespaces
    The first omitted term of the exponential series is $z^{41}/41!$, and the ratio of each subsequent term to its predecessor is at most $z/42$. Hence
    $$E_{40}(z)<e^z<E_{40}(z)+\varepsilon_{40}(z).$$
    By monotonicity of the exponential function and $0<2\kappa_-<2\kappa_+<2$,
    $$E_{40}(2\kappa_-)<e^{2\kappa_{\mathrm{PP}}}<E_{40}(2\kappa_+)+\varepsilon_{40}(2\kappa_+).$$
    Evaluating these two endpoints using rational arithmetic gives the third bound in the lemma.
\end{proof}

\end{document}